\documentclass{aastex}
\usepackage{spr-astr-addons}
\usepackage{url}
\usepackage{graphicx}
\usepackage{longtable}
\usepackage{comment}
\usepackage{natbib}
\usepackage{rotating}
\usepackage{verbatim}
\usepackage{morefloats}
\usepackage{lscape}
\usepackage{threeparttable}
\usepackage{wasysym}
\usepackage{txfonts}
\usepackage{multirow}
\usepackage{upgreek}
 \usepackage{booktabs}
\usepackage[verbose]{placeins}
\usepackage{blindtext}
\usepackage{appendix}
\usepackage{enumitem}
\usepackage{underscore}

\RequirePackage{color}

\newcommand{\emaila}{harold.penaherazo@unito.it}

\newcommand{\wse}{{\it WISE}}

\newcommand{\fer}{{\it Fermi}}

\newcommand{\mywidth}{9.0cm}

\usepackage{aas_macros}
\usepackage{lscape}

\begin{document}

\title{Optical spectroscopic observations of gamma-ray blazar candidates. IX. Optical archival spectra and further observations from SOAR and OAGH}
\shorttitle{Optical Campaign IX}
\shortauthors{Pe\~na-Herazo et al.}

\author{Pe\~na-Herazo, H. A.\altaffilmark{1,2,3} 
F. Massaro\altaffilmark{2,3,4,5},
V. Chavushyan\altaffilmark{1},
E. J. Marchesini\altaffilmark{2,3,6,7,8}, 
A. Paggi\altaffilmark{2,3,4,5}, 
M. Landoni\altaffilmark{9}, 
N. Masetti\altaffilmark{8,10}, 
F. Ricci\altaffilmark{11}, 
R. D'Abrusco\altaffilmark{12}, 
D. Milisavljevic\altaffilmark{12,13}, 
E. Jim\'enez-Bail\'on\altaffilmark{14}, 
F. La Franca\altaffilmark{15}, 
Howard A. Smith\altaffilmark{12} \& 
G. Tosti\altaffilmark{16}
} 
\email{\emaila}
\email{harold.penaherazo@unito.it}
\altaffiltext{}{H. A. Pe\~na-Herazo \\ harold.penaherazo@unito.it}
\altaffiltext{1}{Instituto Nacional de Astrof\'{i}sica, \'Optica y Electr\'onica, Apartado Postal 51-216, 72000 Puebla, M\'exico}
\altaffiltext{2}{Dipartimento di Fisica, Universit\`a degli Studi di Torino, via Pietro Giuria 1, I-10125 Torino, Italy}
\altaffiltext{3}{Istituto Nazionale di Fisica Nucleare, Sezione di Torino, I-10125 Torino, Italy}
\altaffiltext{4}{INAF-Osservatorio Astrofisico di Torino, via Osservatorio 20, 10025 Pino Torinese, Italy.}
\altaffiltext{5}{Consorzio Interuniversitario per la Fisica Spaziale (CIFS), via Pietro Giuria 1, I-10125, Torino, Italy.}
\altaffiltext{6}{Facultad de Ciencias Astron\'{o}micas y Geof\'{\i}sicas, Universidad Nacional de La Plata, La Plata, Argentina.}
\altaffiltext{7}{Instituto de Astrof\'{\i}sica de La Plata, CONICET-UNLP, CCT La Plata, La Plata, Argentina.}
\altaffiltext{8}{INAF -- Osservatorio di Astrofisica e Scienza dello Spazio, via Gobetti 93/3, I-40129, Bologna, Italy}
\altaffiltext{9}{INAF-Osservatorio Astronomico di Brera, Via Emilio Bianchi 46, I-23807 Merate, Italy}
\altaffiltext{10}{Departamento de Ciencias F\'isicas, Universidad Andr\'es Bello, Fern\'andez Concha 700, Las Condes, Santiago, Chile}
\altaffiltext{11}{Instituto de Astrof\'{\i}sica and Centro de Astroingenier\'{\i}a, Facultad de F\'{\i}sica, Pontificia Universidad Catolica de Chile, Casilla 306, Santiago 22, Chile}
\altaffiltext{12}{Center for Astrophysics $\mid$ Harvard \& Smithsonian, 60 Garden Street, Cambridge, MA 02138, USA}
\altaffiltext{13}{Department of Physics and Astronomy, Purdue University, 525 Northwestern Avenue, West Lafayette, IN 47907, USA}
\altaffiltext{14}{Instituto de Astronom\'{\i}a, Universidad Nacional Aut\'onoma de M\'exico, Apdo. Postal 877, Ensenada, 22800 Baja California, M\'exico}
\altaffiltext{15}{Dipartimento di Matematica e Fisica, Universit\`a Roma Tre, via della Vasca Navale 84, I-00146, Roma, Italy}
\altaffiltext{16}{Dipartimento di Fisica, Universit\`a degli Studi di Perugia, 06123 Perugia, Italy}

\begin{abstract}
\end{abstract}
Nearly one third of the sources in the \emph{Fermi}-LAT catalogs lacks a lower energy counterpart, hence being referred as unidentified/unassociated gamma-ray sources (UGSs). In order to firmly classify them, dedicated multifrequency follow-up campaigns  are necessary. These will permit to unveil their nature and identify the fraction that could belong to the class of active galaxies known as blazars that is the largest population of extragalactic $\gamma$-ray sources. In Fermi-LAT catalogs there are also gamma-ray sources associated with multifrequency blazar-like objects known as Blazars Candidates of Uncertain type (i.e., BCUs) for which follow up spectroscopic campaigns are mandatory to confirm their blazar nature. Thus, in 2013 we started an optical spectroscopic campaign to identify blazar-like objects potential counterparts of UGSs and BCUs. Here we report the spectra of 31 additional targets observed as part of our follow up campaign. Thirteen of them are BCUs for which we acquired spectroscopic observations at Observatorio Astrof\'isico Guillermo Haro (OAGH) and at Southern Astrophysical Research Observatory (SOAR) telescopes, while the rest has been identified thanks to the archival observations available from the Sloan Digital Sky Survey (SDSS). We confirm the blazar nature of all BCUs: three of them are in blazar of quasar type (BZQs) while the remaining ones can be spectroscopically classified as BL Lac objects (BZBs). Then we also discovered 18 BL Lac objects lying within the positional uncertainty regions of UGSs that could be their potential counterparts.

\keywords{galaxies: active - galaxies: BL Lacertae objects - quasars: general - surveys - radiation mechanisms: non-thermal}

\section{Introduction}
\label{sec:intro}
Since the launch of the first $\gamma$-ray satellites, as the Energetic Gamma Ray Experiment Telescope (EGRET, \citeauthor{thompson93}~\citeyear{thompson93}) on board the Compton Gamma-ray Observatory, the positional uncertainty of $\gamma$-ray sources was orders of magnitude larger than that at lower frequencies, making nearly $\sim$60\% of the sources listed in the third EGRET catalog lacking low energy counterparts \citep{hartman99}. Thus source identification were and are still among the most challenging tasks in modern $\gamma$-ray astronomy making the analysis of the unidentified/unassociated $\gamma$-ray sources (UGSs) still of one the main scientific objectives for $\gamma$-ray missions \citep[see e.g.,][]{atwood09}.

Nowadays, thanks to i) the better spatial resolution of the \emph{Fermi} - Large Area Telescope \citep[\emph{Fermi}-LAT; see e.g.,][]{atwood09}, ii) new methods (see e.g, \citeauthor{mirabal09}~\citeyear{mirabal09}; \citeauthor{abdo10a}~\citeyear{abdo10a}; \citeauthor{abdo10b}~\citeyear{abdo10b}; \citeauthor{ackermann12}~\citeyear{ackermann12};
\citeauthor{hassan13}~\citeyear{hassan13};
\citeauthor{doert14}~\citeyear{doert14}; \citeauthor{wibrals}~\citeyear{wibrals}) and procedures developed and iii) extensive observational campaigns from radio to the X-ray energy ranges (see e.g. \citeauthor{schinzel17}~\citeyear{schinzel17}; \citeauthor{paggi13}~\citeyear{paggi13};  \citeauthor{paiano17c}~\citeyear{paiano17c};
\citeauthor{kovalev09}~\citeyear{kovalev09};
\citeauthor{hovatta12}~\citeyear{hovatta12};
\citeauthor{cheung12}~\citeyear{cheung12};
\citeauthor{petrov11}~\citeyear{petrov11};
\citeauthor{stroh13}~\citeyear{stroh13};
\citeauthor{shaw13}~\citeyear{shaw13};
\citeauthor{kataoka12}~\citeyear{kataoka12};
\citeauthor{ugs2}~\citeyear{ugs2};
\citeauthor{massaro13}~\citeyear{massaro13}; \citeauthor{landoni15b}~\citeyear{landoni15b}), our capabilities to address this source association problem significantly improved.

In all \emph{Fermi} source catalogs the fraction of UGSs is almost constant. In the First \emph{Fermi}-LAT source catalog (1FGL) UGSs are 43\% (\citeauthor{abdo10a}~\citeyear{abdo10a}); this fraction reduces to 35\% in the Second \emph{Fermi}-LAT source catalog (2FGL; \citeauthor{nolan12}~\citeyear{nolan12}). However, given the large amount of efforts of several groups and teams worldwide (\citeauthor{schinzel17}~\citeyear{schinzel17}; \citeauthor{shaw13b}~\citeyear{shaw13b}; \citeauthor{hovatta14}~\citeyear{hovatta14}; \citeauthor{takahashi12}~\citeyear{takahashi12}; \citeauthor{maeda11}~\citeyear{maeda11}; \citeauthor{petrov13}~\citeyear{petrov13}; \citeauthor{review}~\citeyear{review}; \citeauthor{sandrinelli13}~\citeyear{sandrinelli13}) to search for low energy counterparts of UGSs, the fraction of UGSs listed in the first releases of Fermi catalogs (i.e., 1FGL and 2FGL) are now less than 10\% of the detected sources. On the other hand, in the latest release of the third \emph{Fermi}-LAT source catalog (3FGL; \citeauthor{acero15}~\citeyear{acero15}), listing more than 3033 sources, the UGS fraction is again back to $\sim$33\%. This motivates the continuous need of additional follow up observations and the development of new methods to search for UGSs counterparts.

It is worth noting that blazars, a peculiar class of radio-loud active galactic nuclei (AGNs), are the largest known population of $\gamma$-ray emitters, being 85\% of the associated sources listed in 3FGL and $\sim$36\% of the whole catalog. Thus, we could expect that a significant fraction of UGSs, at least at high Galactic latitudes (i.e. $|b|>20^{\circ}$), could be then associated with blazar-like sources, as previously occurred.

Therefore, in the last decade multifrequency observational campaigns developed for searching blazar-like counterparts of UGSs were successfully carried out: as at radio frequencies (\citeauthor{kovalev09}~\citeyear{kovalev09}; \citeauthor{petrov13}~\citeyear{petrov13}; \citeauthor{schinzel15}~\citeyear{schinzel15}; \citeauthor{schinzel17}~\citeyear{schinzel17}) or thanks to infrared colors of $\gamma$-ray blazar candidates (\citeauthor{paper1}~\citeyear{paper1}; \citeauthor{paper2}~\citeyear{paper2}; \citeauthor{dabrusco16}~\citeyear{dabrusco16}) as well as using optical and X-ray observations (\citeauthor{paggi13}~\citeyear{paggi13}; \citeauthor{takeuchi13}~\citeyear{takeuchi13}). Candidate counterparts of UGSs were then confirmed thanks to extensive optical spectroscopic campaigns (\citeauthor{quest16}~\citeyear{quest16}, \citeauthor{paiano17a}~\citeyear{paiano17a}, \citeauthor{paiano17b}~\citeyear{paiano17b}, \citeauthor{paiano17c}~\citeyear{paiano17c}, \citeauthor{landoni18}~\citeyear{landoni18}, \citeauthor{marchesi18}~\citeyear{marchesi18}).

UGSs are not the only class of $\gamma$-ray sources with unknown/unconfirmed nature. In particular, among the associated sources there are the Blazar candidates of uncertain type (BCUs). These are radio, infrared or X-rays sources, associated with \emph{Fermi}-LAT sources, but showing a multifrequency behavior resembling that of blazars, as, for example, having a flat radio spectrum and/or a double-bumped Spectral Energy Distributions (SED). However all BCUs lack optical spectroscopic observations that could unequivocally confirm their blazar nature \citep[see][for more details]{ackermann15a}. 

According to the AGN unification scenario, blazar emission, mainly non-thermal, is due to particles accelerated in a relativistic jet pointing at a small angle with respect to our line of sight (\citeauthor{blandford78}~\citeyear{blandford78}; \citeauthor{urry95}~\citeyear{urry95}). Blazar radiation spans the entire electromagnetic spectrum. Their SEDs consists of two bumps, the low energy one peaking between the infrared and the optical band while the high-energy one in the X-ray or even $\gamma$-ray energies (\citeauthor{giommi95}~\citeyear{giommi95}; \citeauthor{fossati98}~\citeyear{fossati98}; \citeauthor{abdo10b}~\citeyear{abdo10b}; \citeauthor{abdo10c}~\citeyear{abdo10c}).

Blazars are classified on the basis of rest frame equivalent width ($\mathrm{EW}$) of their optical emission/absorption lines, whenever detectable, referring to BL Lac objects when this is less than 5 \AA$\,$ and to Flat Spectrum Radio Quasars (FSRQs) when they show a typical quasar-like optical spectrum \citep{stickel91}. In the present work we adopt the nomenclature of the fifth edition of the Roma-BZCAT, labelling BL Lac objects as BZBs and FSRQ as BZQs \citep{massaro09,massaro15b}. It is then worth noting that given weak optical features, sometimes present in BL Lacs spectra that also shows a variable continuum on timescales of hours to minutes, obtaining their redshift estimation is always challenging task (\citeauthor{refined}~\citeyear{refined}).

The blazar subclass of BZBs is the largest within the detected classes of \fer\ sources and its number is continuously growing in the latest releases of \fer\ catalogs (\citeauthor{nolan12}~\citeyear{nolan12}; \citeauthor{acero15}~\citeyear{acero15}). Thus UGSs for which a BZB is discovered within their positional uncertainty regions have high chance to be associated in future release of the \fer\ catalog. Thus one of the main aims of the analysis presented here is to search for spectroscopically confirmed BL Lac objects lying within the $\gamma$-ray positional ellipses of Fermi UGSs that, being included in the future release of the Roma-BZCAT, could be then result as associated counterparts of \fer\ sources \cite{ackermann15a}. As secondary goal we also report here new optical spectroscopic observations of a selected sample of BCUs aiming at classifying them. 

In summary the current paper present an analysis of optical spectroscopic observations, both using archival, taken from the latest data release of the Sloan Digitized Sky Survey Data Release 14 (SDSS DR14; \citeauthor{dr14}~\citeyear{dr14}) and follow up observations carried out with SOAR and OAGH. These analysis could permit us: i) confirm the nature of a selected sample of BCUs and ii) search for blazar-like potential candidates of UGSs listed in the \emph{Fermi}-LAT 8-year Source List~\footnote{https://fermi.gsfc.nasa.gov/ssc/data/access/lat/fl8y/} (FL8Y), based on the first eighth science years of \emph{Fermi}-LAT. The FL8Y list will be soon replaced by the 4FGL catalog. Thus our analysis will be timely used to associate \fer\ sources in the next release of the \emph{Fermi} source catalog.

The paper is organized as follows: \S~\ref{sec:campaign} we present the current status of our observational campaign and the main results achieved to date. In \S~\ref{sec:sample} we describe the sample selection while section~\ref{sec:datared} is devoted to methods and data reduction procedures adopted int eh current analysis. \S~\ref{sec:results} report all results achieved thanks to our analysis while \S~\ref{sec:summary} is dedicated to Summary \& Conclusions.

We used cgs units unless stated otherwise, and spectral indices, $\alpha$, are defined by flux density, S$_{\nu}\propto\nu^{-\alpha}$, considering sources with a flat spectrum, mainly in the radio band, those having  $\alpha<$0.5. WISE magnitudes in the [3.4], [4.6], [12], and [22] $\mu$m nominal filters are in the Vega system, these are not corrected for the Galactic extinction since such correction is generally negligible at Galactic latitudes above 10$^\circ$ being of the order of a few percent \citep{ugs1}.

\section{State-of-art of our spectroscopic campaign}
\label{sec:campaign}
The goal of our spectroscopic campaign is to i) confirm the blazar nature of BCUs associated in the \emph{Fermi}-LAT source catalogs and ii) discover potential counterparts of the UGSs selected with different procedures based on radio, infrared and X-ray observations. Here we provide a brief summary of all results recently achieved thanks to our follow up spectroscopic observations. Our spectroscopic campaign is still ongoing and is among several worldwide efforts to reduce the percentage of UGSs and BCUs in the \fer\-LAT catalogs.

We remark that to mitigate the impact on telescopes schedules we tend to observe small samples of BCUs and UGS candidate counterparts during each observing run. Our observations are also combined with an extensive search in optical databases as the SDSS \citep{alam15} and/or the Six-Degree Field Galaxy Survey \citep{jones09} (see e.g., \citeauthor{massaro14}\citeyear{massaro14}; \citeauthor{refined}\citeyear{refined}; \citeauthor{crespo16c}\citeyear{crespo16c}; \citeauthor{quest16}\citeyear{quest16}).

Since the beginning of our campaign, in 2013, we observed a total of 329 unique targets, 122 already known as blazars and listed in the Roma-BZCAT. The remaining ones contains: 68 UGSs; 132 BCUs; 4 AGUs, former nomenclature for BCUs; 2 sources classified as BL Lac and 1 as FSRQ but for which we could not find the optical spectrum in the literature. We found 254 sources classified as BZBs, in addition to 34 BZQs, 17 BZGs and 23 quasars that lacking radio information were not confirmed BZQs but showed the same infrared colors of known Fermi blazars. We obtained a firm redshift measurement for 36 sources given absorption/emission features detected in their optical spectra and additional 51 uncertain 
estimates. Two BZBs, for which the $z$ was estimated, lie at $z>1$, being among the most distant known BL Lacs objects confirmed to date (\citeauthor{pena17}\citeyear{pena17}; \citeauthor{marchesini19}\citeyear{marchesini19}). As result, all spectra of previous observations of our follow-up campaign were published to date (e.g. \citeauthor{paggi14}\citeyear{paggi14}; \citeauthor{massaro15c}\citeyear{massaro15c}; \citeauthor{landoni15}\citeyear{landoni15}; \citeauthor{ricci15}\citeyear{ricci15}; \citeauthor{crespo16a}\citeyear{crespo16a}; \citeauthor{crespo16b}\citeyear{crespo16b}; \citeauthor{pena17}\citeyear{pena17}; \citeauthor{marchesini19}\citeyear{marchesini19}).

\section{Sample description}
\label{sec:sample}
Sources analyzed here are divided in two main samples. The first lists all BCUs visible during the observing nights available to our group in 2018 at SOAR and OAGH telescopes, while the second lists all UGSs, lying in the SDSS footprint, having at least an optical source with a spectrum available. More details are provided below.

\subsection{Unidentified Gamma-ray Sources}
We first selected a sample of UGSs focusing on searching BL Lac objects lying within their positional uncertainty regions. The same strategy, already successfully carried out in \citeauthor{crespo16c}~(\citeyear{crespo16c}), helped us to find new BZBs that were and will probably be associate in the future releases of \fer\ catalogs. To carry out the present search we used the latest releases of the SDSS archival observations (i.e., DR14) in combination with the latest list of \fer\ sources, correspondent to the FL8Y, both unavailable when we performed our previous analyses. 

Thus we selected all 357 UGSs lying in the footprint of the SDSS and then for all of them we searched for sources having an optical spectrum available. Our final sample of UGSs inspected lists 166 UGSs.

\subsection{Blazar Candidates of Uncertain type}
BCUs, as described in the \fer\ catalogs, is a tentative classification for associated counterparts that i) shows a blazar-like multifrequency behavior or ii) are classified as Blazars of Uncertain type in the Roma-BZCAT. They are divided in three types \citep{ackermann15a}, namely: BCUs I, those with a counterpart with available optical spectrum but without a good signal-to-noise ratio to classify them either as BZBs or BZQs; BCUs II, lacking optical spectrum in the literature but having an estimate of the peak energy of their plausible synchrotron component; BCU III, blazar-like multifrequency emission and a flat radio spectrum but lacking an optical spectrum and for which it was not possible to estimate the synchrotron peak frequency.

Thirteen BCUs, previously associated with radio, infrared and/or X-ray counterparts have been pointed with SOAR and OAGH. All of them belong to the FL8Y list,
with six of them in 3FGL
and their selection was mainly driven by visibility and expected magnitude (i.e. they are all brighter than 18.5 mag in B band). Observation logs are reported in Table~\ref{tab:main}.

\section{Methods and data reduction procedures}
\label{sec:datared}

\subsection{SDSS spectroscopic data of UGSs}
We visually inspected all spectra of SDSS sources lying within the positional uncertainty regions, at 95\% level of confidence, of our UGS selected sample. Then, according to our previous analyses, we selected only those optical sources with blue featureless-like optical spectrum, facilitating its classification as BZBs (i.e. sources with blue optical spectrum with spectral features having $EW < 5$ \AA). We did not consider those sources with optical spectral dominated by their host galaxy. We measured the $EW$ of all emission/absorption lines to confirm their nature.

\begin{figure*}{}
\begin{center}
\includegraphics[width=15cm]{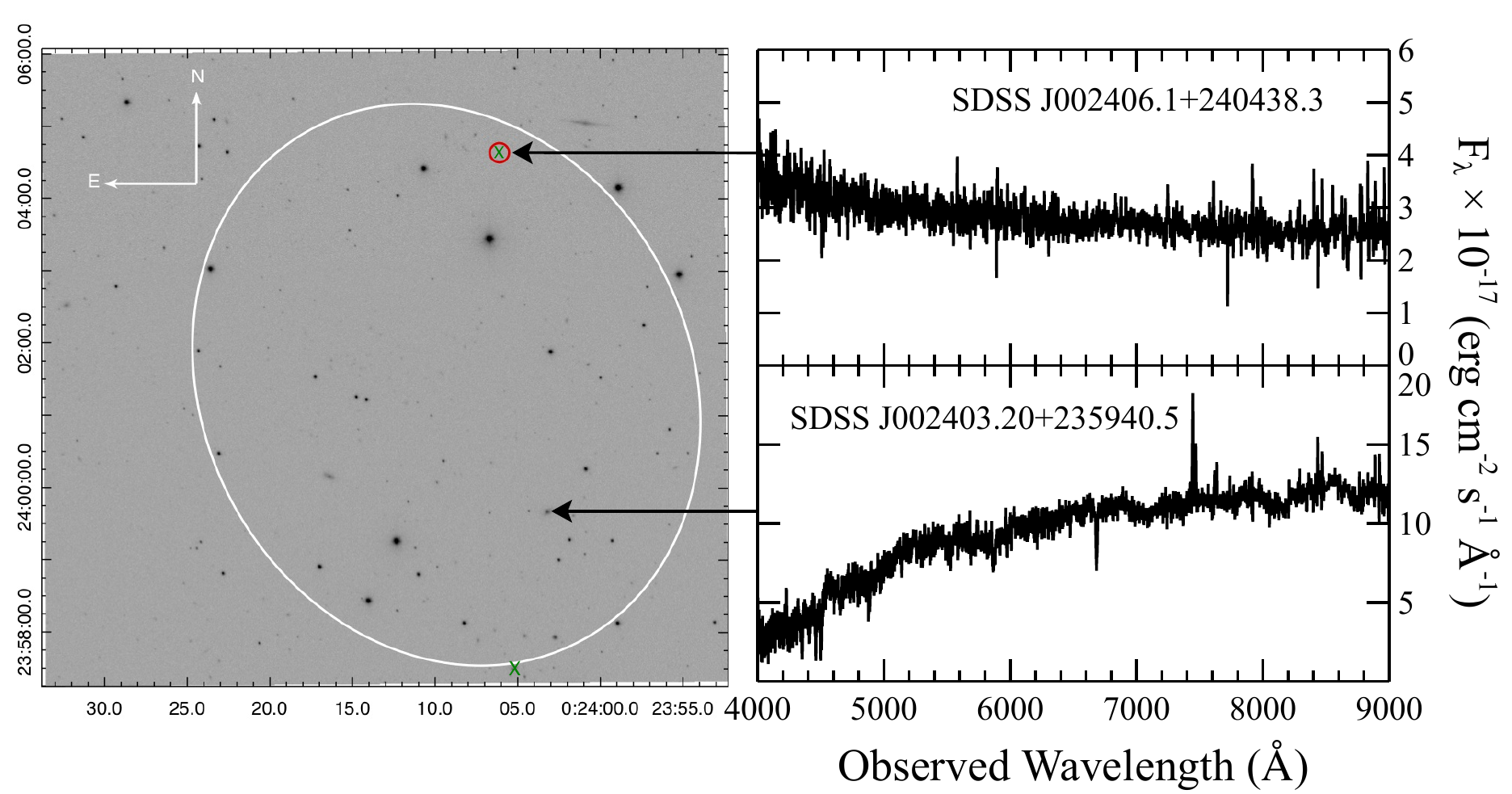} 
\end{center}
\caption{We show the optical image of the UGS: FL8Y J0024.1+2401 taken in the $i$ filter and available in the SDSS archive.
The white ellipse marks the \emph{Fermi}-LAT positional uncertainty region at a 95\% confidence. The red circle indicates the position of the BZB SDSS J002406.1+240438.3 identified and classified thanks to our analysis. Green crosses point all radio sources reported in NVSS catalog. Insets shows all the optical spectra of SDSS sources inspected during our analysis, SDSS J002406.1+240438.3 in the top panel and that of a normal elliptical galaxy in the bottom one.}
\label{fig:sdss}
\end{figure*}

In Figure~\ref{fig:sdss} we show the case of FL8Y J0024.1+2401, for which we analyzed all spectra available within the \fer\ -LAT positional uncertainty region of the FL8Y list including stars, a late type galaxy and the, previously unknown, BZB. We also report here the location of all NVSS sources present in the field, where only a single radio source lies within the positional uncertainty region at 95\% level of confidence highlighted with the white ellipse. This radio source is the counterpart of the optical object having the typical spectrum of a BL Lac, namely: SDSS J002406.1+240438.3. This strengthen our  classification, since almost all BL Lacs known to date have radio counterparts with only a few exception (see e.g. \citeauthor{radioweak}~\citeyear{radioweak}). Thus we claim that SDSS J002406.1+240438.3 could be a potential counterpart of the UGS, which association could be confirmed in the upcoming release of the \emph{Fermi} catalog.

It is worth noting that we only searched for BZBs dominated by non-thermal emission in the optical band, thus radio sources with spectra dominated by their host elliptical galaxy in the visible light (i.e., as for BZG, see Roma-BZCAT for details on their classification), were discarded in the present analysis. Searching for BZQs in optical surveys is less efficient since they all appear as normal quasars in the optical band and the lack of radio observations, necessary to verify if their spectral index is flat, prevent us to confirm their nature with a similar analysis. All newly discovered optical BZBs lying within the $\gamma$ uncertainty regions of UGS could be their potential counterparts.  

\subsection{Newly acquired optical spectroscopic observations of BCUs}
Here we report the results of pointed spectroscopic observations carried out in 2018 at SOAR and OAGH for which we only observed a BCUs. In Table~\ref{tab:main} we report the log of all observations while in the following we provide all basic details of data acquisition and reduction procedures.

Optical spectroscopic observations of four sources were collected at Southern Astrophysical Research Telescope (SOAR) 4.1m telescope, at Cerro Pach{\'o}n, Chile in remote observing mode  on May 31$^{st}$ of 2018.  The observed targets are: NVSS J160005-252439, PKS 2043-682, PMN J2103-6232 and PMN J2211-7039. We used single, long slit mode of the Goodman High Throughput spectrograph (\citeauthor{clemens04}~\citeyear{clemens04}), with a slit width of $1\farcs 0$ and the 400 l mm$^{-1}$ grating giving a dispersion of $\sim 2$ \AA$\,$ pixel$^{-1}$. The observed spectral range was 4000 to 7000 \AA$\,$ and finally we obtained a resolution of 6 \AA.

Nine sources were observed with the 2.1 m telescope at OAGH in Cananea, Mexico, using the Boller \& Chivens spectrograph between April 14$^{th}$ and April 16$^{th}$, 2018. Collected spectra have a wavelength range from 4000 \AA$\,$ to 7000 \AA$\,$ and were acquired with a slit width of $2\farcs5$, the final resolution is of 14 \AA.

We acquired three different exposures for all our targets. Data reduction was then carried out using IRAF standard reduction packages \citep{tody86}. We performed bias and flat fielding corrections, and cosmic rays removal according to our standard procedure (see e.g. \citeauthor{pena17}~\citeyear{pena17}). Cosmic rays removal was performed using L.A. Cosmic IRAF algorithm \citeauthor{vandockum01}~\citeyear{vandockum01}. Furthermore, spectra were also calibrated using a Hg-Ar or He-Ne-Ar lamp acquired for each source. 

In addition, during each night, we observed at least one standard spectrophotometric star to accomplish relative flux calibration.

All spectra presented in the current analysis were corrected by galactic extinction using the reddening law of \citeauthor{cardelli89} \citeyearpar{cardelli89} and values of $E_{\bv}$ computed by \citeauthor{schlegel98} \citeyearpar{schlegel98} and reddening and extinction maps of \citeauthor{schlafly11}~\citeyear{schlafly11}. Finally, we performed box smoothing for visual representation and, to highlight  spectroscopic features, we normalized the spectra to the local continuum.

Additional details on the data reduction procedures can be found in previous analyses of our follow up campaign (see. e.g~\citeauthor{landoni15}\citeyear{landoni15}; \citeauthor{ricci15}\citeyear{ricci15}; \citeauthor{crespo16a}\citeyear{crespo16a}; \citeauthor{pena17}\citeyear{pena17}; \citeauthor{marchesini19}\citeyear{marchesini19}).

\section{Results}
\label{sec:results}
Results of current analysis are reported below for each sample separately (i.e.,  UGSs first and BCUs later). In Figure Figure~\ref{fig:J1113} and from Figure~\ref{fig:J002} to Figure~\ref{fig:J2244} we show the optical spectra of UGSs potential counterparts in SDSS and from Figure~\ref{fig:J1046} to Figure~\ref{fig:PMNJ2103} we show the optical spectra and finding chart for those targets observed at OAGH and SOAR, all ordered by right ascension. Results are then summarized in Table~\ref{tab:main}, with spectral line measurements in Table~\ref{tab:measures}.

\subsection{Searching for Gamma-ray BL Lac candidates using archival observations}
\label{sec:sdss}
To carry out this investigation we searched in the footprint of SDSS DR14 all sources having an optical spectrum and lying with the positional uncertainty of UGSs. The total number of UGSs lying in the SDSS footprint is 357, however only 166 of them have at least an optical source with a spectroscopic observation within the positional uncertainty region of \fer\ .  For only 18 out of 166 we found that one of the spectroscopic sources is classifiable as BZB. For the resting 148 UGSs we does not exclude the possibility of having a blazar  within the \fer\ positional uncertainty  region.

All these spectroscopic identifications are based on the measurements of the $\mathrm{EW}$ of emission/absorption features found in the optical spectra or on the lack of them. 
We obtained a $z$ estimate for SDSS J111346.03+152842.9 that appear to lie at $z=0.2589$. Our result is based on the identification of Ca II H\&K ($EW$=8-7 $\pm 2$ \AA), G band ($EW$=9 $\pm 2$ \AA), H$\beta$ ($EW$=13 $\pm 2$ \AA) and Mg I$\lambda5174$ ($EW$=7 $\pm 2$ \AA) absorption lines. However, the BZB SDSS J111346.03+152842.9 spectrum also shows spectral features of a foreground star, as shown in Figure~\ref{fig:J111346}. Additionally, we found an lower limit redshift estimation of SDSS J150316.57+165117.7 by identification of Mg II and Fe II multiplets of an intervening absorption system at $z=0.972$. Finally, we find a tentative estimate of the redshift of SDSS J223704.78+184055.9 at $z=0.724$ by the possible identification of Ca II H\&K ($EW$=3-2  $\pm 1$ \AA) and G band ($EW$=2 $\pm 1$ \AA) absorption lines.

\begin{figure*}{}
\begin{center}$
\begin{array}{cc}
\includegraphics[width=\mywidth]{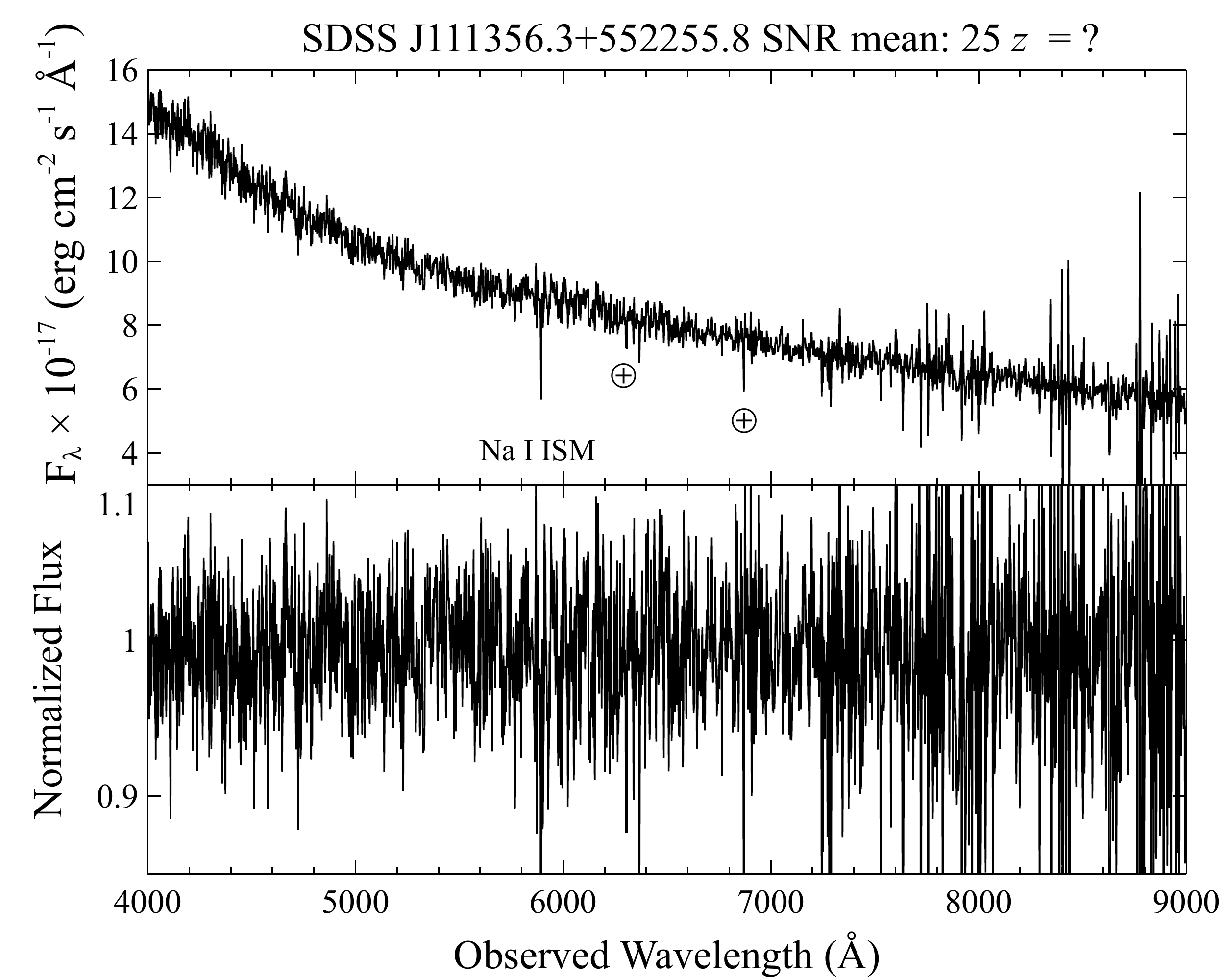} &
\includegraphics[clip=true, width=7cm]{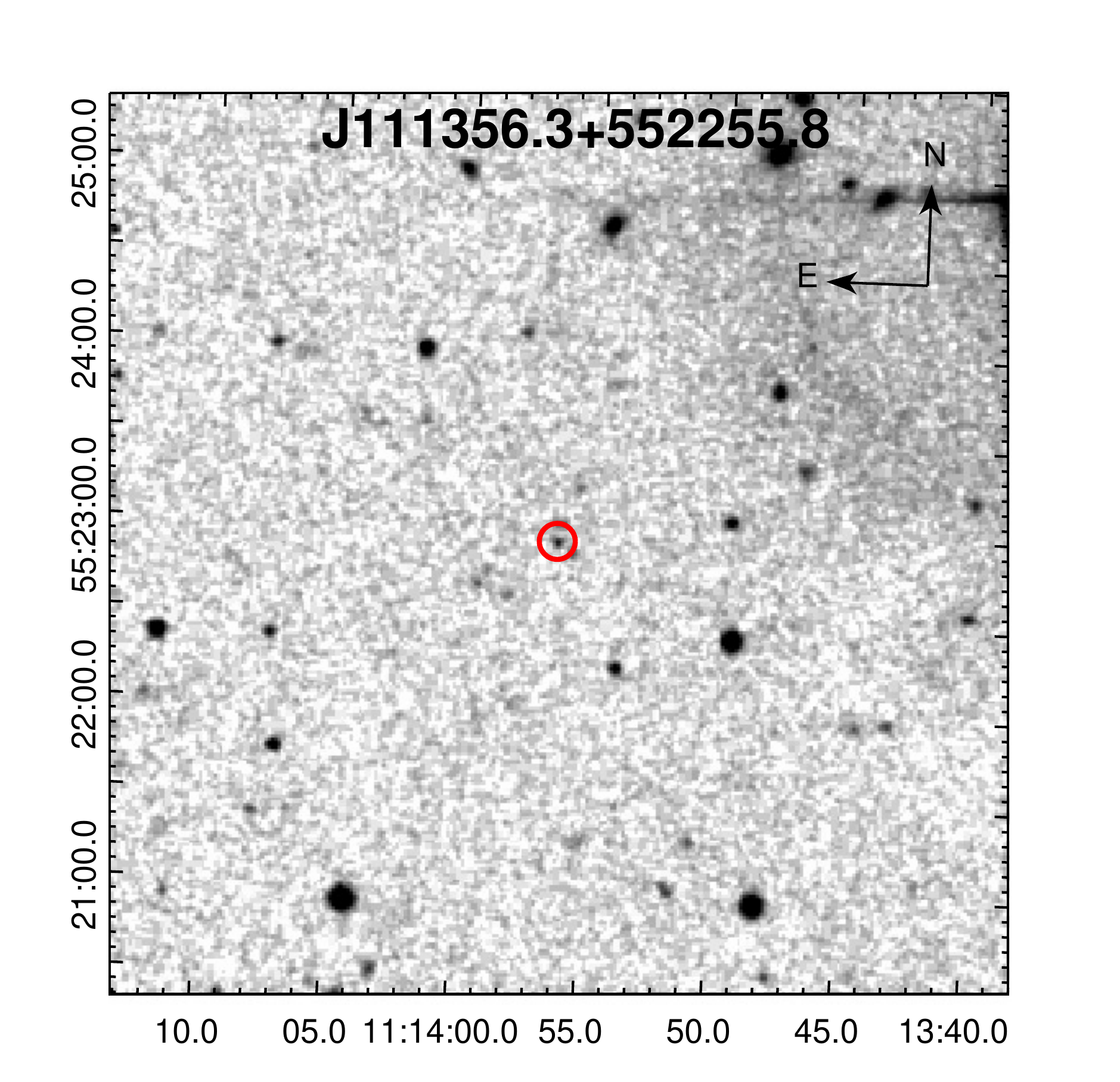} \\
\end{array}$
\end{center}
\caption{(Left panel) Optical spectrum of  SDSS J111356.3+552255.8 potential candidate of the UGS FL8Y J1113.8+5524, in the upper part it is shown the Signal-to-Noise Ratio of the spectrum. (Right panel) The finding chart ( $5'\times 5'$ ) retrieved from the Digitized Sky Survey (DSS) highlighting the location of the potential counterpart:  SDSS J111356.3+552255.8 (red circle).}
\label{fig:J1113}
\end{figure*}

Additional detailed information for all 18 BZBs are available in Table~\ref{tab:main} and Table~\ref{tab:measures} while their optical spectra are shown in Figure~\ref{fig:J1113} and from Figures \ref{fig:J002} to \ref{fig:J2244}.

Since the SDSS has the same footprint of the FIRST survey, we confirm that all newly discovered BZBs have a radio counterpart, as expected for this radio-loud AGN population. We also confirm that all BZBs newly discovered have a classical double bumped SED. In addition all 18 targets have both IR colors of gamma-ray BL Lacs (\citeauthor{connect}~\citeyear{connect}) but also those in the optical band are consistent with the BZB distribution (\citeauthor{massaro12b}~\citeyear{massaro12b}).

To provide additional evidence that our newly discovered BZBs are potential counterparts of \fer\ sources we also computed their \wse\ infrared colors. Association radius adopted to search the \wse\ counterpart of the optical SDSS source was set to $3\farcs3$ as reported in \citep{ugs1}.

\begin{figure}[ht]
\begin{center}
\includegraphics[width=8cm]{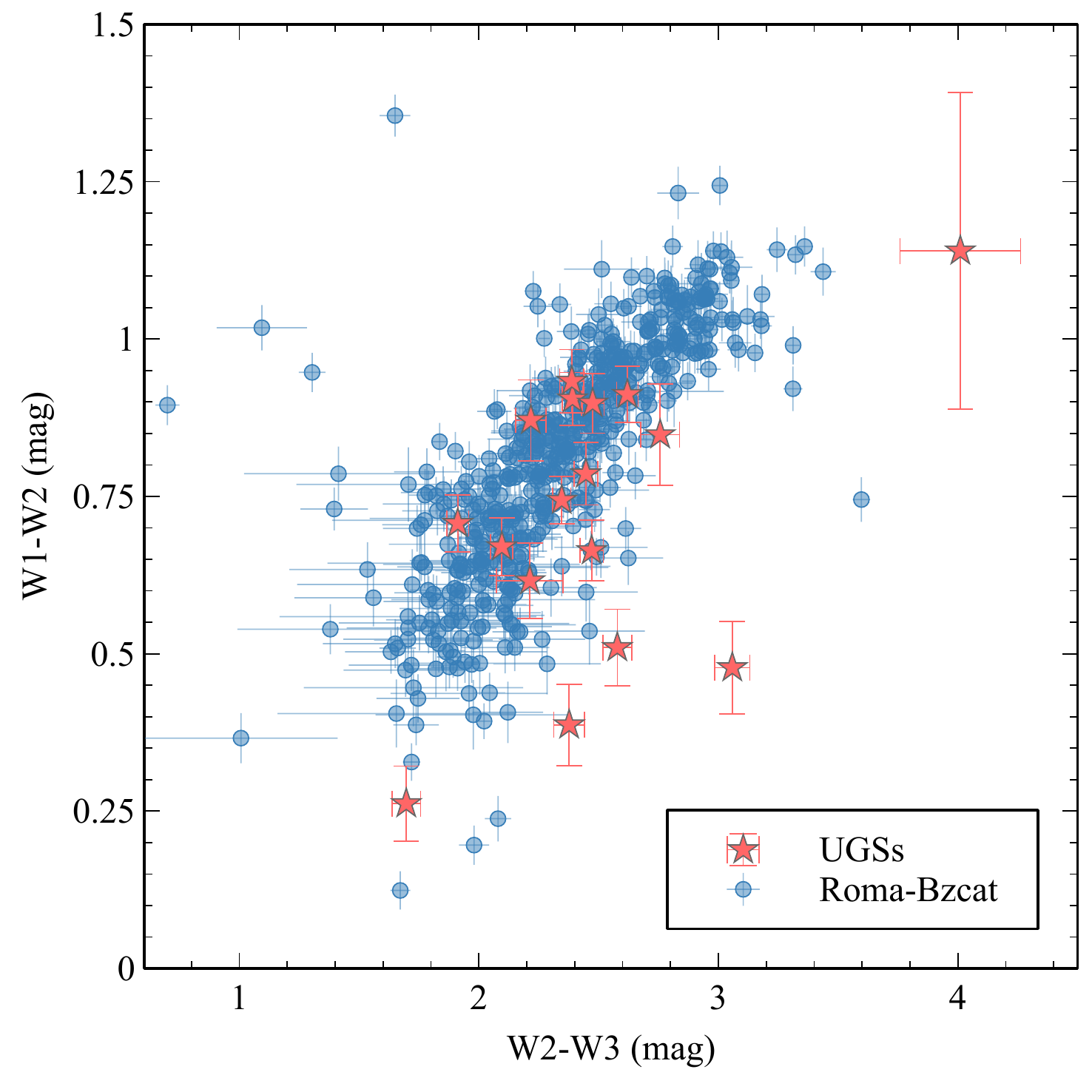} 
\end{center}
\caption{The comparison between the mid-IR WISE colors of the Fermi BZBs associated from the Roma-BZCAT and the newly discovered BZBs located within the positional uncertainty region of the selected UGSs.}
\label{fig:wisecol}
\end{figure}

Motivated by the fact that \fer\ blazars occupy a distinctive area in the WISE color-color diagram (\citeauthor{paper1}~\citeyear{paper1}), we verified that the location of the 18 newly identified BZBs, in the WISE color-color diagram $W1[3.4 \mu m]- W2[4.6\mu m]-W3[12\mu m]$, is consistent with the well-known \wse\ Gamma-ray Strip \citep{massaro12c}. In Figure~\ref{fig:wisecol} we show that 17 out of 18 newly discovered BZBs, all UGS potential counterparts, clearly overlay with the \wse\ Gamma-ray Strip, having \wse\ colors consistent with those of BZBs listed the Roma-BZCAT associated in the 3FGL. The BZB SDSS J220652.9+221722.2 is not shown on the mid-IR color diagram since it lacks a \wse\ counterpart within $3\farcs3$. The location of the newly discovered BZBs suggests that infrared emission of these potential counterparts is dominated by non-thermal radiation, as expected for $\gamma$-ray BZBs, thus strengthening our results.

As additional test, we also compared the distribution of the angular separation between the position of the optical source and that of the \fer\ UGS for all new BZBs and all BZBs, belonging to the Roma-BZCAT and associated in the 3LAC. 

\begin{figure}[ht]
\begin{center}
\includegraphics[width=8cm]{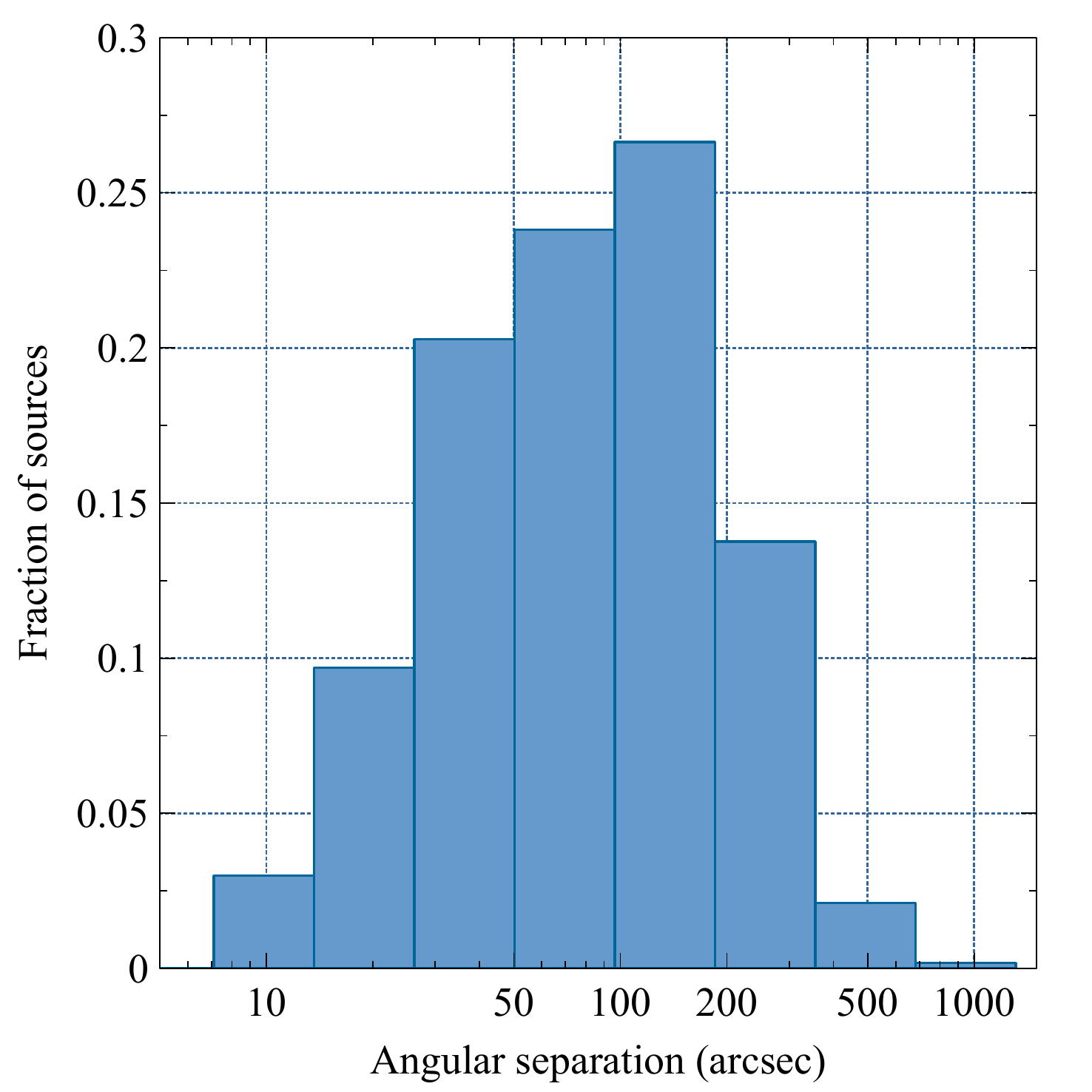} 
\end{center}
\caption{Distribution of the angular separation between \fer\ -LAT and associate counterpart position of the Roma-BZCAT BZBs.}
\label{fig:angsep}
\end{figure}

In Figure~\ref{fig:angsep} we show the two distributions. It is worth noting that all 18 BZBs, lying within the positional uncertainty ellipses of UGSs, have angular separation within the 98th percentile of the previous distribution with the only exception of SDSS J172100.07+251249.7 for which this angular separation is $15.5\arcmin$.

\subsection{Optical spectroscopic observations of Blazar Candidates of Uncertain type}
\label{sec:bcus}
Thanks to our optical spectra collected at SOAR and OAGH, we are also able to confirm the blazar nature of all 13 BCUs in our selected sample. 

In particular, three BCUs shows a clear quasar-like optical spectrum, leading to a BZQ classification, namely: VCS J1157+1638 counterpart of FL8Y J1157.5+1639, having a broad MgII$\lambda2798$ ($EW$=27 $\pm 6$ \AA) emission line and FeII$\lambda2964$ ($EW$=7 $\pm 5$ \AA) at a redshift of 1.058; AT20G J131938-004939 associated with FL8Y J1319.5-0046 showing MgII$\lambda2798$ ($EW$=60 $\pm 4$ \AA), FeII$\lambda2964$ ($EW$=5 $\pm 3$ \AA) emission lines and a redshift estimation of 0.890 and Q 1326-0516 counterpart of FL8Y J1329.4-0530 with an MgII$\lambda2798$ ($EW$=33 $\pm 7$ \AA) and an FeII$\lambda2964$ ($EW$=3 $\pm 5$ \AA) emission line at a redshift of $0.575$.

The remaining 10 BCUs were indeed all classified as BZBs given their quasi-featureless spectra. We were able to estimate redshifts for 3 of them. In particular, CRATES J104630+544953 associated with FL8Y J1046.1+5449 lies at z=0.249 and shows Ca II H\&K emission lines ($EW$=7-10 $\pm 2$ \AA) and [OIII] doublet emission lines ($EW$=4-6 $\pm 3$ \AA); RFC J1231+3711 counterpart of FL8Y J1231.1+3711 shows [OII] emission line ($EW$=8 $\pm 3$ \AA) and the absorption lines of Ca II H\&K and G band leading to a redshift estimate of $0.219$ and RFC J1808+3520 counterpart of FL8Y J1808.9+3522 for which the optical spectrum presents Ca II H\&K ($EW$=5-4 $\pm 3$ \AA), G band ($EW$=4 $\pm 2$ \AA) and H$\beta$ ($EW$=19 $\pm 3$ \AA) absorption lines allowing us to obtain a redshift estimate of $0.142$. 

Sources with redshift estimates were all acquired at OAGH and their redshift uncertainties are of the order of 0.004.

\section{Summary and Conclusions}
\label{sec:summary}
This work present a spectroscopic analysis of i) a selected sample of UGSs listed in FL8Y to search for blazar-like potential counterparts lying within their positional uncertainty regions along with ii) new observations acquired with SOAR and OAGH telescopes to confirm the nature of a selected sample of BCUs. In this paper, the ninth of our optical campaign, we present a total of 31 optical spectra analyzed, consisting in i) data archive retrieval of 18 potential counterparts of UGSs and ii) pointed spectroscopic observations of 13 BCUs.

We highlight that previous results of our optical spectroscopic campaign were already extensively used to i) increase the number of sources listed in the Roma-BZCAT and ii) associate $\gamma$-ray sources in the lat releases of \emph{Fermi}-LAT source catalogs.

Our results are summarized as follows: 
\begin{itemize}
\item For the UGS sample, selected out the preliminary list of the forthcoming \emph{Fermi}-LAT 4FGL catalog, we discovered 18 new BZBs out of 166
UGSs inspected, all lying in the SDSS DR14 footprint and within the $\gamma$-ray positional uncertainty regions at 95\% level of confidence.
\item We analyzed spectra of 13 BCUs observed in OAGH and SOAR telescope. Three of them shows quasar-like spectra, namely: VCS J1157+1638, AT20G J131938-004939 and Q 1326-0516, thus being BCU in the latest release of the \fer\ catalog can be all classified as BZQs. The remaining 10 sources have all optical spectra featureless as generally occurs for BZBs, with only three exceptions for which we obtained a redshift estimate, namely: CRATES J104630+544953 at z=0.249; RFC J1231+3711 at $z=0.219$; RFC J1808+3520 at $z=0.142$.
\end{itemize}

In particular, within the sample of newly discovered BZBs we estimated the redshift of SDSS J111346.03+152842.9 at $z=0.2589$ and for SDSS J223704.78+184055.9 we report a tentative redshift of $z=0.724$ given the signal-to-noise ratio of the spectrum. Additionally, we obtained a lower limit estimate of the redshift for SDSS J150316.57+165117.7 at $z\geq 0.972$.

We also verified that mid-IR properties of these new BZBs are all in agreement with those of known and associated \emph{Fermi} BZBs listed in Roma-BZCAT. Then seventeen out of 18 newly discovered BZBs have an angular separation within 98th percentile of the \emph{Fermi} BZB distribution. These additional evidence strengthen our results that the new BZBs could be all potential counterparts of the UGSs.

The present analysis provides new spectroscopic identifications of potential counterparts of UGSs and BCUs. In particular we remark that most of BCUs analyzed here are BZBs supporting the trend that unclassified \fer\ sources tend to be BZBs since BL Lacs are still the most elusive class of $\gamma$-ray sources.

Our optical spectroscopic follow up campaign is still ongoing with observing nights coming in the current and next semesters, expecting to release new results as soon as collected.

Finally we highlight that, while carrying out our analysis two sources presented here, namely the BCU FL8Y J1541.7+1413 associated with the source RFC J1541+1414 has been classified as a BL Lac at $z=$0.223 and the same potential counterpart of the UGS FL8Y J2244.6+2502, i.e., SDSS J224436.7+250342.6, was classified as a blazar at redshift $z=$0.65 \citep{paiano18}, in agreement with our results/classifications but providing redshift estimates thanks to the better spectroscopic observations available to the other group of colleagues.

\acknowledgements
PH acknowledges support from the CONACyT program for Ph.D. studies. The work of F.M. and A.P. is partially supported by the ``Departments of Excellence 2018-2022'' Grant awarded by the Italian Ministry of Education, University and Research (MIUR) (L. 232/2016) and made use of resources provided by the Compagnia di San Paolo for the grant awarded on the BLENV project (S1618\_L1\_MASF\_01) and by the Ministry of Education, Universities and Research for the grant MASF\_FFABR\_17\_01. F.M. also acknowledges financial contribution from the agreement ASI-INAF n.2017-14-H.0 while A.P. the financial support from the Consorzio Interuniversitario per la fisica Spaziale (CFIS) under the agreement related to the grant MASF\_CONTR\_FIN\_18\_02. F.R. acknowledges support from FONDECYT Postdoctorado 3180506 and CONICYT project Basal AFB-170002. This work was partially supported from CONACyT research grant No. 280789. 
We thank the staff at the Observatorio Astrof\'isico \, Guillermo~Haro (OAGH) for all their help during the observation runs.
Based on observations obtained at the Southern Astrophysical Research (SOAR) telescope, which is a joint project of the Minist\'{e}rio da Ci\^{e}ncia, Tecnologia, e Inova\c{c}\~{a}o (MCTI) da Rep\'{u}blica Federativa do Brasil, the U.S. National Optical Astronomy Observatory (NOAO), the University of North Carolina at Chapel Hill (UNC), and Michigan State University (MSU).
Part of this work is based on archival data, software or on-line services provided by the ASI Science Data Center. 
This publication makes use of data products from the Wide-field Infrared Survey Explorer, 
which is a joint project of the University of California, Los Angeles, and 
the Jet Propulsion Laboratory/California Institute of Technology, 
funded by the National Aeronautics and Space Administration.
This research makes use of SDSS DR14 archival data. Funding for SDSS-III has been provided by the Alfred P. Sloan Foundation, the Participating Institutions, the National Science Foundation, and the U.S. Department of Energy Office of Science. The SDSS-III web site is http://www.sdss3.org/. SDSS-III is managed by the Astrophysical Research Consortium for the Participating Institutions of the SDSS-III Collaboration including the University of Arizona, the Brazilian Participation Group, Brookhaven National Laboratory, Carnegie Mellon University, University of Florida, the French Participation Group, the German Participation Group, Harvard University, the Instituto de Astrofisica de Canarias, the Michigan State/Notre Dame/JINA Participation Group, Johns Hopkins University, Lawrence Berkeley National Laboratory, Max Planck Institute for Astrophysics, Max Planck Institute for Extraterrestrial Physics, New Mexico State University, New York University, Ohio State University, Pennsylvania State University, University of Portsmouth, Princeton University, the Spanish Participation Group, University of Tokyo, University of Utah, Vanderbilt University, University of Virginia, University of Washington, and Yale University.
This research has made use of the USNOFS Image and Catalogue Archive
operated by the United States Naval Observatory, Flagstaff Station
(http://www.nofs.navy.mil/data/fchpix/).
TOPCAT\footnote{\underline{http://www.star.bris.ac.uk/$\sim$mbt/topcat/}} 
\citep{taylor05} for the preparation and manipulation of the tabular data and the images.

\onecolumn{
\tiny
\begin{center}
\begin{landscape}

\begin{longtable}{lcccccccccc}
\caption{Observing Log and Summary of Results} \label{tab:main}\\
\tablecaption{Summary of targets} 
\tablewidth{\textwidth} \\
\multicolumn{1}{c}{Fermi Name} & 
\multicolumn{1}{c}{Name} &
\multicolumn{1}{c}{R.A. (J2000)} &
\multicolumn{1}{c}{Dec. (J2000)} &
\multicolumn{1}{c}{WISE Counterpart} & 
\multicolumn{1}{c}{Obs. Date} &
\multicolumn{1}{c}{Telescope} &
\multicolumn{1}{c}{S/N} & 
\multicolumn{1}{c}{Exp. Time} &
\multicolumn{1}{c}{z} &
\multicolumn{1}{c}{Class} \\
\multicolumn{1}{c}{ } &
\multicolumn{1}{c}{ } &
\multicolumn{1}{c}{hh:mm:ss} &
\multicolumn{1}{c}{dd:mm:ss} &
\multicolumn{1}{c}{ } &
\multicolumn{1}{c}{yyy-mm-dd} &
\multicolumn{1}{c}{ } &
\multicolumn{1}{c}{ } &
\multicolumn{1}{c}{$\left( s \right)$} &
\multicolumn{1}{c}{ } &
\multicolumn{1}{c}{ } \\
\hline
 \endhead

\multicolumn{11}{c}{\small \textsc{BCUs}} \\ 
\hline
 FL8Y J1046.1+5449 &  CRATES J104630+544953    & 10:46:28.80 & +54:49:44.39 & J104628.24+544944.8 & 2018-04-16 & OAGH & 4 & 1800 & 0.249 & BZB\\
 FL8Y J1129.4+3033 &  WISE J112937.30+303634.4 & 11:29:37.30 & +30:36:34.45 & J112937.30+303634.4 & 2018-04-14 & OAGH & 6 & 1800 & ? & BZB\\
 FL8Y J1157.5+1639 &  VCS J1157+1638           & 11:57:34.84 & +16:38:59.65 & J115734.83+163859.6 & 2018-04-14 & OAGH & 16 & 1800 & $1.058$ & BZQ\\
 FL8Y J1231.1+3711 &  RFC J1231+3711           & 12:31:24.10 & +37:11:02.04 & J123124.08+371102.2 & 2018-04-14 & OAGH & 5 & 1800 & $0.219$ & BZB\\
 FL8Y J1319.5-0046 &  AT20G J131938-004939     & 13:19:38.84 & -00:49:39.29 & J131938.76-004939.9 & 2018-04-16 & OAGH & 24 & 1800 & $0.890$ & BZQ\\
 FL8Y J1329.4-0530 &  Q 1326-0516              & 13:29:28.61 & -05:31:36.12 & J132928.62-053135.7 & 2018-04-16 & OAGH & 61 & 1200 & $0.575$ & BZQ\\
 FL8Y J1331.7-0647 &  WISE J133146.84-064633.1 & 13:31:46.85 & -06:46:33.17 & J133146.84-064633.1 & 2018-04-01 & OAGH & 9 & 1800 & ? & BZB\\
 FL8Y J1541.7+1413 &  RFC J1541+1414           & 15:41:50.10 & +14:14:37.58 & J154150.09+141437.6 & 2018-04-16 & OAGH & 9 & 1200 & ? & BZB\\
 FL8Y J1559.8-2525 &  NVSS J160005-252439      & 16:00:05.40 & -25:24:39.46 & J160005.35-252439.7 & 2018-05-31 & SOAR & 4 & 900 & ? & BZB\\
 FL8Y J1808.9+3522 &  RFC J1808+3520           & 18:08:49.69 & +35:20:42.68 & J180849.69+352042.7 & 2018-04-16 & OAGH & 18 & 1200 & $0.142$ & BZB\\
 FL8Y J2048.6-6804 &  PKS 2043-682             & 20:48:23.83 & -68:04:51.70 & J204823.99-680451.9 & 2018-05-31 & SOAR & 14 & 800 & ? & BZB\\
 FL8Y J2103.8-6233 &  PMN J2103-6232           & 21:03:38.39 & -62:32:25.92 & J210338.38-623225.8 & 2018-05-31 & SOAR & 22 & 400 & ? & BZB\\
 3FGL J2212.3-7039 &  PMN J2211-7039           & 22:11:56.23 & -70:39:14.82 & J221156.23-703914.8 & 2018-05-31 & SOAR & 5 & 800 & ? & BZB\\
\hline
\multicolumn{11}{c}{\small \textsc{UGSs}} \\ 
\hline
 FL8Y J0024.1+2401 & J002406.10+240438.3      & 00:24:06.10 & +24:04:38.39 & J002406.10+240438.6 &  2015-09-21 & SDSS & 9  & 3600 & ? & BZB\\
 FL8Y J0112.1-0320 & J011205.78-031753.6      & 01:12:05.79 & -03:17:53.63 & J011205.78-031753.7 &  2014-11-19 & SDSS & 5  & 7201 & ? & BZB\\
 FL8Y J0829.7+5106 & J082948.08+510827.9      & 08:29:48.08 & +51:08:27.96 & J082948.08+510827.6 &  2010-01-13 & SDSS & 17 & 3601 & ? & BZB\\
 FL8Y J1113.9+1527 & J111346.03+152842.9      & 11:13:46.03 & +15:28:42.96 & J111345.94+152843.9 &  2005-01-09 & SDSS & 5  & 3300 & $0.2589$& BZB\\
 FL8Y J1113.8+5524 & J111356.3+552255.8       & 11:13:56.31 & +55:22:55.85 & J111356.24+552255.3 &  2014-01-02 & SDSS & 25 & 4500 & ? & BZB\\
 FL8Y J1137.2+0534 & J113737.76+053016.5      & 11:37:37.77 & +05:30:16.56 & J113737.76+053016.5 &  2012-01-20 & SDSS & 16 & 5405 & ? & BZB\\
 FL8Y J1243.6+1727 & J124351.76+172644.3      & 12:43:51.76 & +17:26:44.34 & J124351.76+172644.4 &  2012-06-12 & SDSS & 12 & 2702 & ? & BZB\\
 FL8Y J1258.4+6552 & J125733.06+655100.2      & 12:57:33.07 & +65:51:00.25 & J125733.04+655059.9 &  2014-03-08 & SDSS & 19 & 4500 & ? & BZB\\
 FL8Y J1503.3+1651 & J150316.57+165117.7      & 15:03:16.58 & +16:51:17.78 & J150316.56+165117.6 &  2012-04-22 & SDSS & 13 & 2702 & $\geq 0.972$ & BZB\\
 FL8Y J1511.4+0549 & J151100.45+054920.8      & 15:11:00.45 & +05:49:20.82 & J151100.44+054920.5 &  2011-04-01 & SDSS & 8  & 5405 & ? & BZB\\
 FL8Y J1516.3+4353 & J151631.37+434949.5      & 15:16:31.37 & +43:49:49.58 & J151631.37+434949.7 &  2012-05-25 & SDSS & 12 & 3603 & ? & BZB\\
 FL8Y J1544.9+3218 & J154433.19+322149.1      & 15:44:33.19 & +32:21:49.14 & J154433.19+322148.6 &  2011-07-05 & SDSS & 27 & 3603 & ? & BZB\\
 FL8Y J1721.3+2529 & J172100.07+251249.7      & 17:21:00.08 & +25:12:49.79 & J172100.05+251249.7 &  2011-05-28 & SDSS & 13 & 7207 & ? & BZB\\
 FL8Y J2207.1+2222 & J220652.9+221722.2       & 22:06:52.90 & +22:17:22.20 &                     &  2014-10-24 & SDSS & 4  & 4500 & ? & BZB\\
 FL8Y J2207.1+2222 & J220704.10+222231.4      & 22:07:04.11 & +22:22:31.40 & J220704.11+222231.5 &  2014-10-25 & SDSS & 11 & 4500 & ? & BZB\\
 FL8Y J2228.5+2211 & J222839.49+221125        & 22:28:39.49 & +22:11:25.01 & J222839.49+221125.0 &  2014-10-30 & SDSS & 17 & 4500 & ? & BZB\\
 FL8Y J2236.9+1840 & J223704.78+184055.9      & 22:37:04.78 & +18:40:55.99 & J223704.83+184055.5 &  2015-06-22 & SDSS & 10 & 4500 & 0.724? & BZB\\
 FL8Y J2244.6+2502 & J224436.7+250342.6       & 22:44:36.70 & +25:03:42.62 & J224436.66+250343.1 &  2015-10-14 & SDSS & 28 & 5401 & ? & BZB\\
\hline
\end{longtable}
\begin{flushleft}
Column description: 
 (1) \emph{Fermi}-LAT 8-year Source List Name;
 (2) Associated counterpart; 
 (3) R.A. (Equinox J200);  
 (4) Dec. (Equinox J200);
 (5) WISE counterpart;
 (6) Observation date;
 (7) Telescope;
 (8) Signal-to-noise ratio;
 (9) Exposure time; 
 (10) Redshift, question marks indicate unknown $z$;
 (11) Source classification.
\end{flushleft}
\end{landscape}
\end{center}}

\twocolumn

\onecolumn{
\begin{center}

\begin{longtable}{lcccc}
\caption{Spectral Line Measurements} \label{tab:measures}\\
\tablewidth{\textwidth} 
\tabletypesize{\footnotesize} \\
\multicolumn{1}{c}{Name} & 
\multicolumn{1}{c}{Line ID} &
\multicolumn{1}{c}{EW \AA} &
\multicolumn{1}{c}{$\lambda_{obs}$ \AA} &
\multicolumn{1}{c}{Type} \\
\hline
\endhead
\hline
CRATES J104630+544953   &Ca II H\&K&$7-10\pm2$&4966-4920&A\\
                        &[O III]&$4-6\pm3$&6192-6253&E\\
                        &G band&$5\pm1$&5376&A\\
                        &Mg I&$10\pm1$&6460&A\\
\hline
SDSS J111346.03+152842.9&Ca II H\&K&$8-7\pm2$&4995-4953&A\\
                        &G band&$90\pm2$&5418&A\\
                        &H$\beta$&$13\pm2$&6123&A\\
                        &Mg I&$7\pm2$&6515&A\\
\hline
VCS J1157+1638          &Mg II&$27\pm6$&5760&E\\
                        &Fe II 2964&$7\pm5$&—&E\\
\hline
RFC J1231+3711          &[O II]&$8\pm3$&4550.0&E\\
                        &Ca II H\&K&$13\pm4$&4841-4794&A\\
                        &G band&$7.2\pm1$&5244&A\\
\hline
AT20G J131938-004939    &Mg II&$60\pm4$&5298&E\\
                        &[Ne V]&$2-4\pm3$&6318-6478&E\\
                        &C II]&$3\pm2$&4394&E\\
\hline
Q 1326-0516             &Mg II&$33\pm7$&4411&E\\
                        &Fe II 2964&$3\pm5$&—&E\\
\hline
SDSS J150316.57+165117.7&Mg II&$2-1\pm1$&5512-5527&A\\
                        &Fe II 2600&$1-1\pm1$&5100-5126&A\\
                        &Fe II 2344&$2\pm1$&4621&A\\
                        &Fe II 2374&$1\pm1$&4681&A\\
\hline
RFC J1808+4520          &Ca II H\&K&$4-5\pm3$&4531-4493&A\\
                        &G band&$4\pm2$&4916&A\\
                        &H$\beta$&$19\pm3$&5552&A\\
                        &Na I&$5\pm3$&6727&A\\
\hline
SDSS J223704.78+184055.9*&Ca II H\&K&$2-3\pm1$&6842-6780&A\\
                        &G band&$2\pm1$&7407&A\\
\hline
\end{longtable}
\begin{flushleft}
Column description: 
 (1) Target Name;
 (2) Identification of the line; 
 (3) Equivalent width in \AA;  
 (4) Observed wavelength in \AA;
 (5) Line type: A for absorption and E for emission.
 
*For SDSS J223704.78+184055.9 we provide measurements of the tentative identification of the reported lines.
\end{flushleft}
\end{center}}
\twocolumn

\bibliographystyle{apj}

\appendix

\section{Unidentified Gamma Ray Sources}
\clearpage
\begin{figure*}{}
\begin{center}$
\begin{array}{cc}
\includegraphics[width=\mywidth]{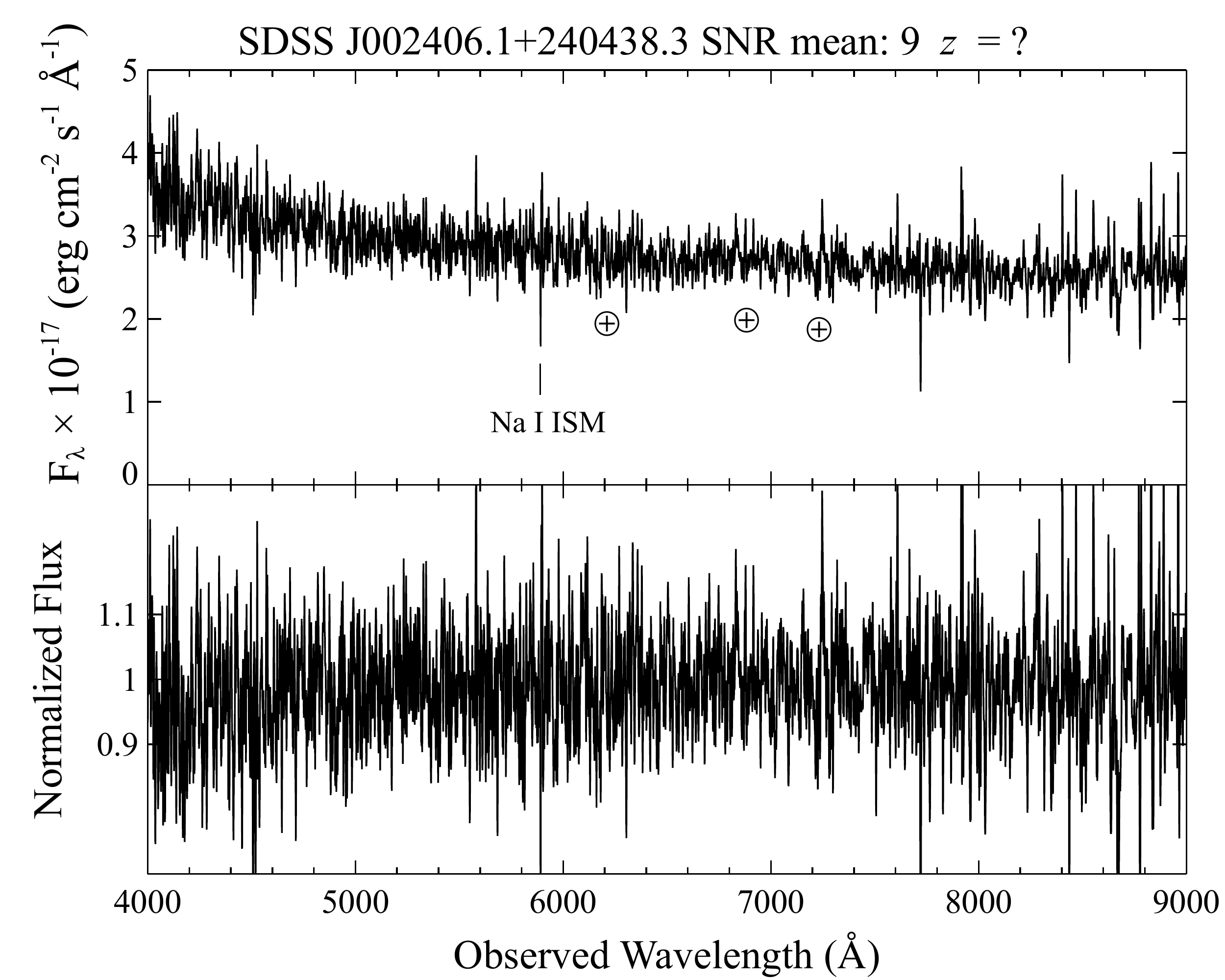} &
\includegraphics[clip=true, width=7cm]{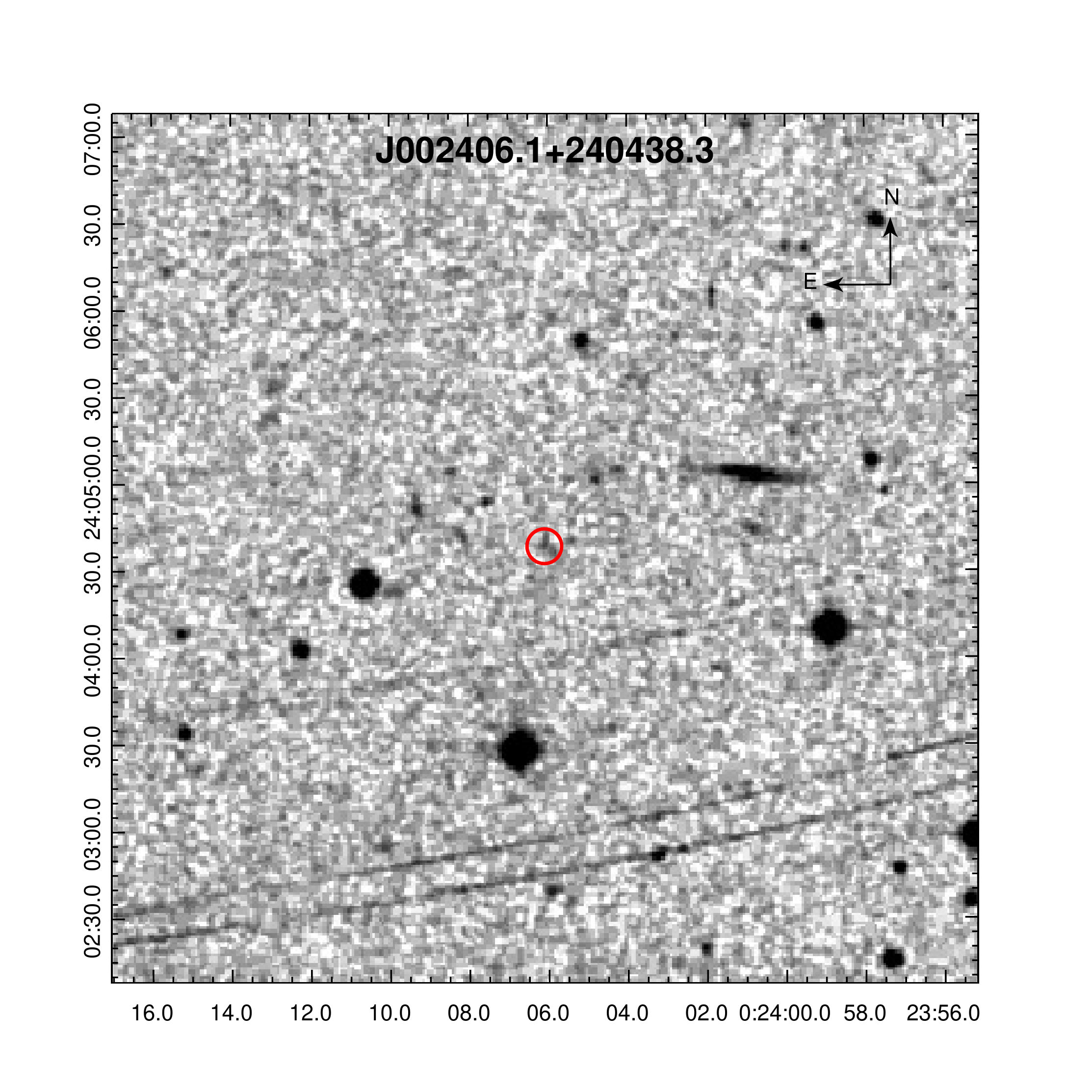} \\
\end{array}$
\end{center}
\caption{(Left panel) Optical spectrum of  SDSS J002406.1+240438.3 potential candidate of the UGS with FL8Y J0024.1+2401, in the upper part it is shown the Signal-to-Noise Ratio of the spectrum. (Right panel) The finding chart ( $5'\times 5'$ ) retrieved from the Digitized Sky Survey (DSS) highlighting the location of the potential counterpart:  SDSS J002406.1+240438.3 (red circle).}
\label{fig:J002}
\end{figure*}

\begin{figure*}{}
\begin{center}$
\begin{array}{cc}
\includegraphics[width=\mywidth]{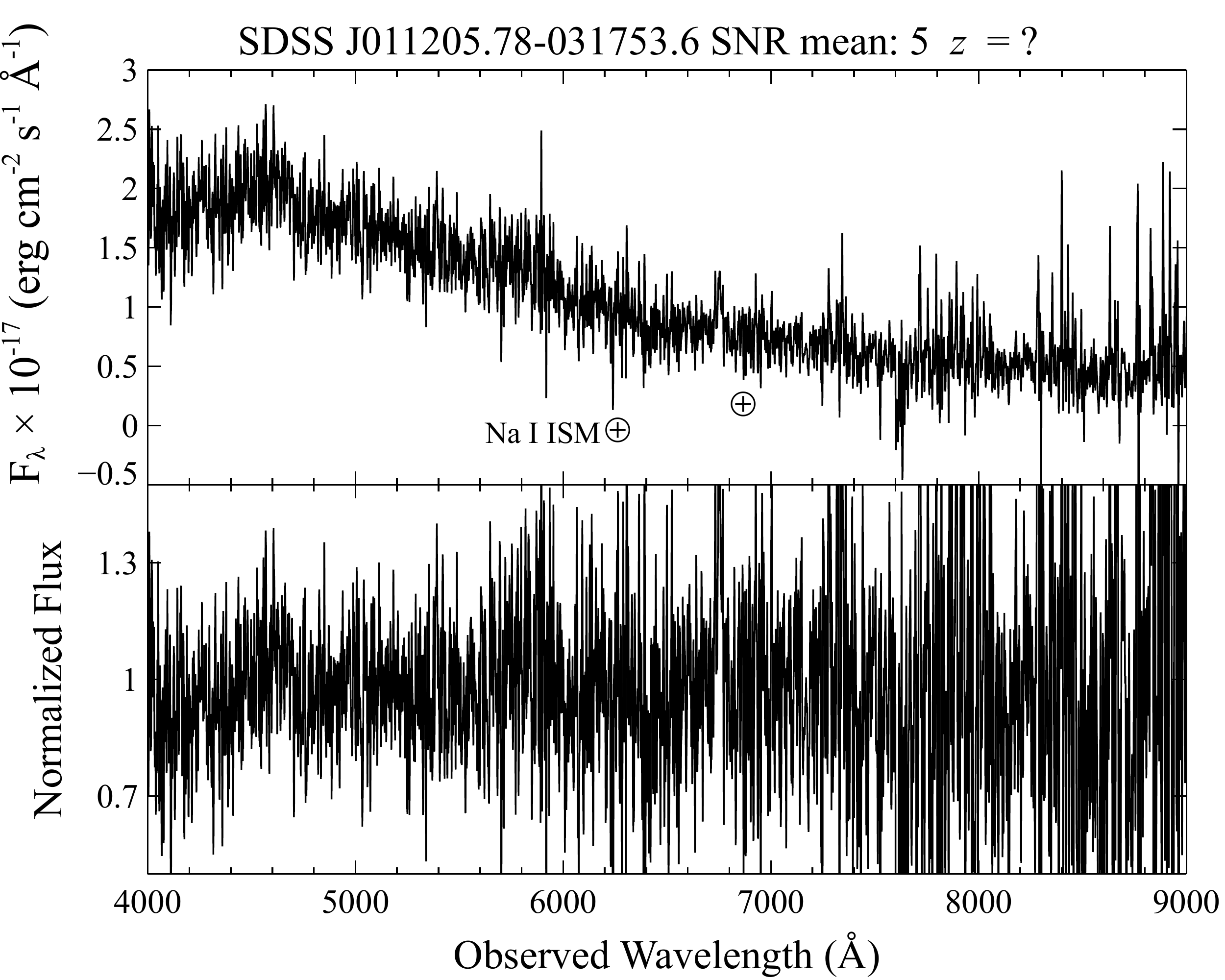} &
\includegraphics[clip=true, width=7cm]{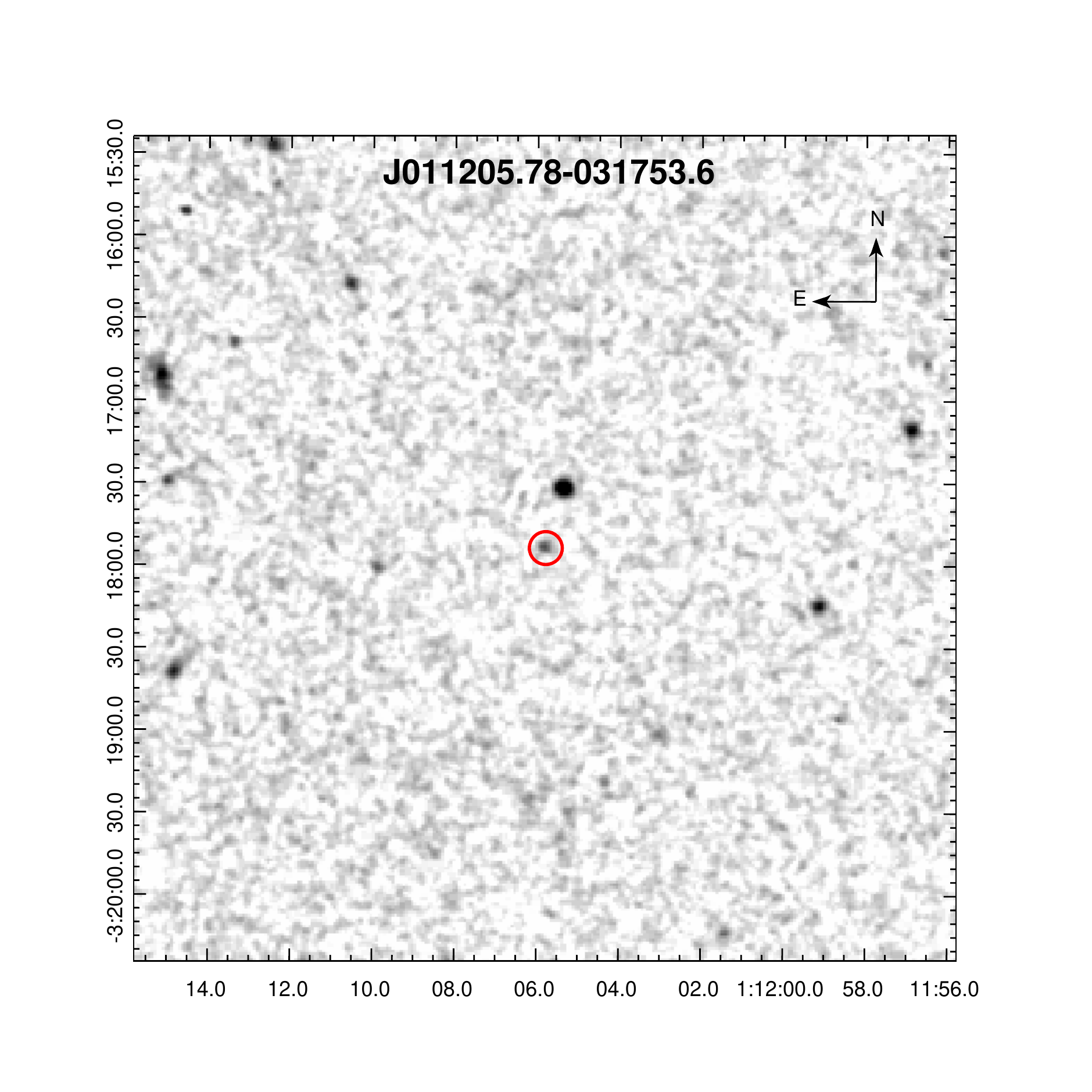} \\
\end{array}$
\end{center}
\caption{(Left panel) Optical spectrum of  SDSS J011205.78-031753.6 potential candidate of the UGS with FL8Y J0112.1-0320, in the upper part it is shown the Signal-to-Noise Ratio of the spectrum. (Right panel) The finding chart ( $5'\times 5'$ ) retrieved from the Digitized Sky Survey (DSS) highlighting the location of the potential counterpart:  SDSS J011205.78-031753.6 (red circle).}
\label{fig:J0112}
\end{figure*}

\begin{figure*}{}
\begin{center}$
\begin{array}{cc}
\includegraphics[width=\mywidth]{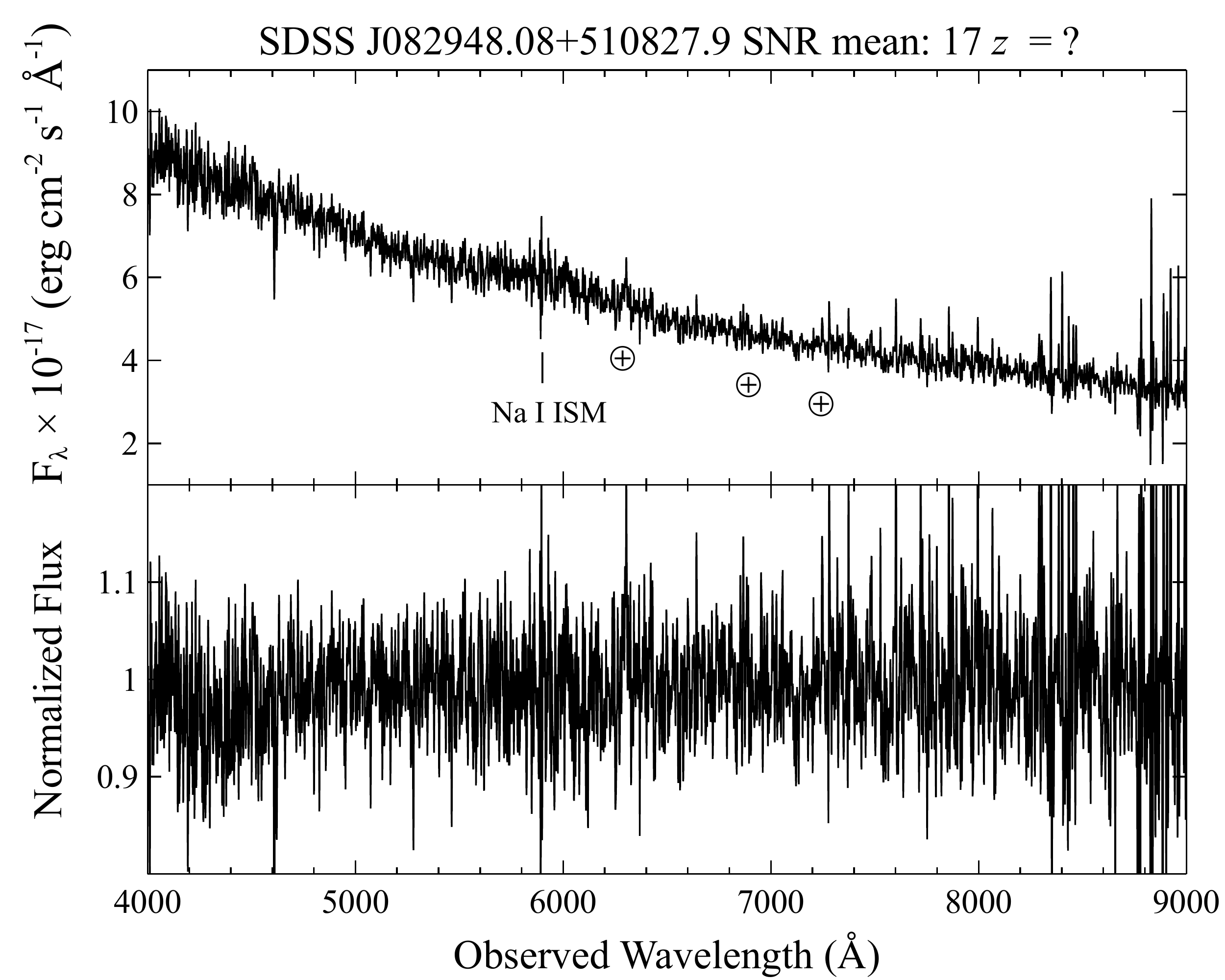} &
\includegraphics[clip=true, width=7cm]{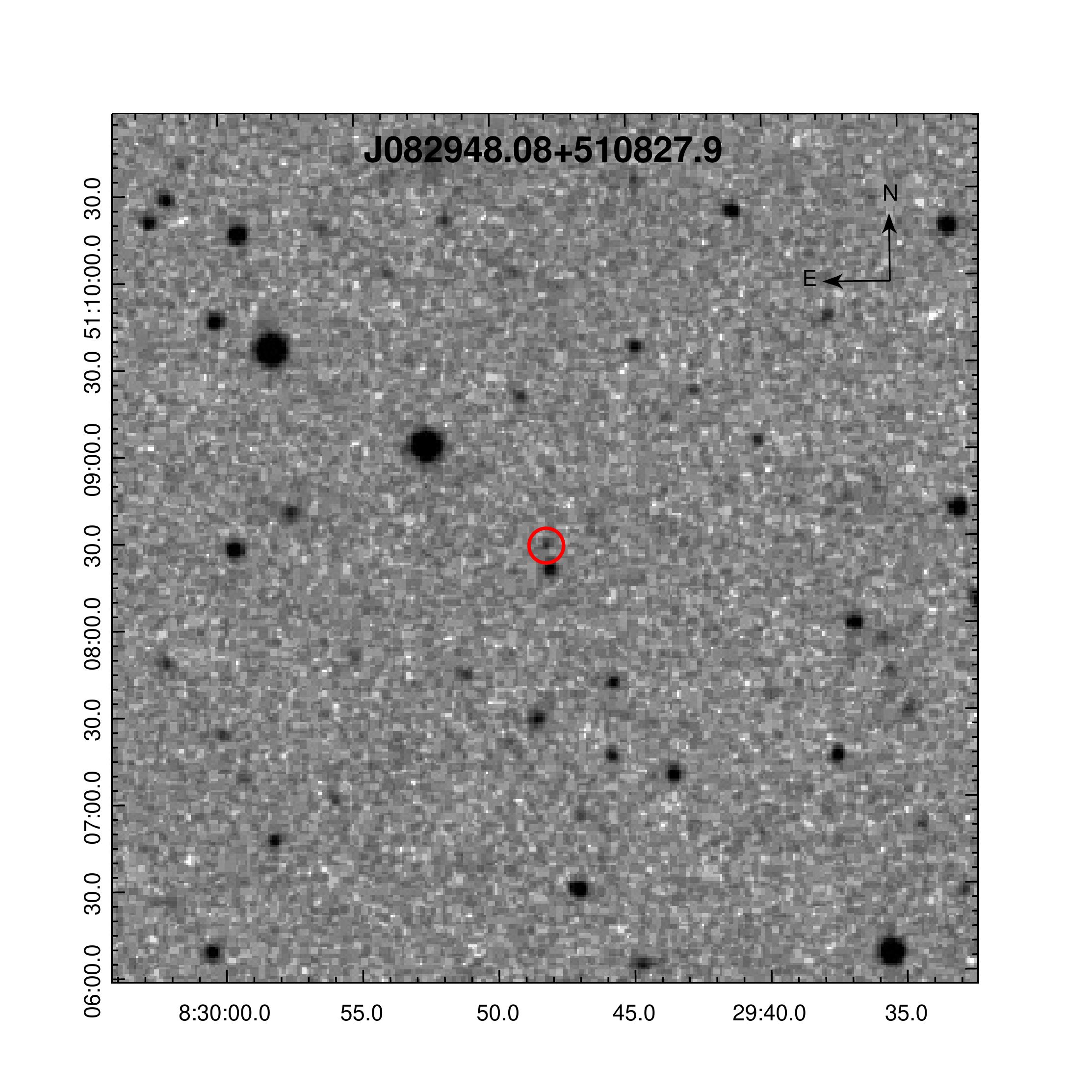} \\
\end{array}$
\end{center}
\caption{(Left panel) Optical spectrum of  SDSS J082948.08+510827.9 potential candidate of the UGS with FL8Y J0829.7+5106, in the upper part it is shown the Signal-to-Noise Ratio of the spectrum. (Right panel) The finding chart ( $5'\times 5'$ ) retrieved from the Digitized Sky Survey (DSS) highlighting the location of the potential counterpart:  SDSS J082948.08+510827.9 (red circle).}
\label{fig:J0829}
\end{figure*}

\begin{figure*}{}
\begin{center}$
\begin{array}{cc}
\includegraphics[width=\mywidth]{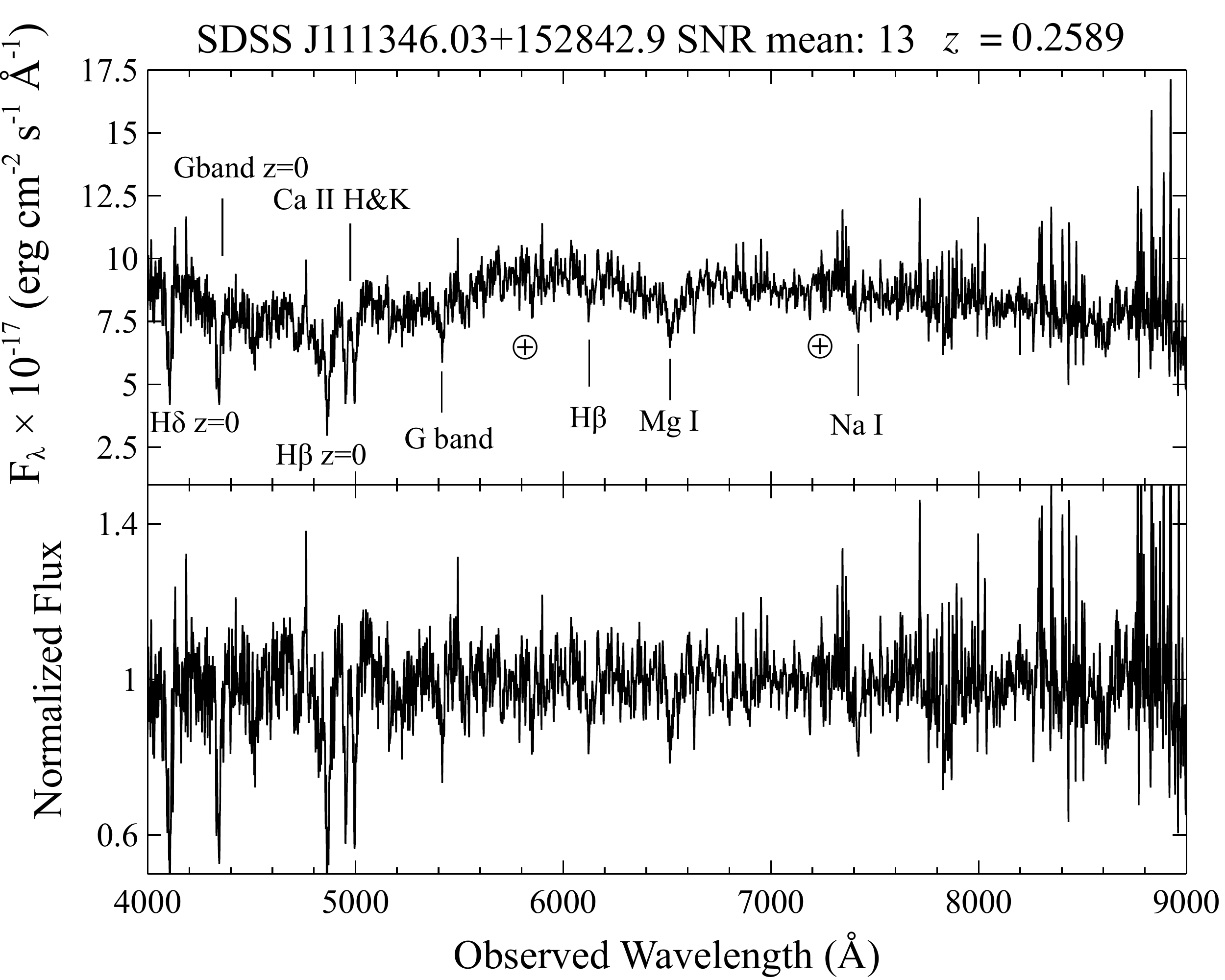} &
\includegraphics[clip=true, width=7cm]{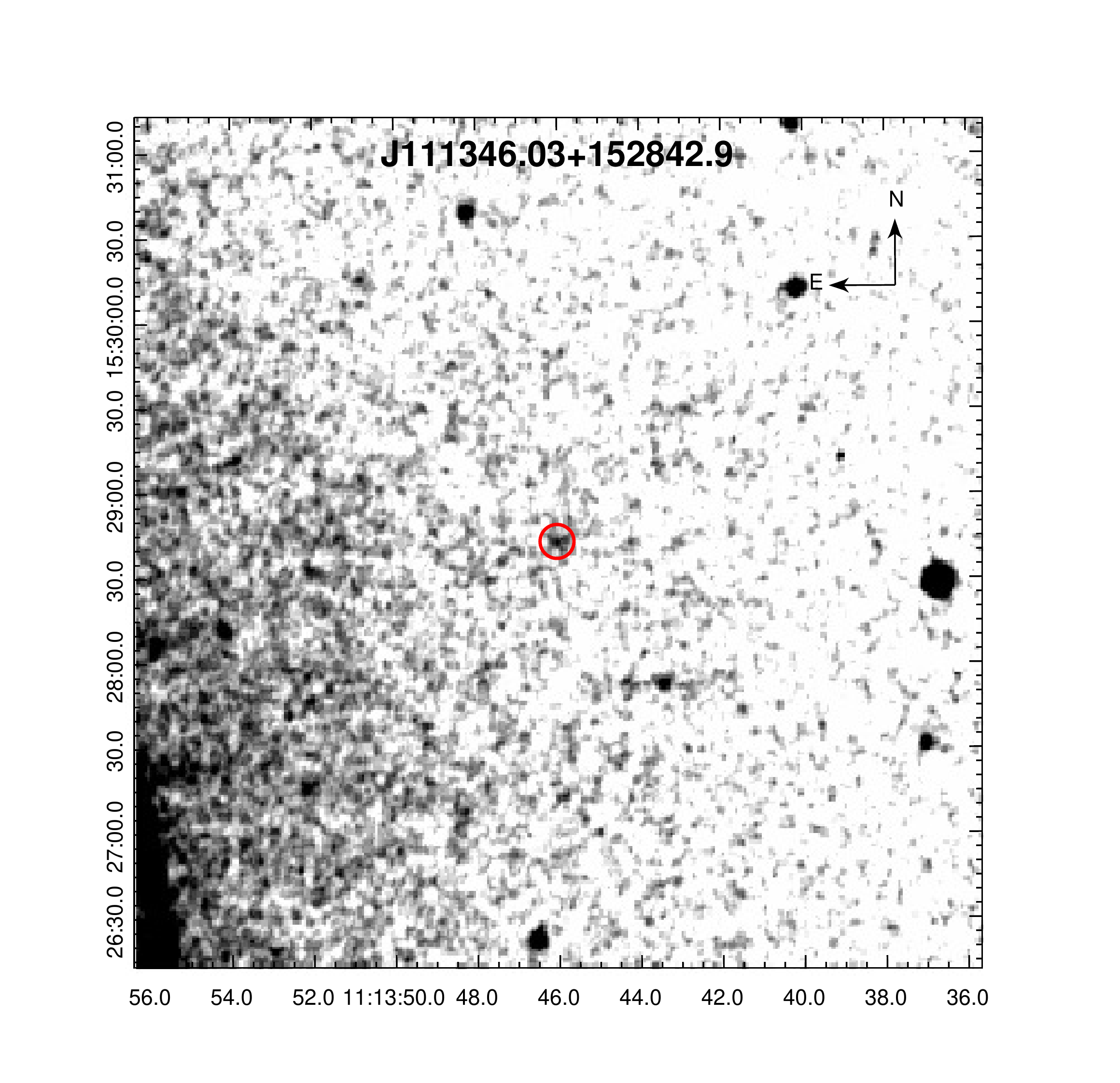} \\
\end{array}$
\end{center}
\caption{(Left panel) Optical spectrum of  SDSS J111346.03+152842.9 potential candidate of the UGS with FL8Y J1113.9+1527, in the upper part it is shown the Signal-to-Noise Ratio of the spectrum. (Right panel) The finding chart ( $5'\times 5'$ ) retrieved from the Digitized Sky Survey (DSS) highlighting the location of the potential counterpart: SDSS J111346.03+152842.9 (red circle).}
\label{fig:J111346}
\end{figure*}

\begin{figure*}{}
\begin{center}$
\begin{array}{cc}
\includegraphics[width=\mywidth]{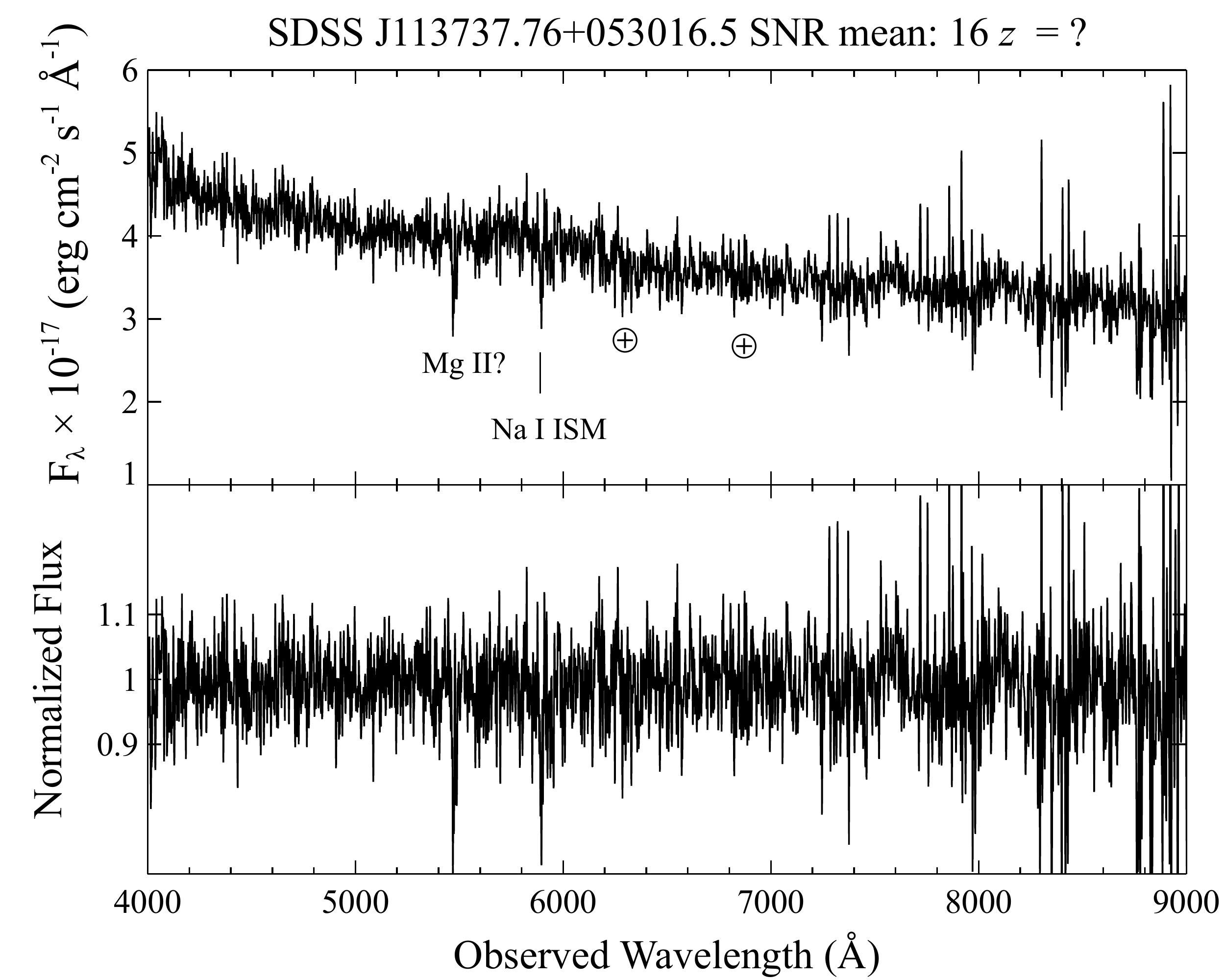} &
\includegraphics[clip=true, width=7cm]{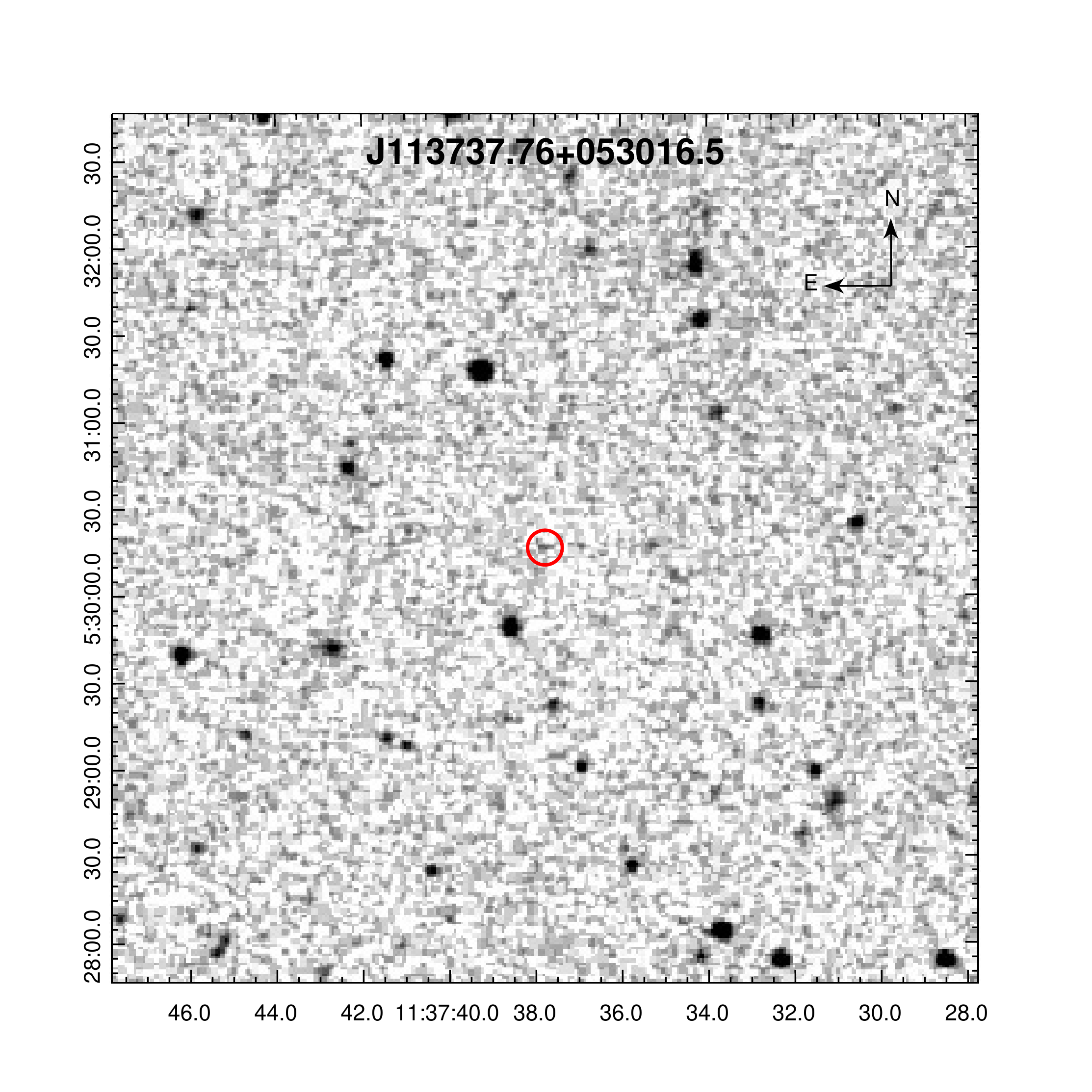} \\
\end{array}$
\end{center}
\caption{(Left panel) Optical spectrum of  SDSS J113737.76+053016.5 potential candidate of the UGS with FL8Y J1137.2+0534, in the upper part it is shown the Signal-to-Noise Ratio of the spectrum. (Right panel) The finding chart ( $5'\times 5'$ ) retrieved from the Digitized Sky Survey (DSS) highlighting the location of the potential counterpart:   SDSS J113737.76+053016.5 (red circle).}
\label{fig:J1137}
\end{figure*}

\begin{figure*}{}
\begin{center}$
\begin{array}{cc}
\includegraphics[width=\mywidth]{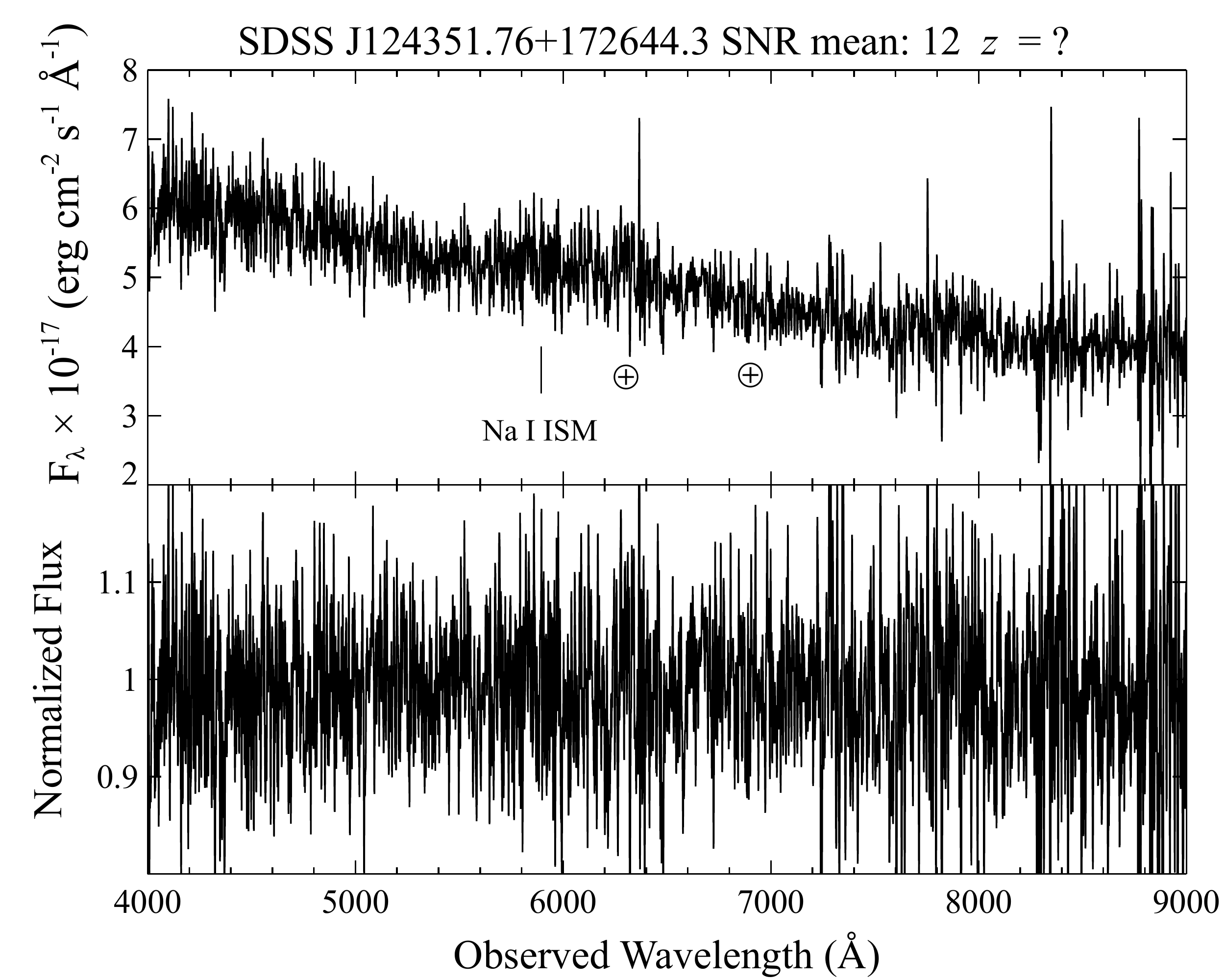} &
\includegraphics[clip=true, width=7cm]{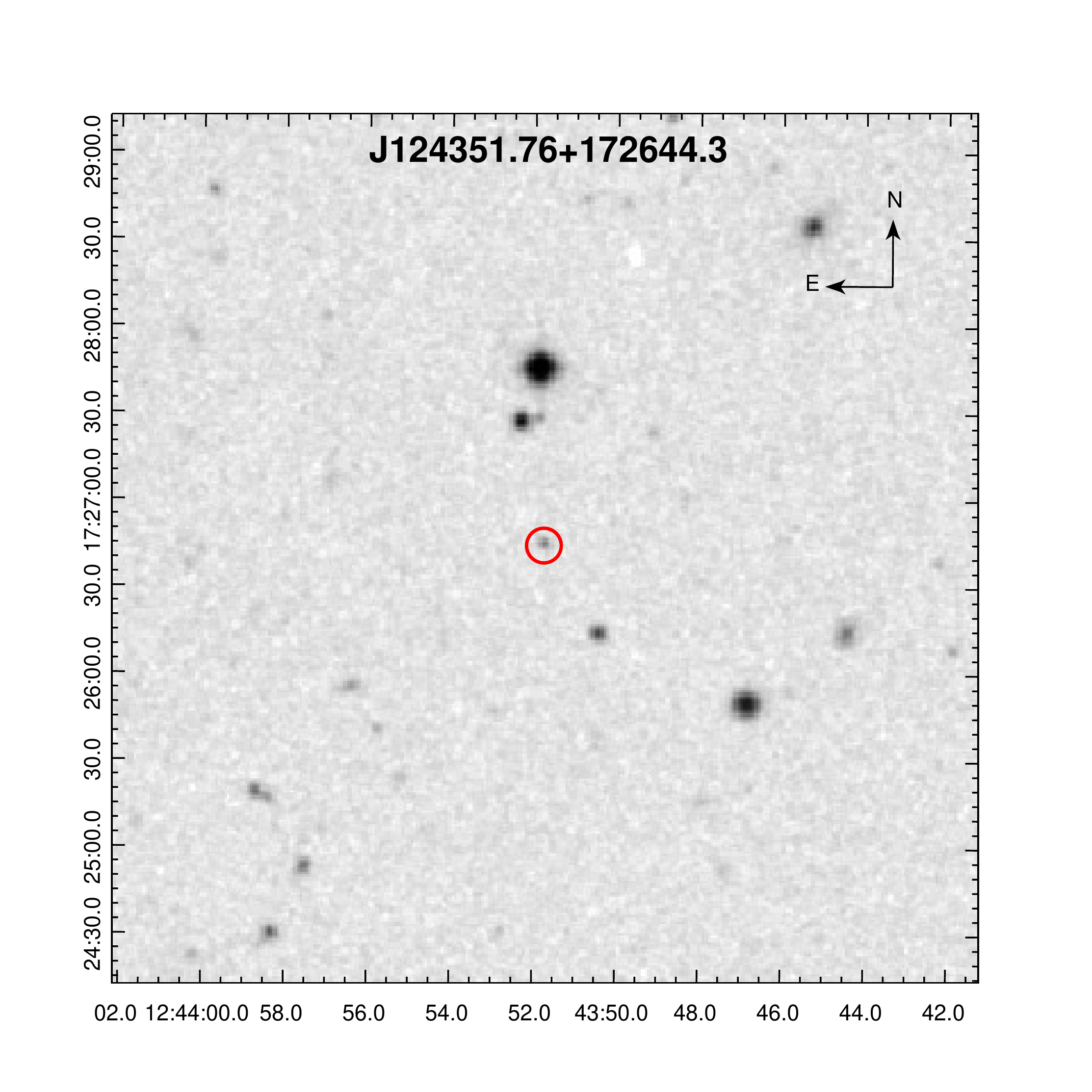} \\
\end{array}$
\end{center}
\caption{(Left panel) Optical spectrum of SDSS J124351.76+172644.3 potential candidate of the UGS with FL8Y J1243.6+1727, in the upper part it is shown the Signal-to-Noise Ratio of the spectrum. (Right panel) The finding chart ( $5'\times 5'$ ) retrieved from the Digitized Sky Survey (DSS) highlighting the location of the potential counterpart:   SDSS J124351.76+172644.3 (red circle).}
\label{fig:J1243}
\end{figure*}

\begin{figure*}{}
\begin{center}$
\begin{array}{cc}
\includegraphics[width=\mywidth]{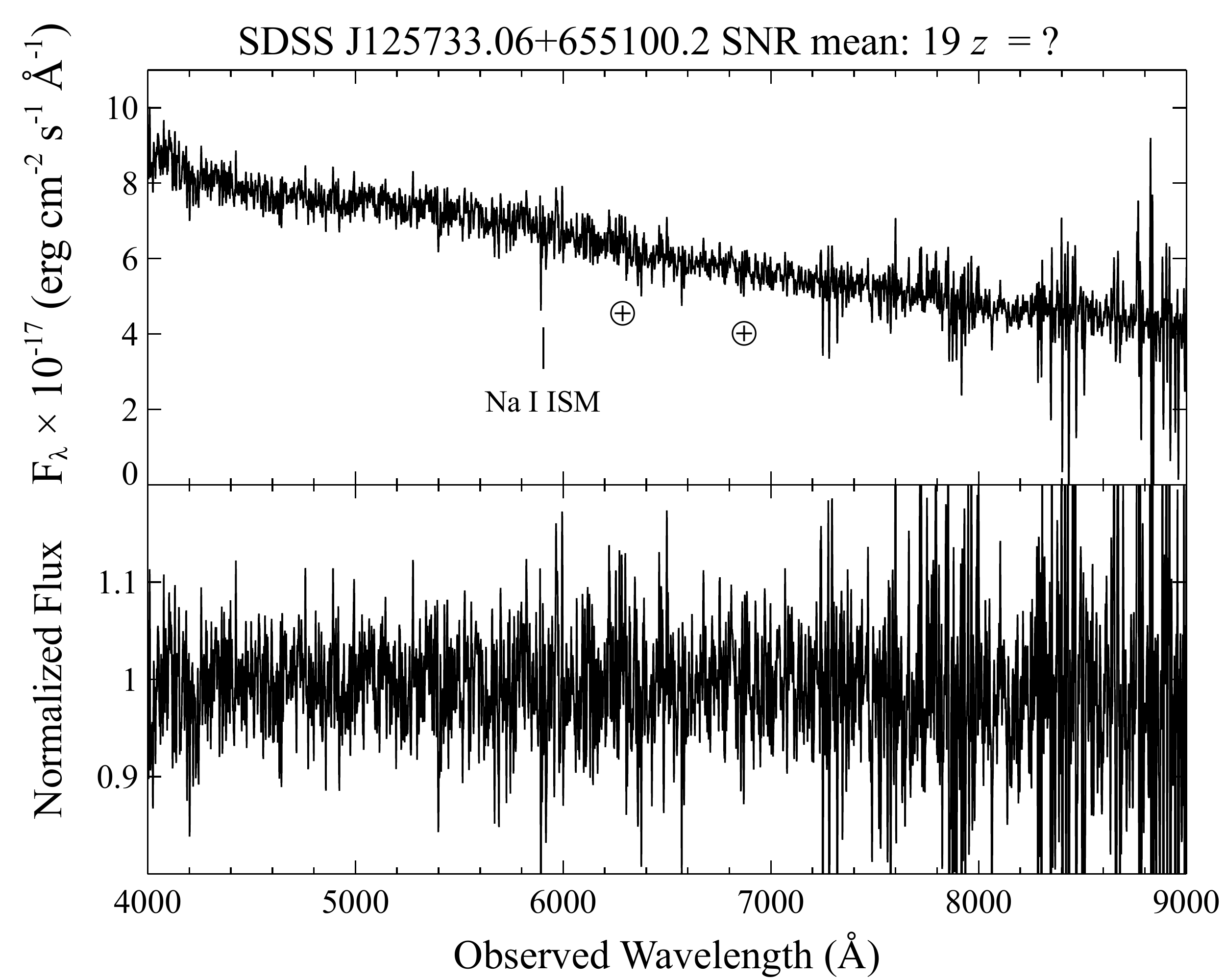} &
\includegraphics[clip=true, width=7cm]{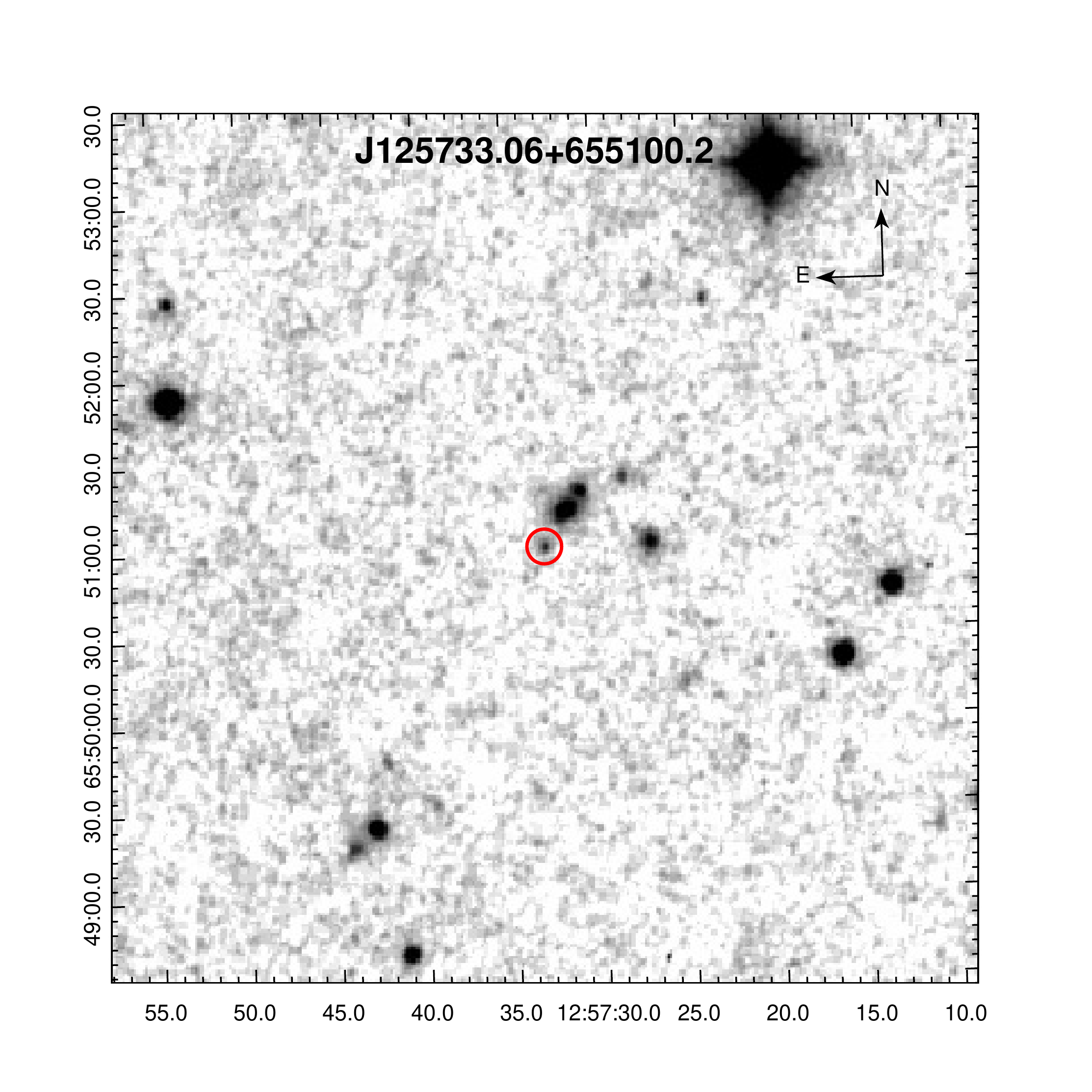} \\
\end{array}$
\end{center}
\caption{(Left panel) Optical spectrum of SDSS J125733.06+655100.2 potential candidate of the UGS with FL8Y J1258.4+6552, in the upper part it is shown the Signal-to-Noise Ratio of the spectrum. (Right panel) The finding chart ( $5'\times 5'$ ) retrieved from the Digitized Sky Survey (DSS) highlighting the location of the potential counterpart:   SDSS J125733.06+655100.2 (red circle).}
\label{fig:J1257}
\end{figure*}

\begin{figure*}{}
\begin{center}$
\begin{array}{cc}
\includegraphics[width=\mywidth]{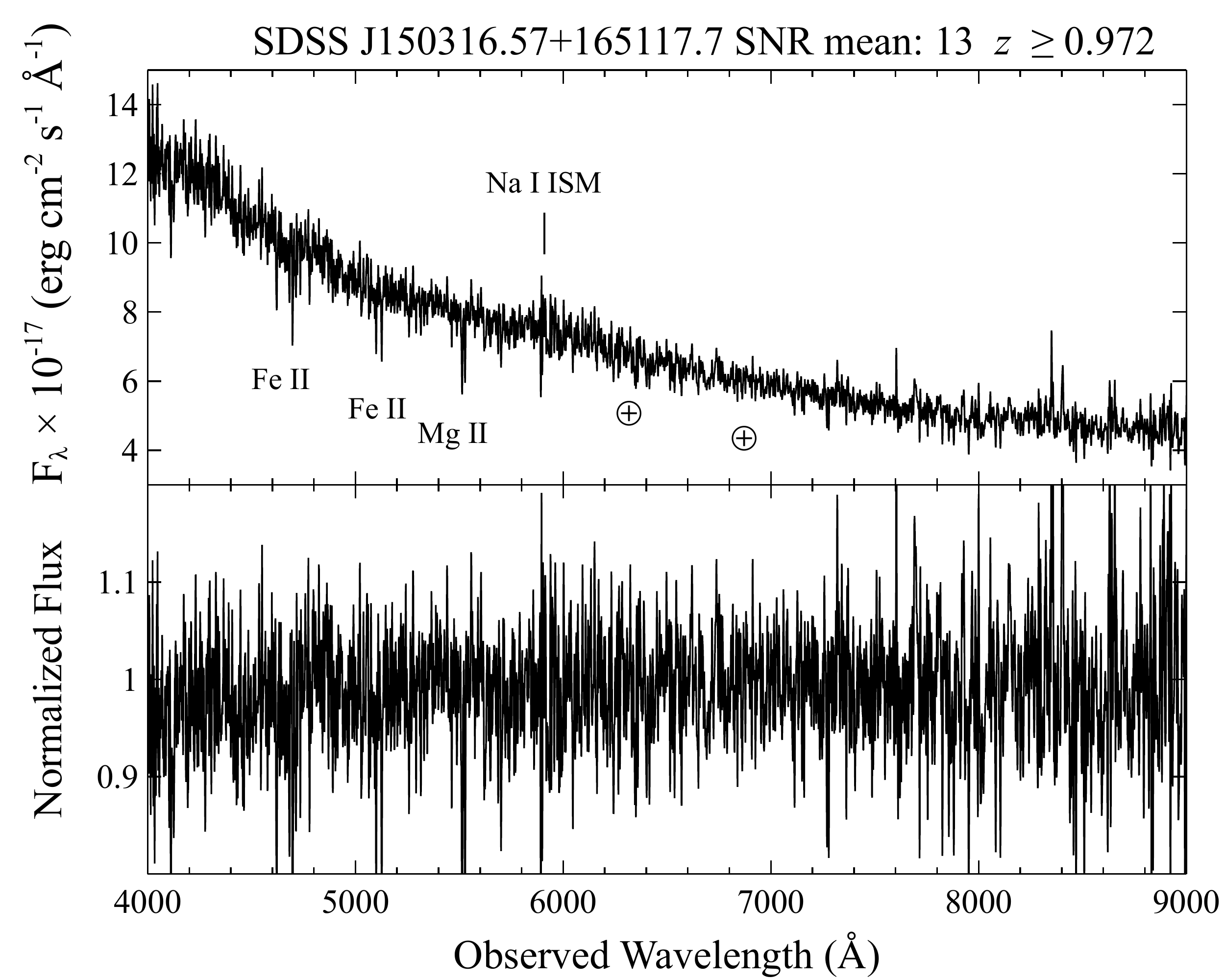} &
\includegraphics[clip=true, width=7cm]{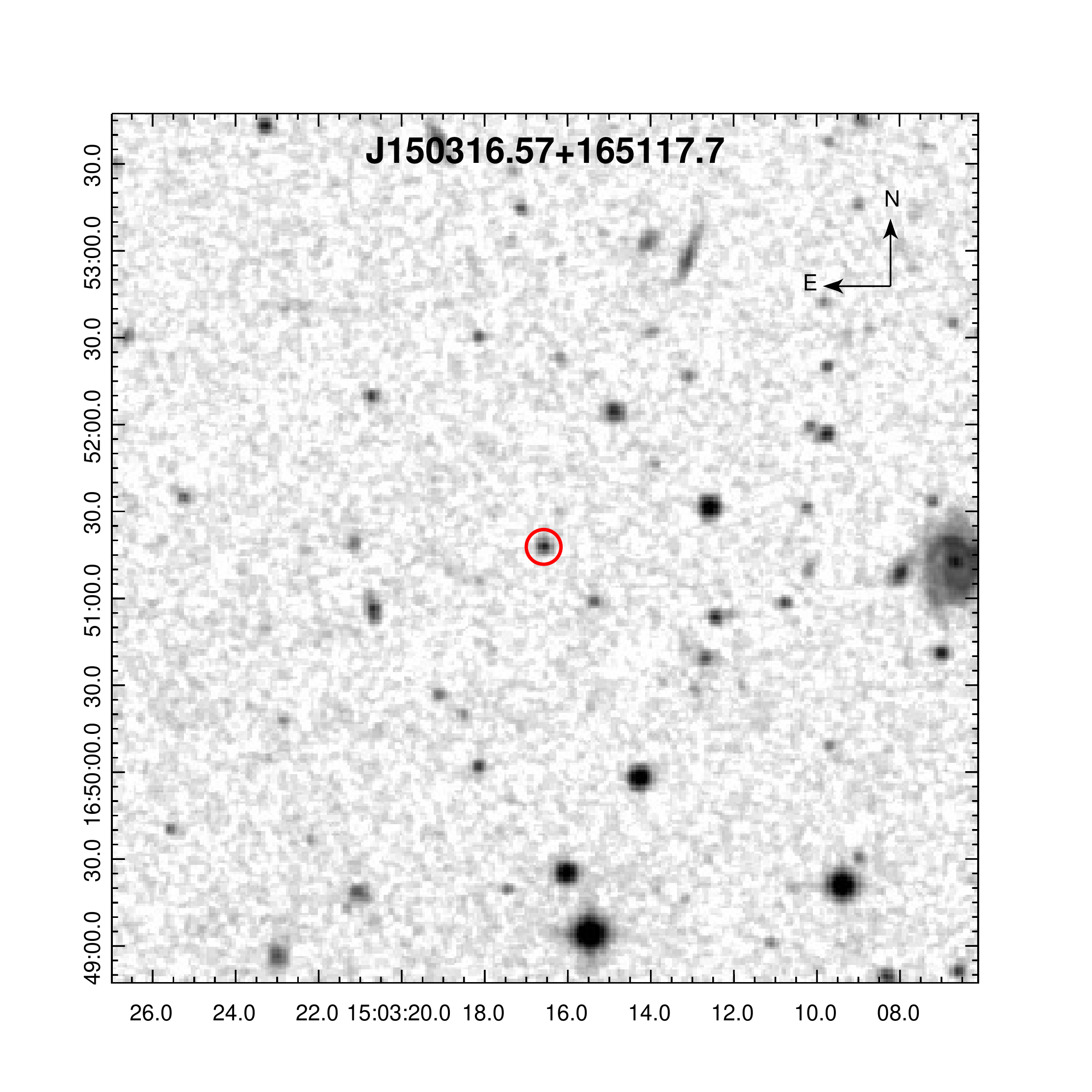} \\
\end{array}$
\end{center}
\caption{(Left panel) Optical spectrum of SDSS J150316.57+165117.7 potential candidate of the UGS with FL8Y J1503.3+1651, in the upper part it is shown the Signal-to-Noise Ratio of the spectrum. (Right panel) The finding chart ( $5'\times 5'$ ) retrieved from the Digitized Sky Survey (DSS) highlighting the location of the potential counterpart:  SDSS J150316.57+165117.7 (red circle).}
\label{fig:J1503}
\end{figure*}

\begin{figure*}{}
\begin{center}$
\begin{array}{cc}
\includegraphics[width=\mywidth]{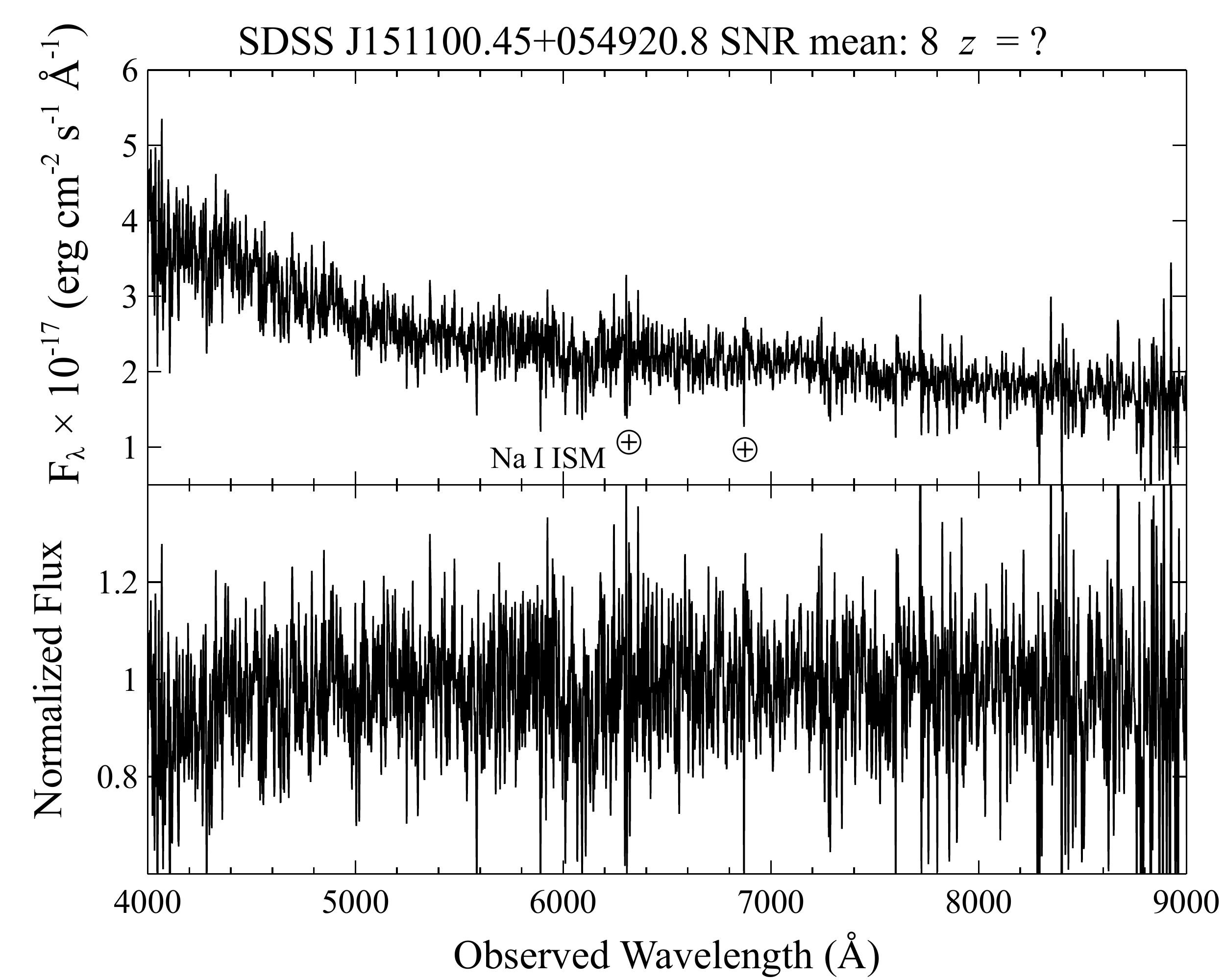} &
\includegraphics[clip=true, width=7cm]{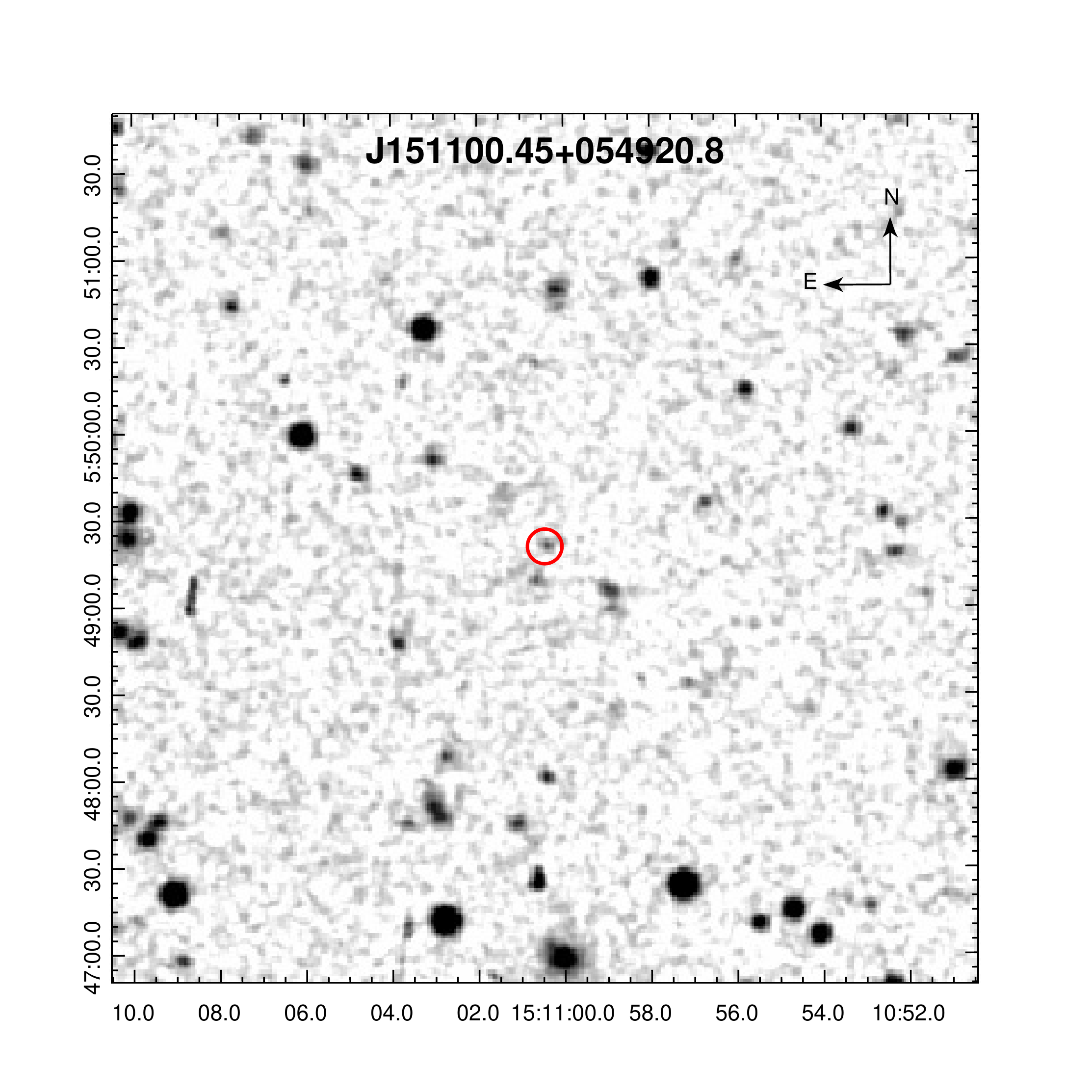} \\
\end{array}$
\end{center}
\caption{(Left panel) Optical spectrum of SDSS J151100.45+054920.8 potential candidate of the UGS with FL8Y J1511.4+0549, in the upper part it is shown the Signal-to-Noise Ratio of the spectrum. (Right panel) The finding chart ( $5'\times 5'$ ) retrieved from the Digitized Sky Survey (DSS) highlighting the location of the potential counterpart:  SDSS J151100.45+054920.8 (red circle).}
\label{fig:J1511}
\end{figure*}

\begin{figure*}{}
\begin{center}$
\begin{array}{cc}
\includegraphics[width=\mywidth]{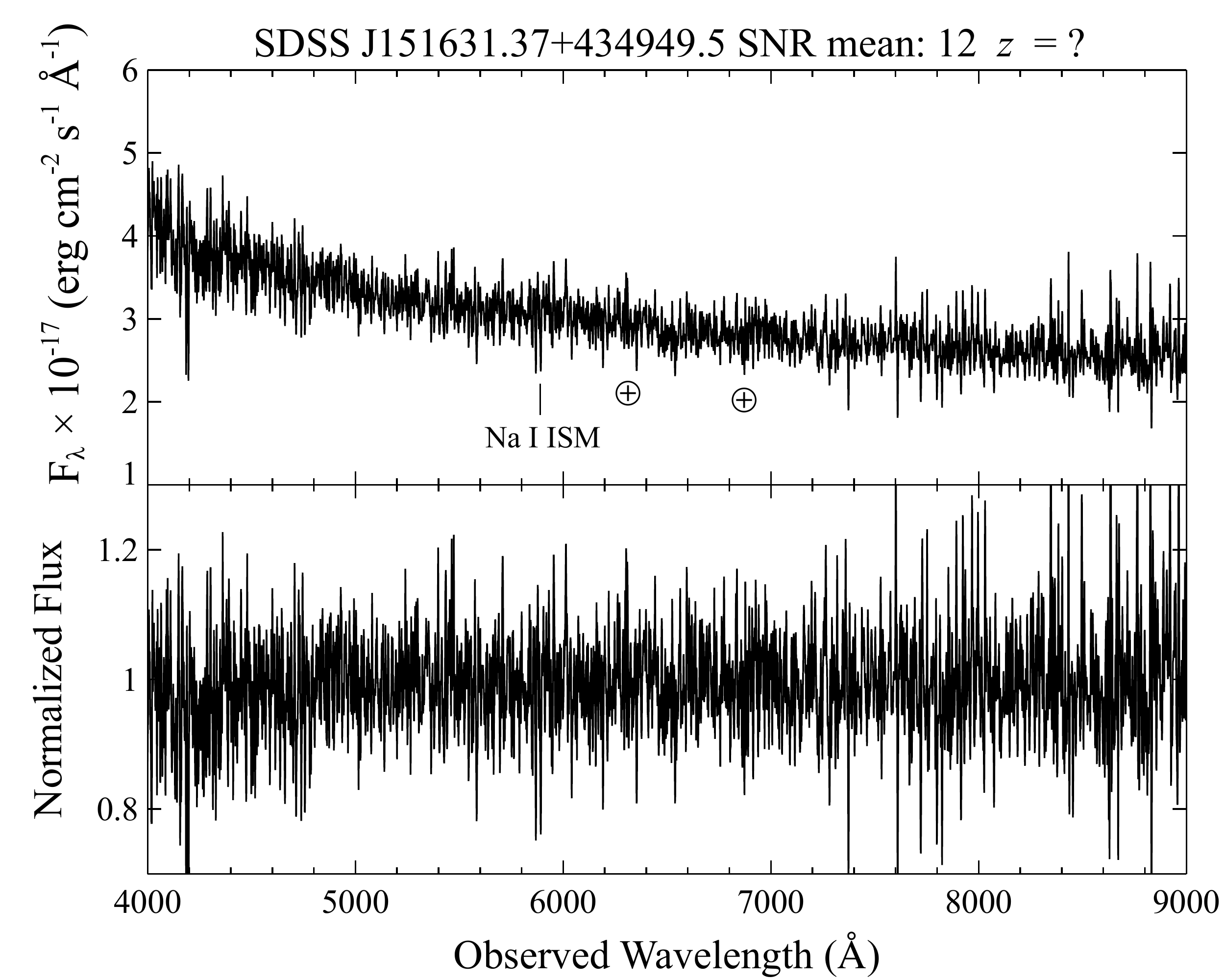} &
\includegraphics[clip=true, width=7cm]{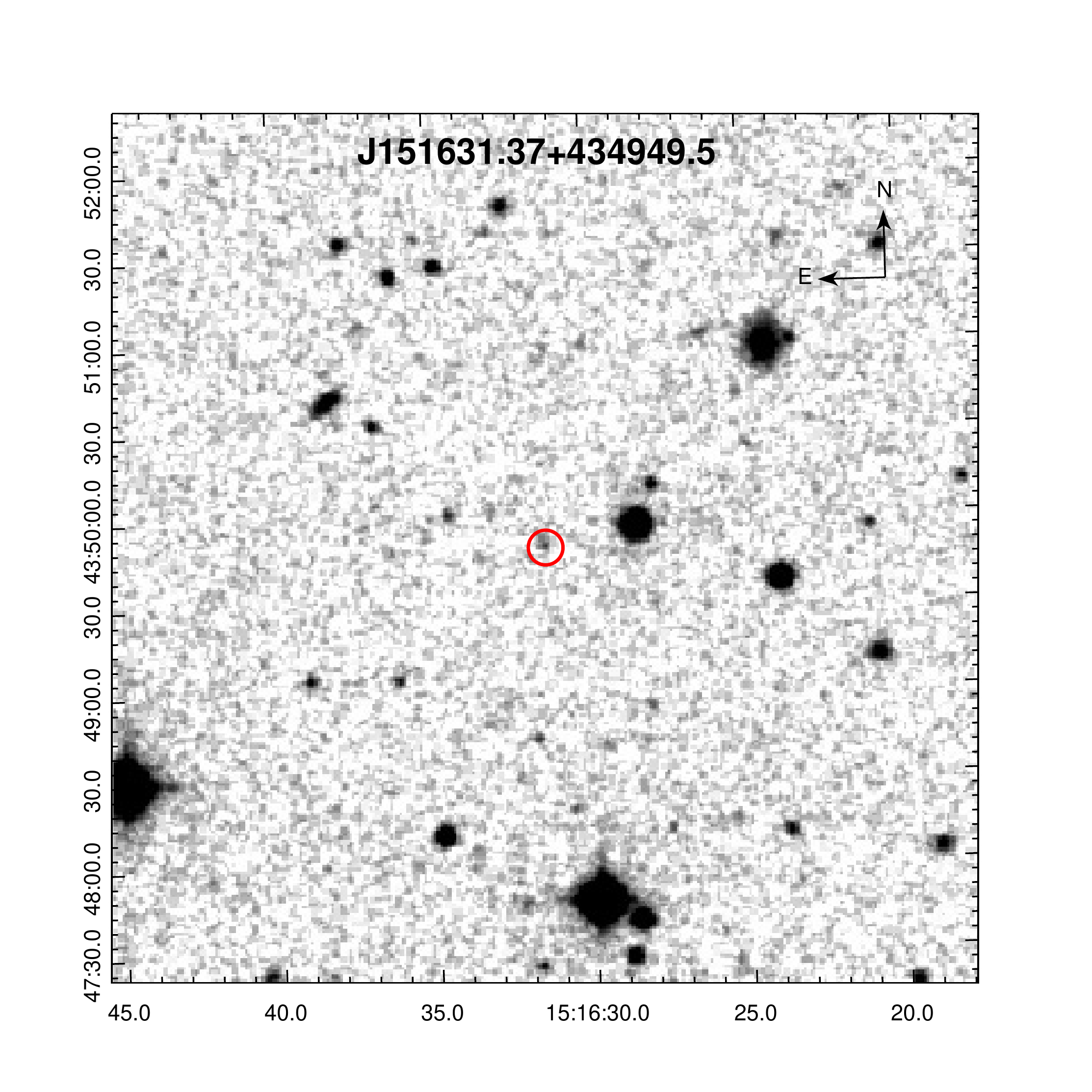} \\
\end{array}$
\end{center}
\caption{(Left panel) Optical spectrum of SDSS J151631.37+434949.5 potential candidate of the UGS with FL8Y J1516.3+4353, in the upper part it is shown the Signal-to-Noise Ratio of the spectrum. (Right panel) The finding chart ( $5'\times 5'$ ) retrieved from the Digitized Sky Survey (DSS) highlighting the location of the potential counterpart:  SDSS J151631.37+434949.5 (red circle).}
\label{fig:J1516}
\end{figure*}

\begin{figure*}{}
\begin{center}$
\begin{array}{cc}
\includegraphics[width=\mywidth]{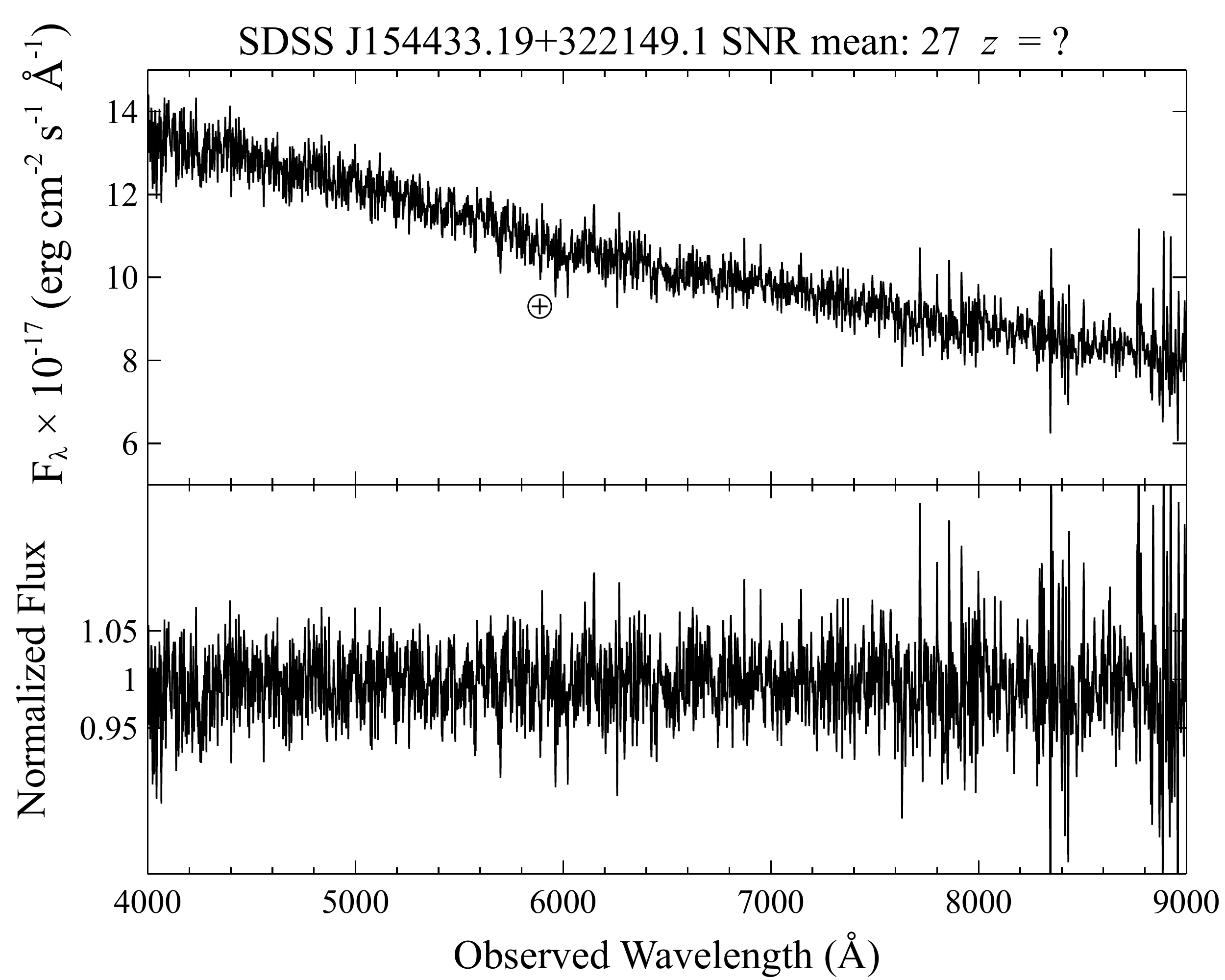} &
\includegraphics[clip=true, width=7cm]{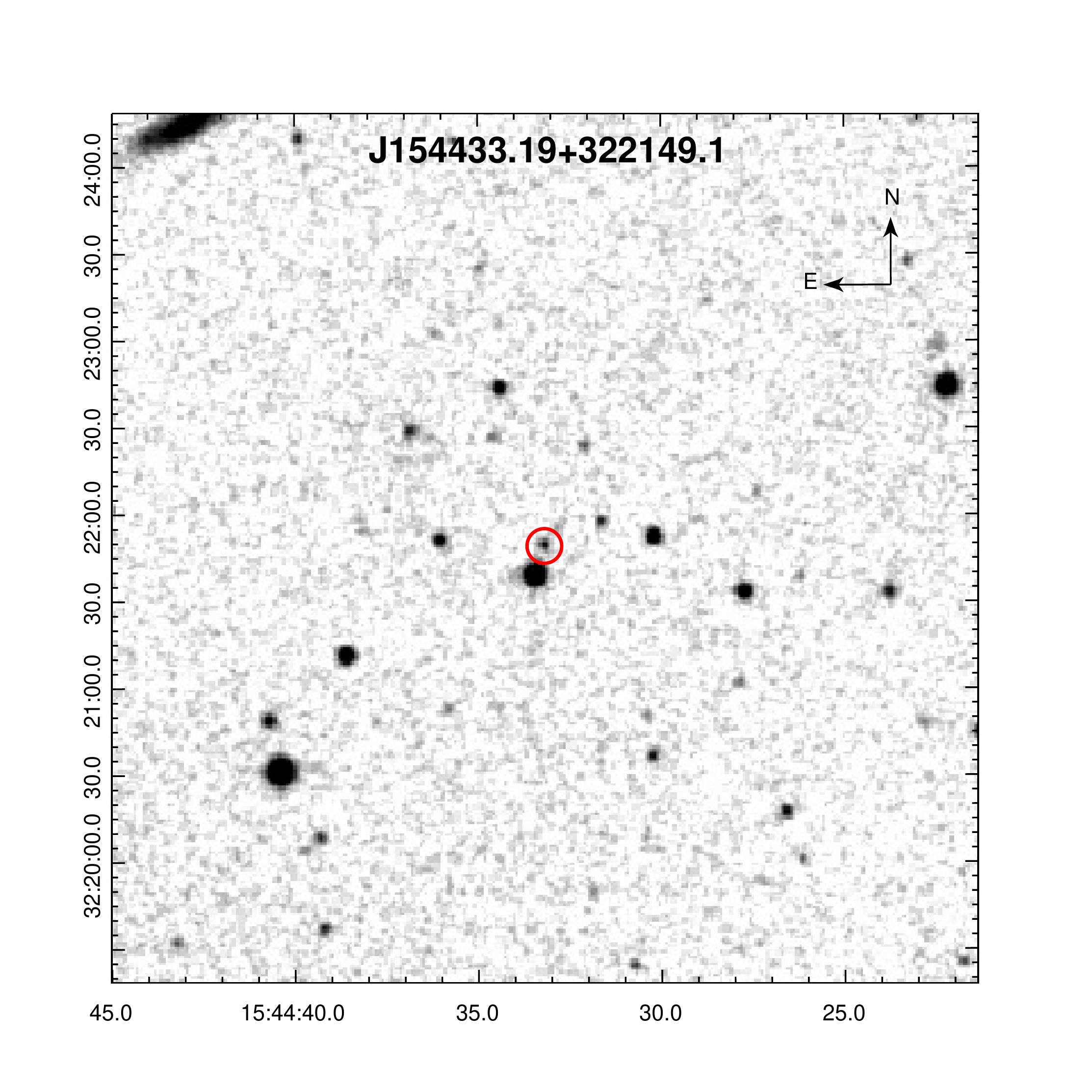} \\
\end{array}$
\end{center}
\caption{(Left panel) Optical spectrum of SDSS J154433.19+322149.1 potential candidate of the UGS with FL8Y J1544.9+3218, in the upper part it is shown the Signal-to-Noise Ratio of the spectrum. (Right panel) The finding chart ( $5'\times 5'$ ) retrieved from the Digitized Sky Survey (DSS) highlighting the location of the potential counterpart: SDSS J154433.19+322149.1 (red circle).}
\label{fig:J1544}
\end{figure*}

\begin{figure*}{}
\begin{center}$
\begin{array}{cc}
\includegraphics[width=\mywidth]{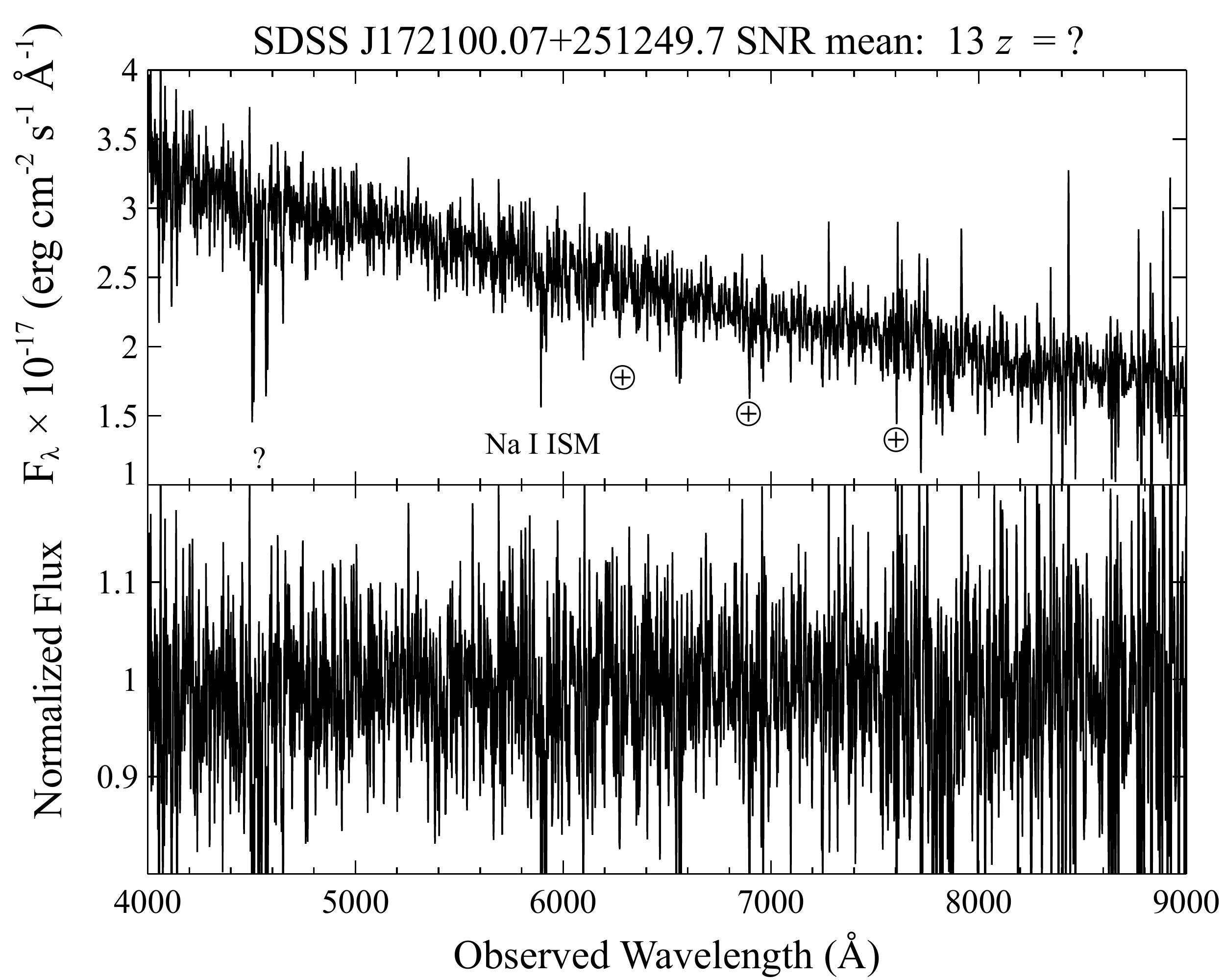} &
\includegraphics[clip=true, width=7cm]{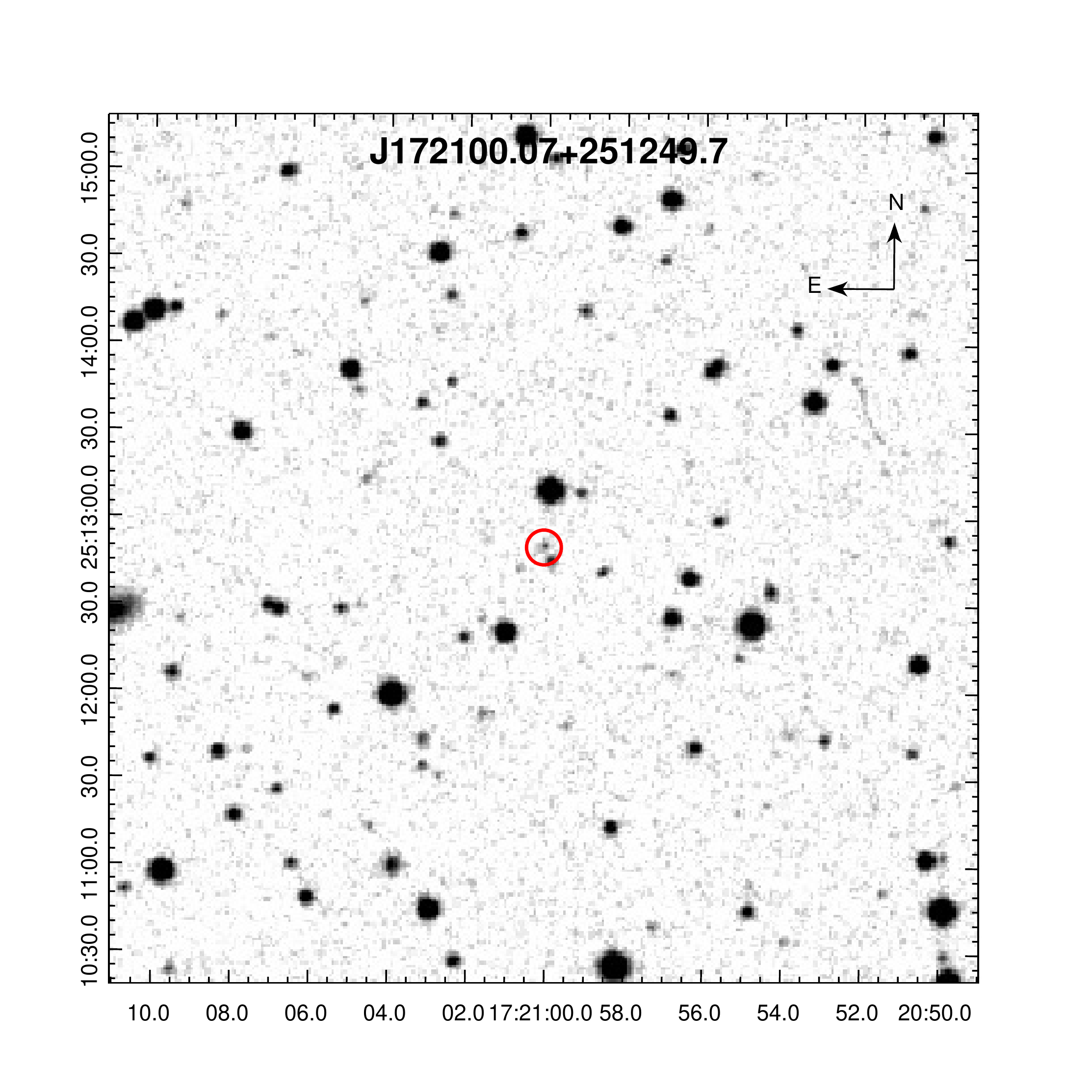} \\
\end{array}$
\end{center}
\caption{(Left panel) Optical spectrum of SDSS J172100.07+251249.7 potential candidate of the UGS with FL8Y J1721.3+2529, in the upper part it is shown the Signal-to-Noise Ratio of the spectrum. (Right panel) The finding chart ( $5'\times 5'$ ) retrieved from the Digitized Sky Survey (DSS) highlighting the location of the potential counterpart: SDSS J172100.07+251249.7 (red circle).}
\label{fig:J1721}
\end{figure*}

\begin{figure*}{}
\begin{center}$
\begin{array}{cc}
\includegraphics[width=\mywidth]{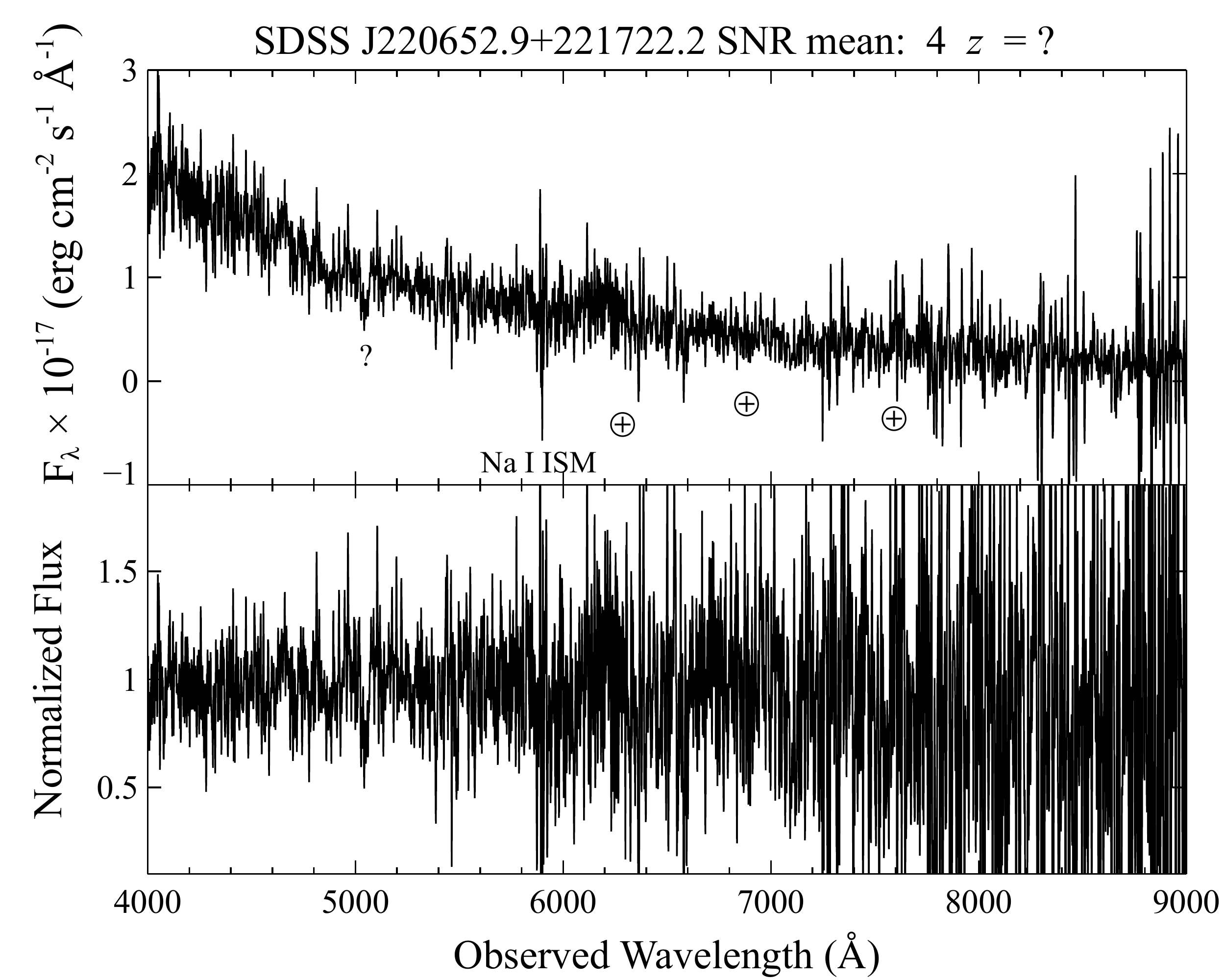} &
\includegraphics[clip=true, width=7cm]{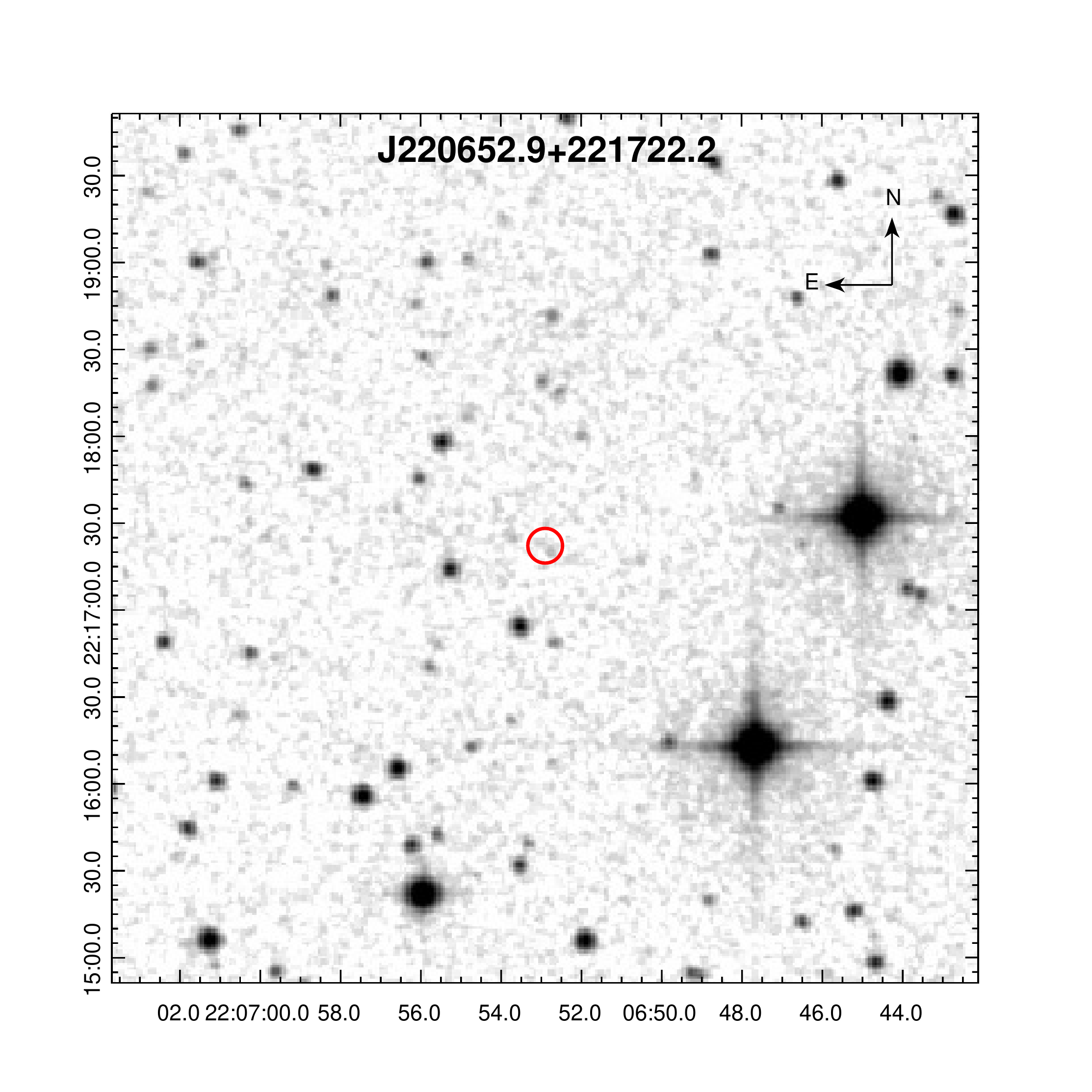} \\
\end{array}$
\end{center}
\caption{(Left panel) Optical spectrum of SDSS J220652.9+221722.2 potential candidate of the UGS with FL8Y J2207.1+2222, in the upper part it is shown the Signal-to-Noise Ratio of the spectrum. (Right panel) The finding chart ( $5'\times 5'$ ) retrieved from the Digitized Sky Survey (DSS) highlighting the location of the potential counterpart: SDSS J220652.9+221722.2 (red circle).}
\label{fig:J2206}
\end{figure*}

\begin{figure*}{}
\begin{center}$
\begin{array}{cc}
\includegraphics[width=\mywidth]{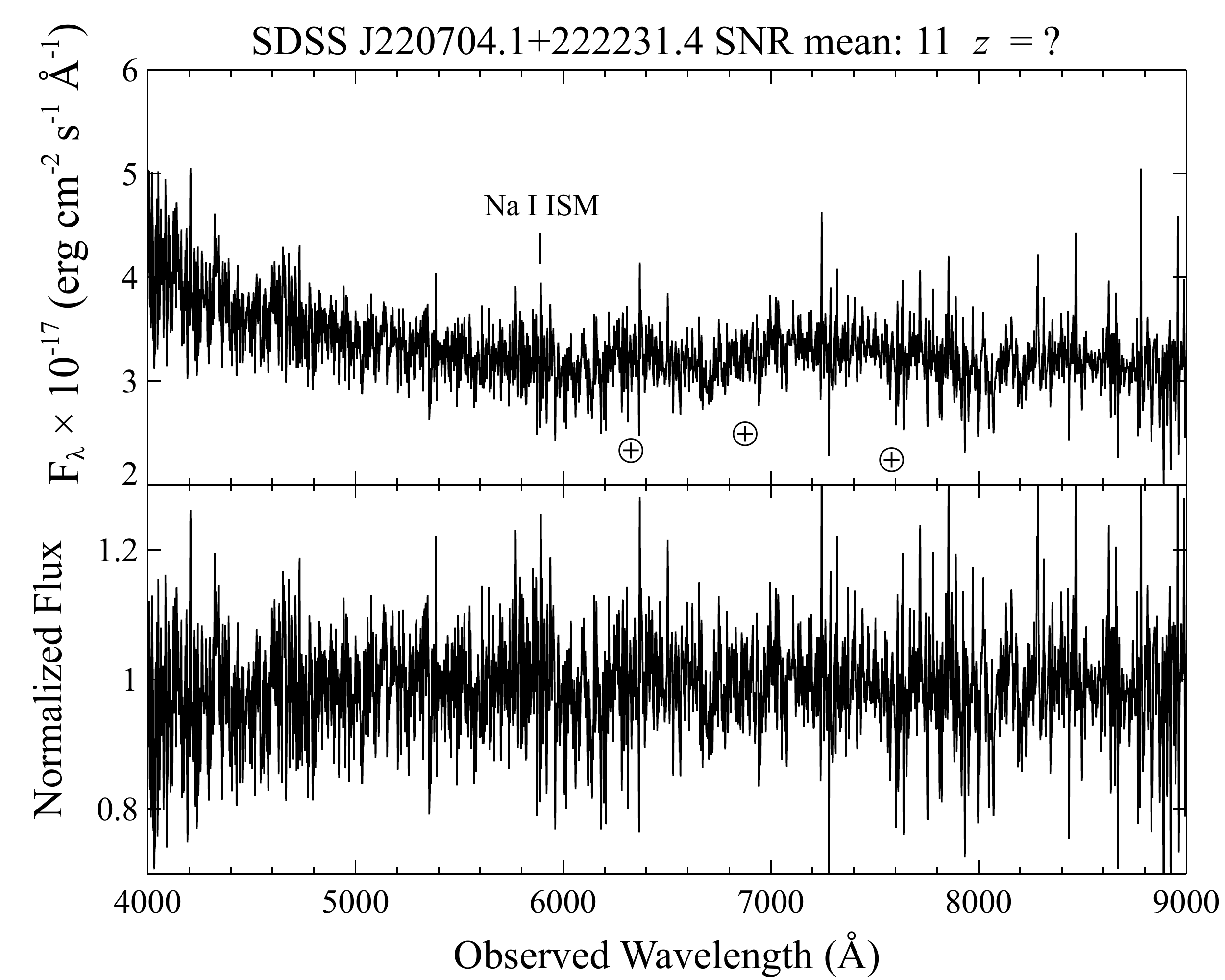} &
\includegraphics[clip=true, width=7cm]{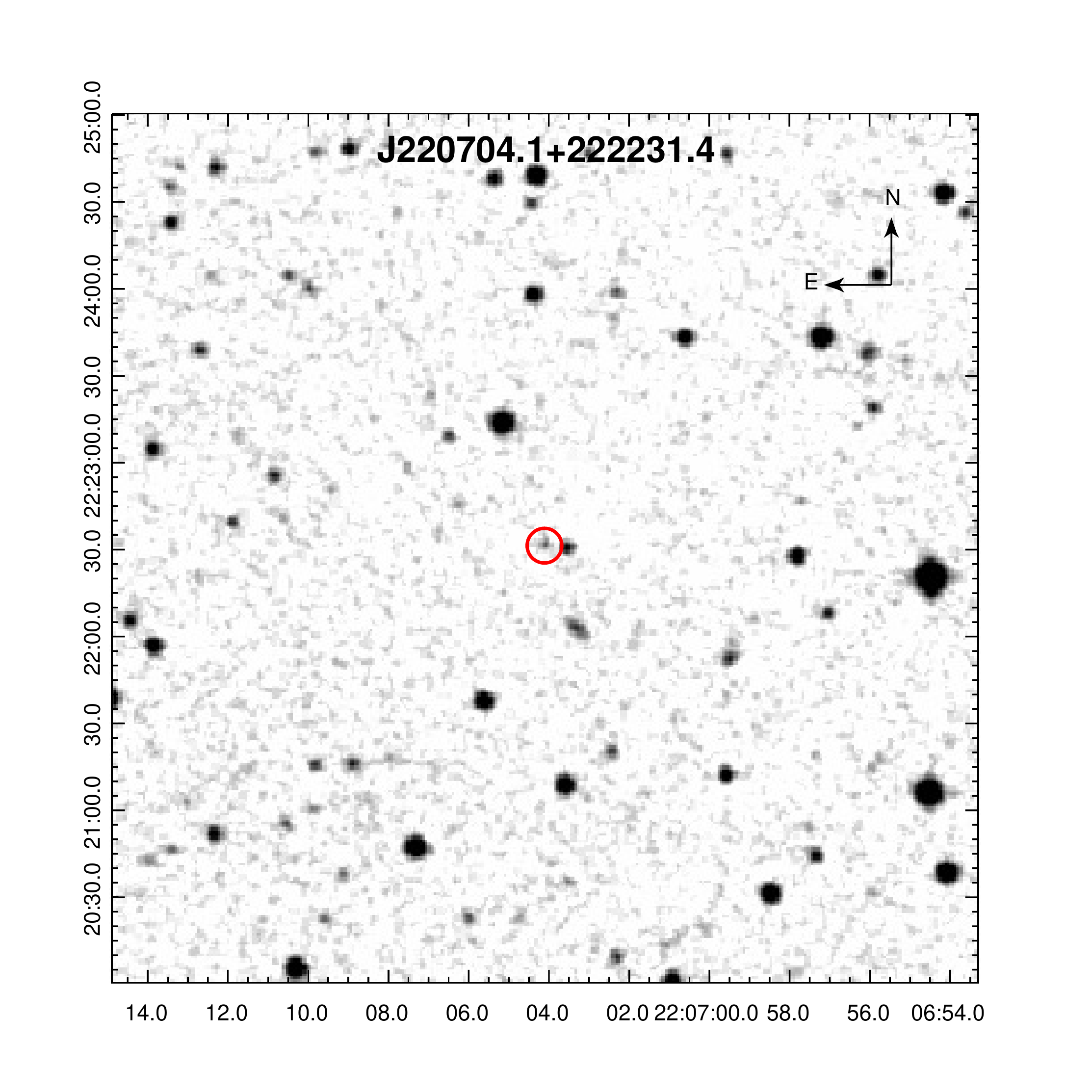} \\
\end{array}$
\end{center}
\caption{(Left panel) Optical spectrum of SDSS J220704.1+222231.4 potential candidate of the UGS with FL8Y J2207.1+2222, in the upper part it is shown the Signal-to-Noise Ratio of the spectrum. (Right panel) The finding chart ( $5'\times 5'$ ) retrieved from the Digitized Sky Survey (DSS) highlighting the location of the potential counterpart: SDSS J220704.1+222231.4 (red circle).}
\label{fig:J2207}
\end{figure*}

\begin{figure*}{}
\begin{center}$
\begin{array}{cc}
\includegraphics[width=\mywidth]{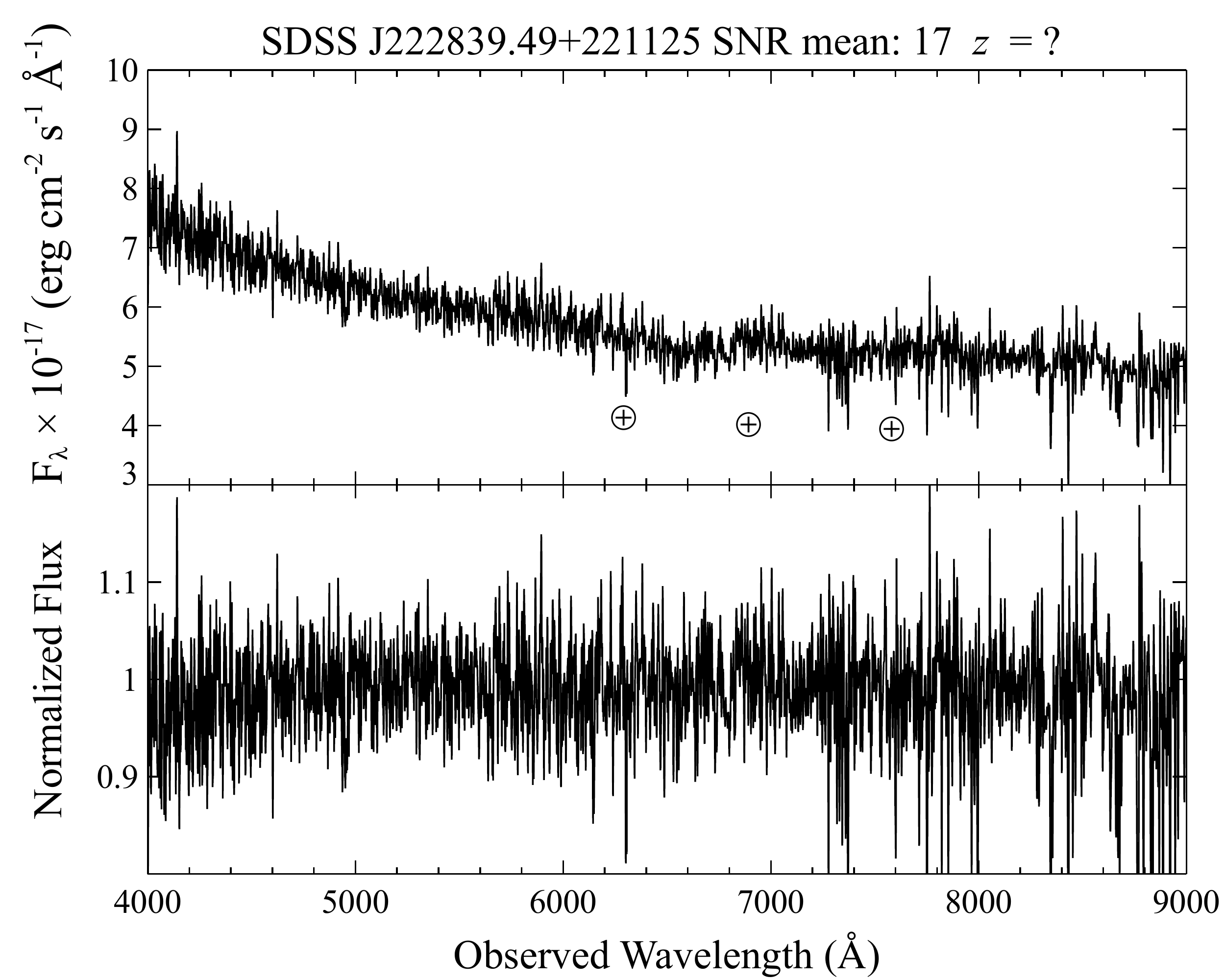} &
\includegraphics[clip=true, width=7cm]{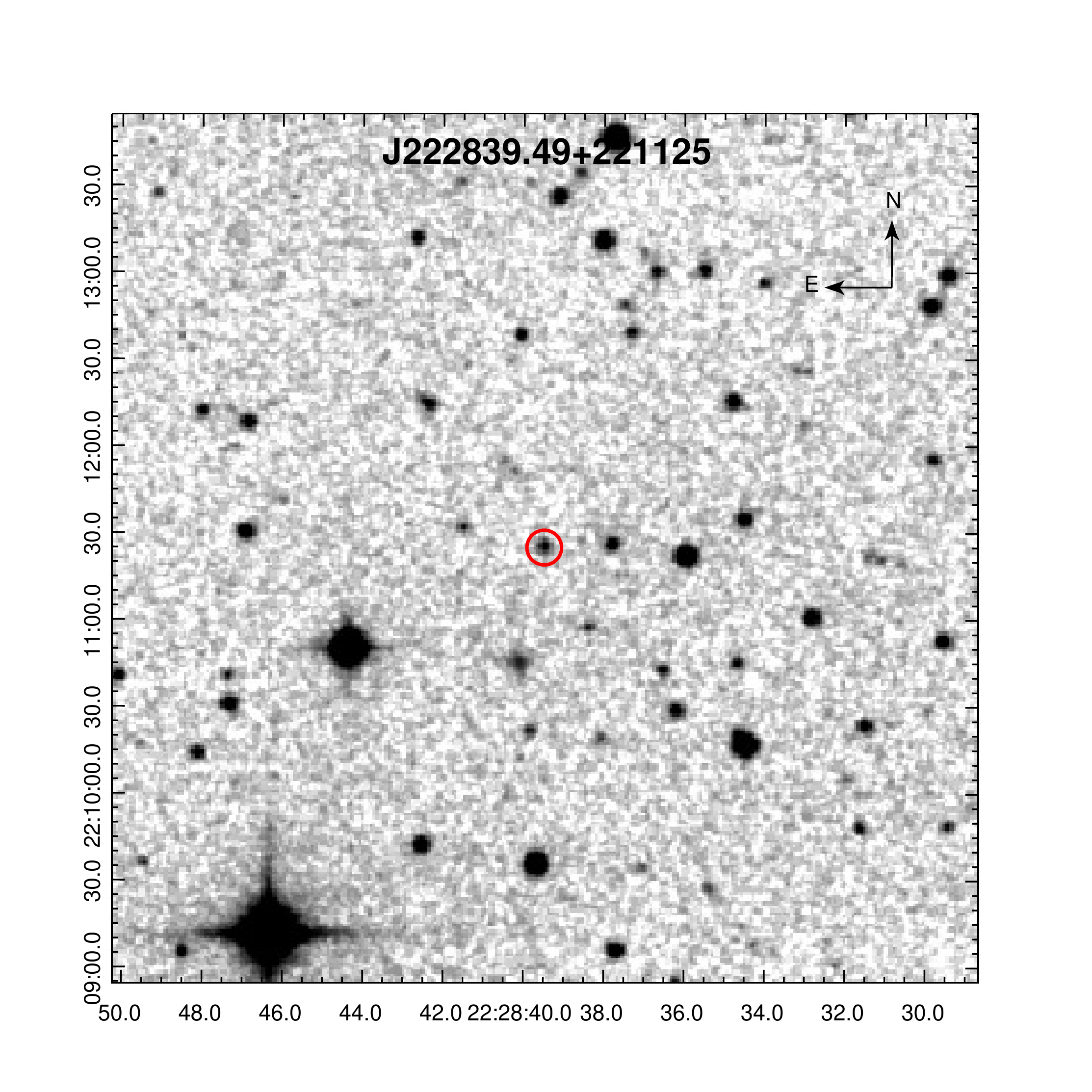} \\
\end{array}$
\end{center}
\caption{(Left panel) Optical spectrum of SDSS J222839.49+221125 potential candidate of the UGS with FL8Y J2228.5+2211, in the upper part it is shown the Signal-to-Noise Ratio of the spectrum. (Right panel) The finding chart ( $5'\times 5'$ ) retrieved from the Digitized Sky Survey (DSS) highlighting the location of the potential counterpart: SDSS J222839.49+221125 (red circle).}
\label{fig:J2228}
\end{figure*}

\begin{figure*}{}
\begin{center}$
\begin{array}{cc}
\includegraphics[width=\mywidth]{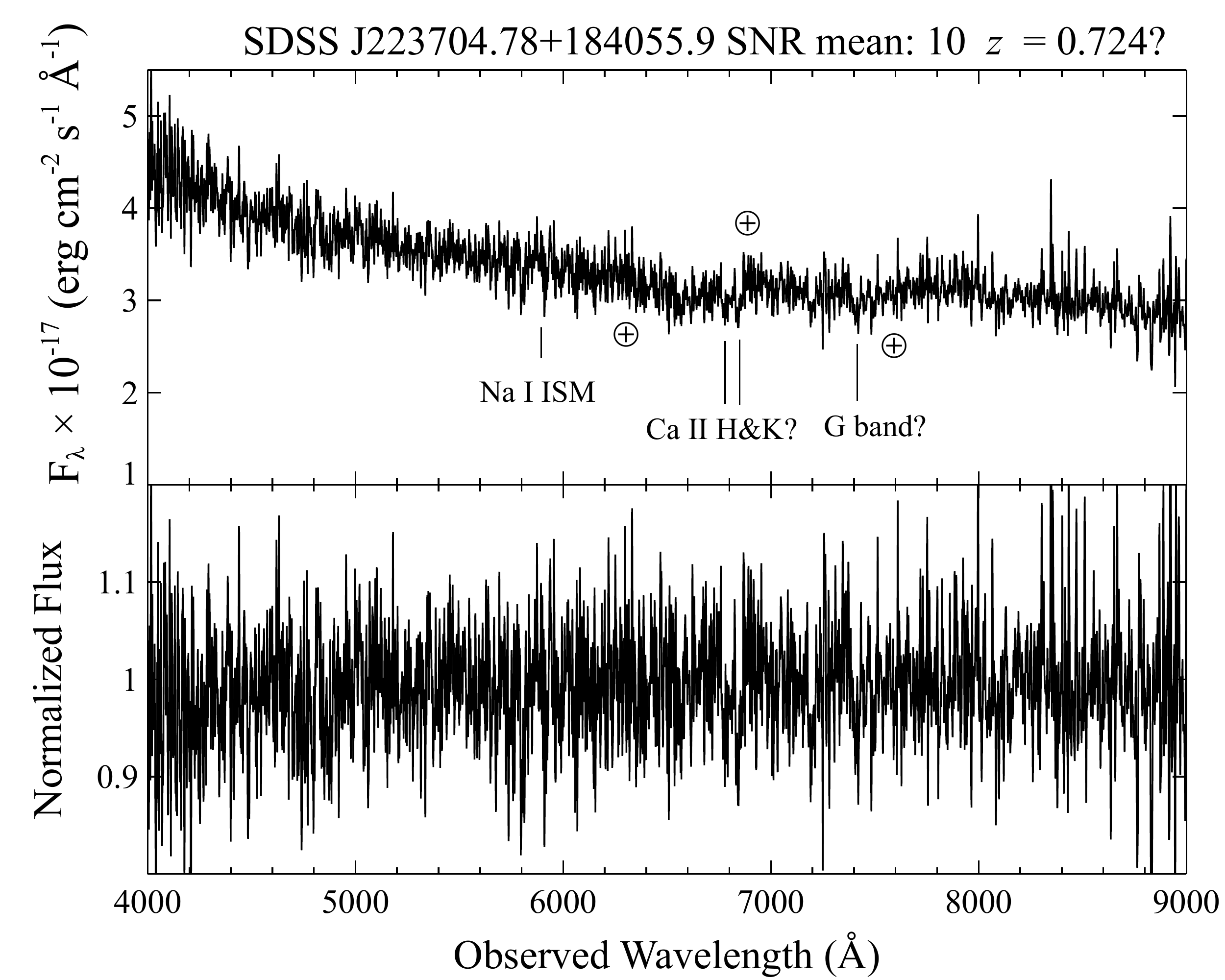} &
\includegraphics[clip=true, width=7cm]{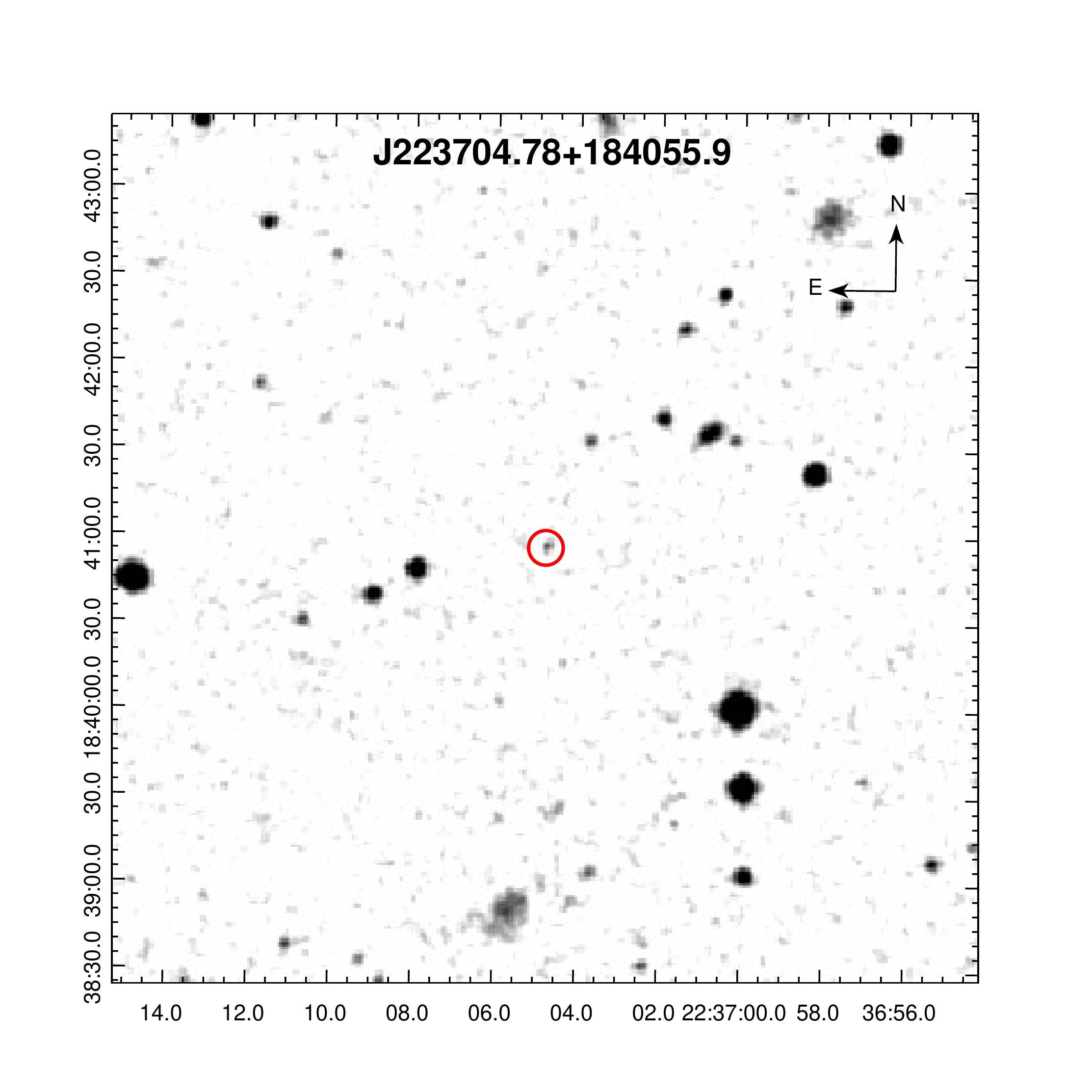} \\
\end{array}$
\end{center}
\caption{(Left panel) Optical spectrum of SDSS J223704.78+184055.9 potential candidate of the UGS with FL8Y 2236.9+1840, in the upper part it is shown the Signal-to-Noise Ratio of the spectrum. (Right panel) The finding chart ( $5'\times 5'$ ) retrieved from the Digitized Sky Survey (DSS) highlighting the location of the potential counterpart: SDSS J223704.78+184055.9 (red circle).}
\label{fig:J2237}
\end{figure*}

\begin{figure*}{}
\begin{center}$
\begin{array}{cc}
\includegraphics[width=\mywidth]{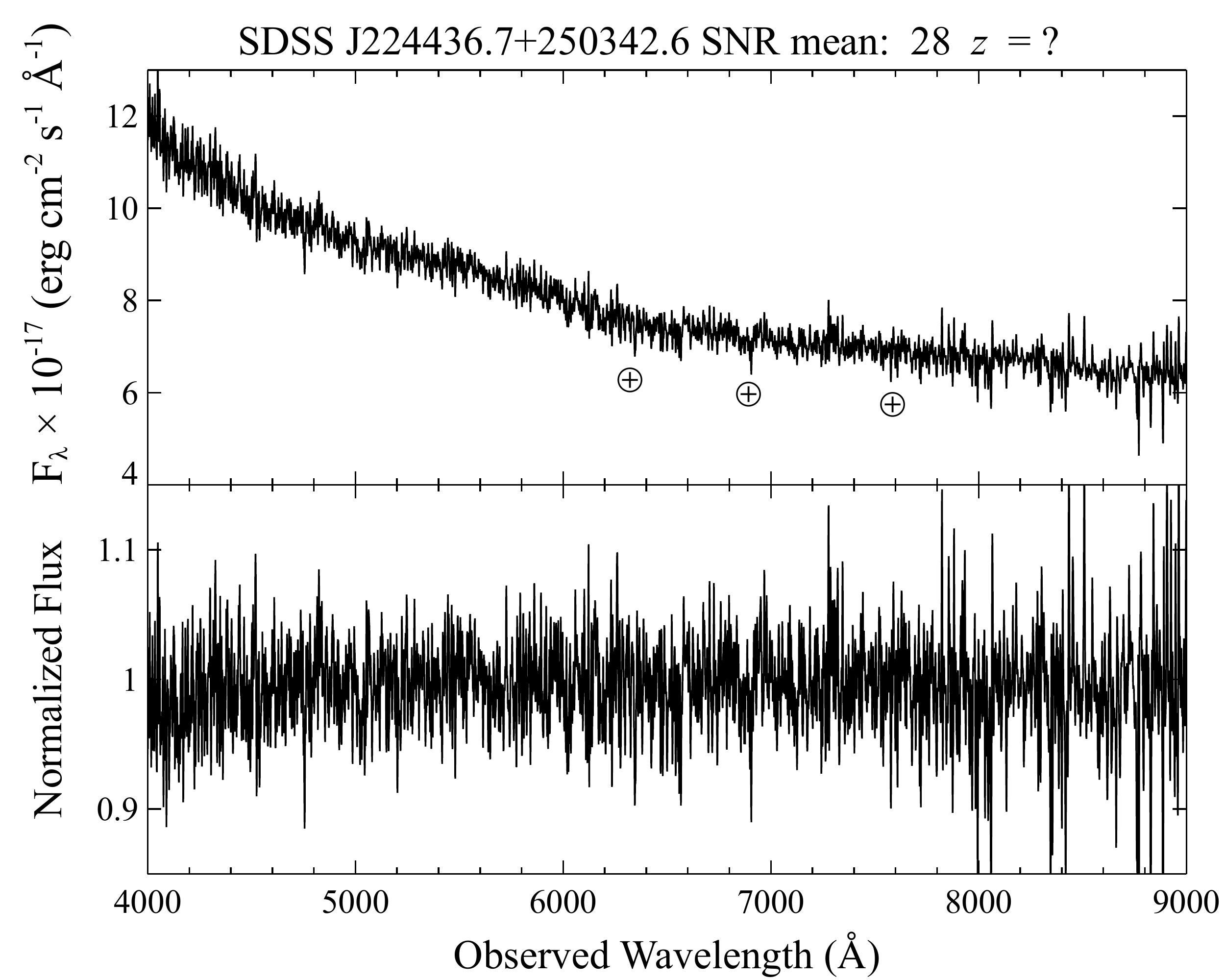} &
\includegraphics[clip=true, width=7cm]{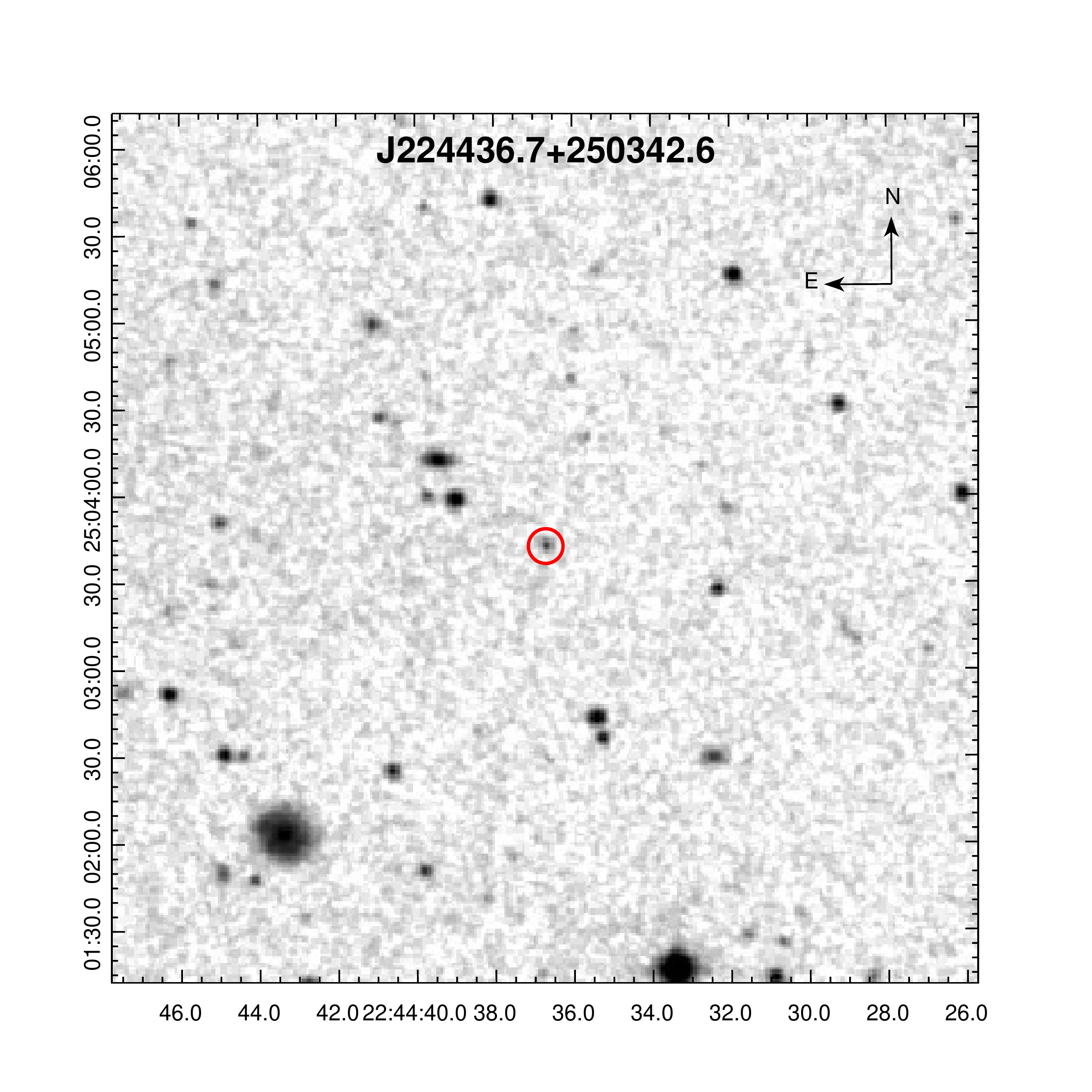} \\
\end{array}$
\end{center}
\caption{(Left panel) Optical spectrum of SDSS J224436.7+250342.6 potential candidate of the UGS with FL8Y J2244.6+2502, in the upper part it is shown the Signal-to-Noise Ratio of the spectrum. (Right panel) The finding chart ( $5'\times 5'$ ) retrieved from the Digitized Sky Survey (DSS) highlighting the location of the potential counterpart: SDSS J224436.7+250342.6 (red circle).}
\label{fig:J2244}
\end{figure*}

\clearpage

\section{Blazar Candidates of Uncertain Type}

\begin{figure*}{}
\begin{center}$
\begin{array}{cc}
\includegraphics[width=\mywidth]{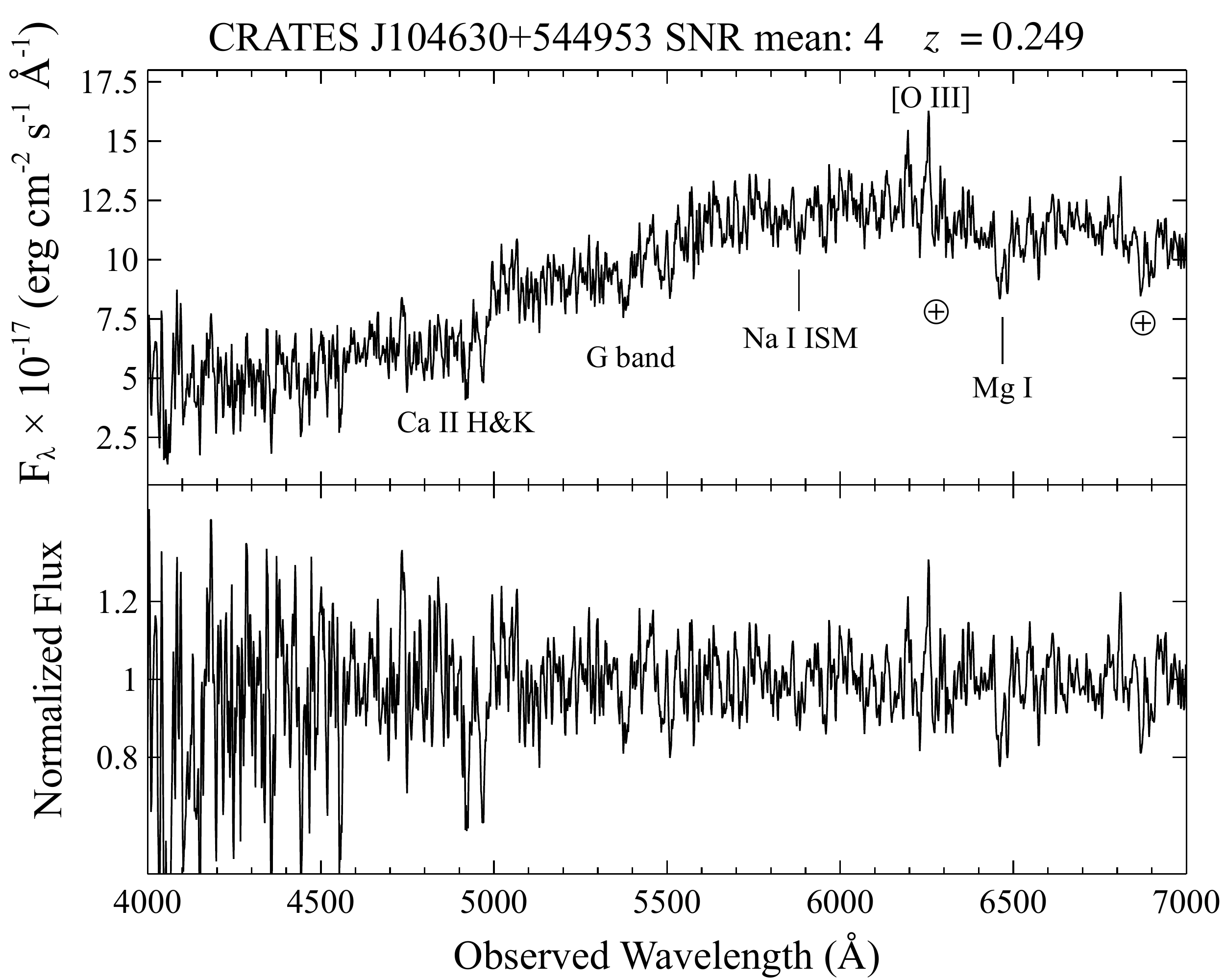} &
\includegraphics[clip=true, width=7cm]{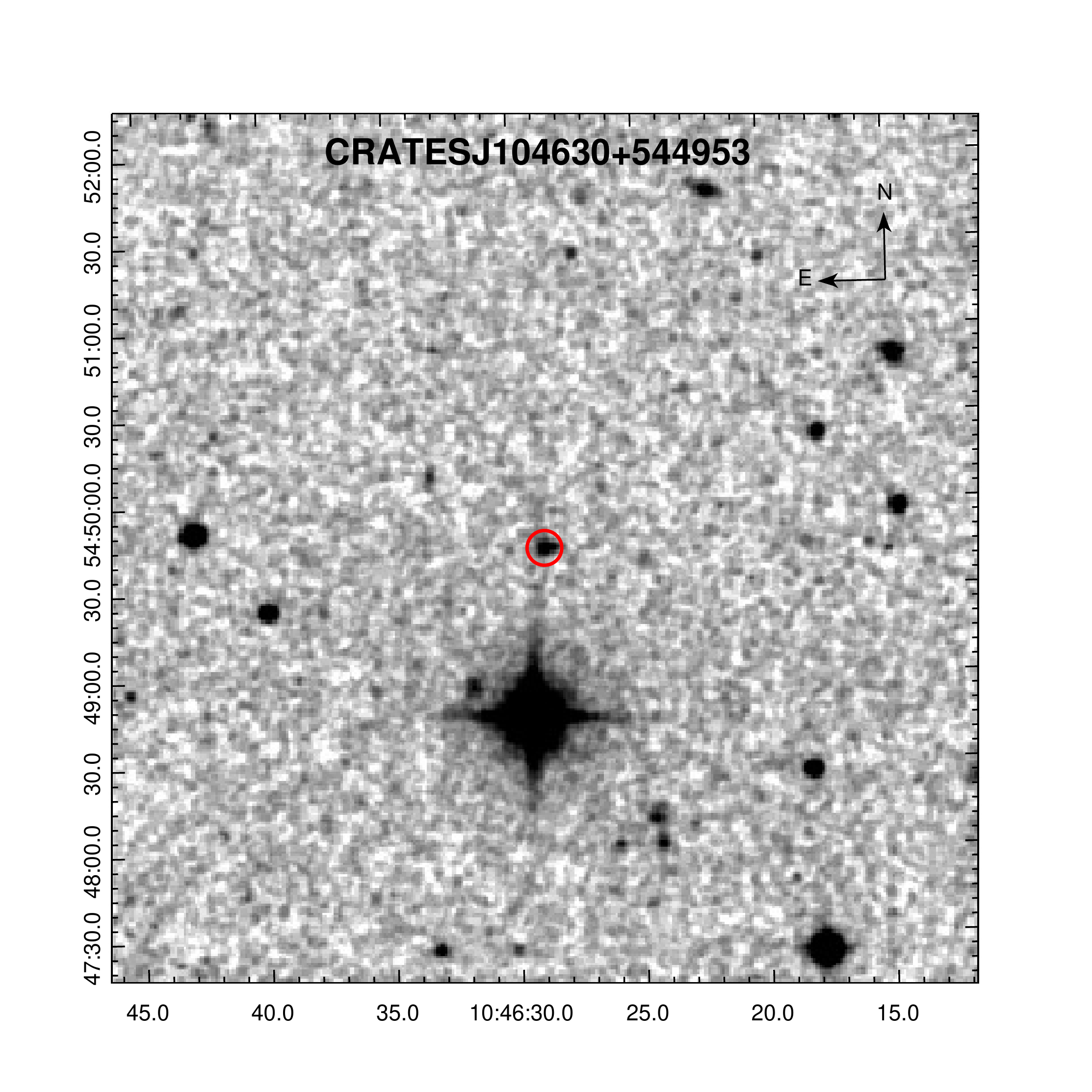} \\
\end{array}$
\end{center}
\caption{(Left panel) Optical spectrum of  CRATES J104630+544953 associated with FL8Y J1046.1+5449, in the upper part it is shown the Signal-to-Noise Ratio of the spectrum. (Right panel) The finding chart ( $5'\times 5'$ ) retrieved from the Digitized Sky Survey (DSS) highlighting the location of the counterpart:  CRATES J104630+544953 (red circle).}
\label{fig:J1046}
\end{figure*}

\begin{figure*}{}
\begin{center}$
\begin{array}{cc}
\includegraphics[width=\mywidth]{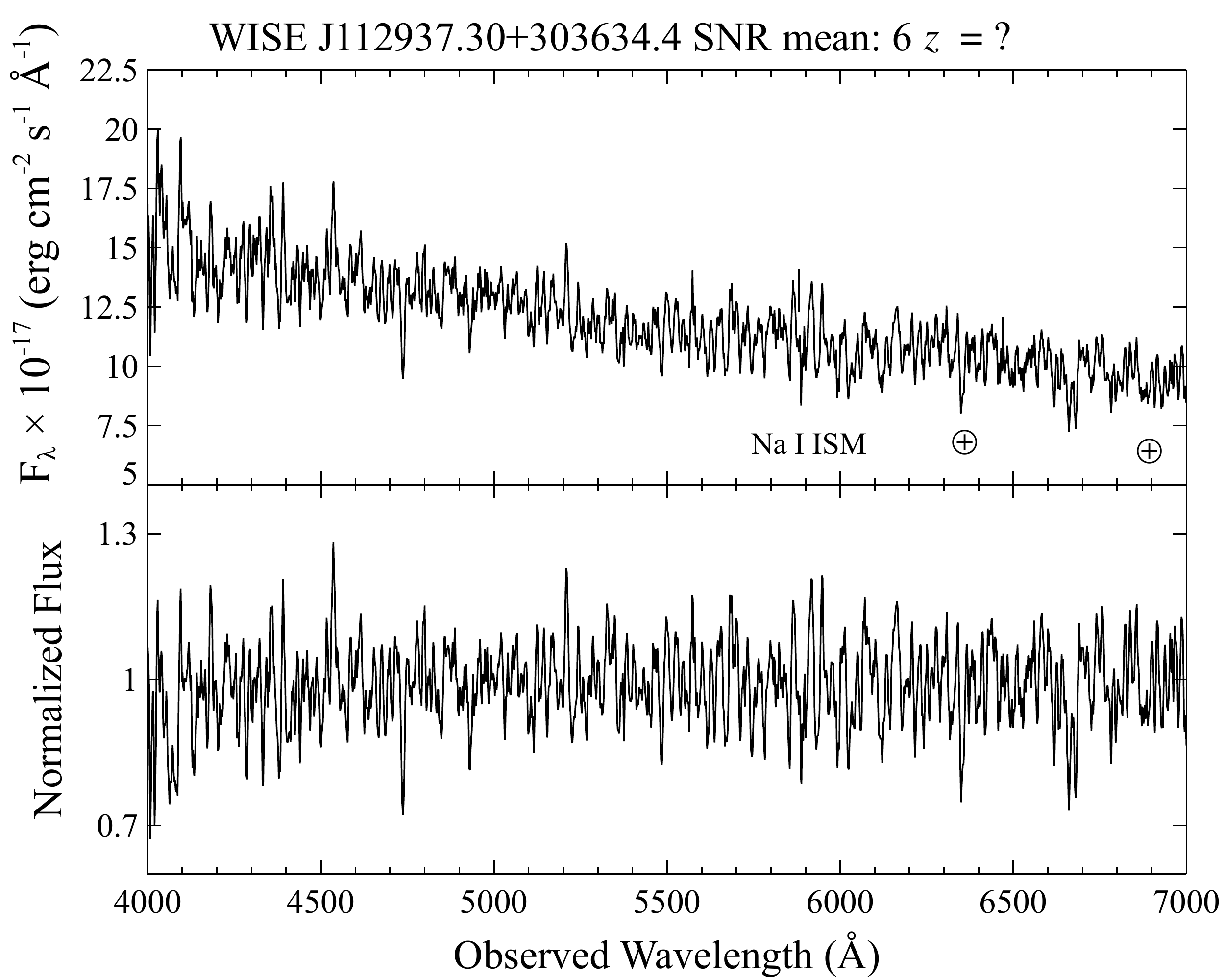} &
\includegraphics[trim=8.25cm 0cm 8.25cm 0cm,clip=true, width=7cm]{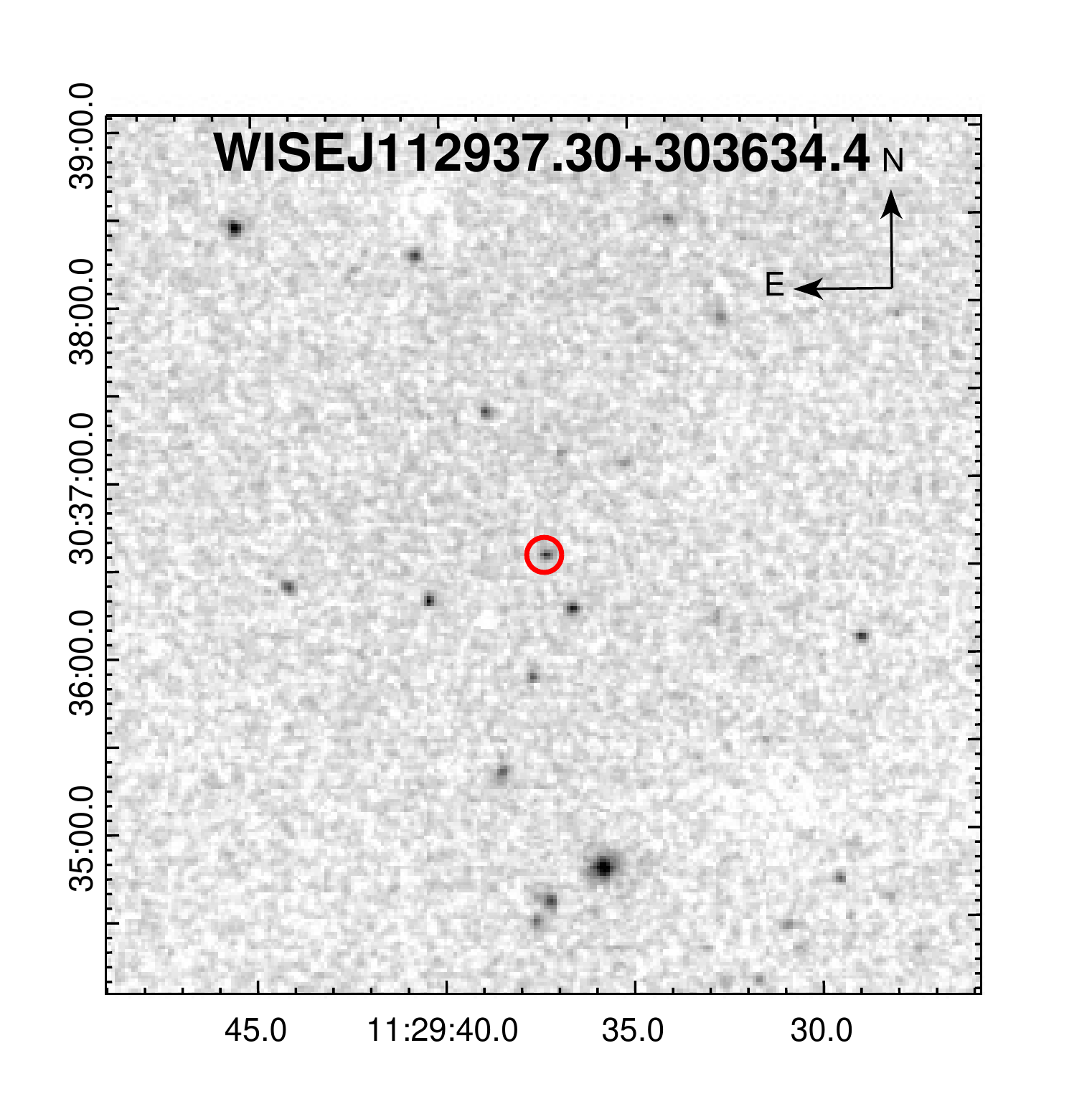} \\
\end{array}$
\end{center}
\caption{(Left panel) Optical spectrum of  WISE J112937.30+303634.4 associated with FL8Y  J1129.4+3033, in the upper part it is shown the Signal-to-Noise Ratio of the spectrum. (Right panel) The finding chart ( $5'\times 5'$ ) retrieved from the Digitized Sky Survey (DSS) highlighting the location of the counterpart:   WISE J112937.30+303634.4 (red circle).}
\label{fig:J1129}
\end{figure*}

\begin{figure*}{}
\begin{center}$
\begin{array}{cc}
\includegraphics[width=\mywidth]{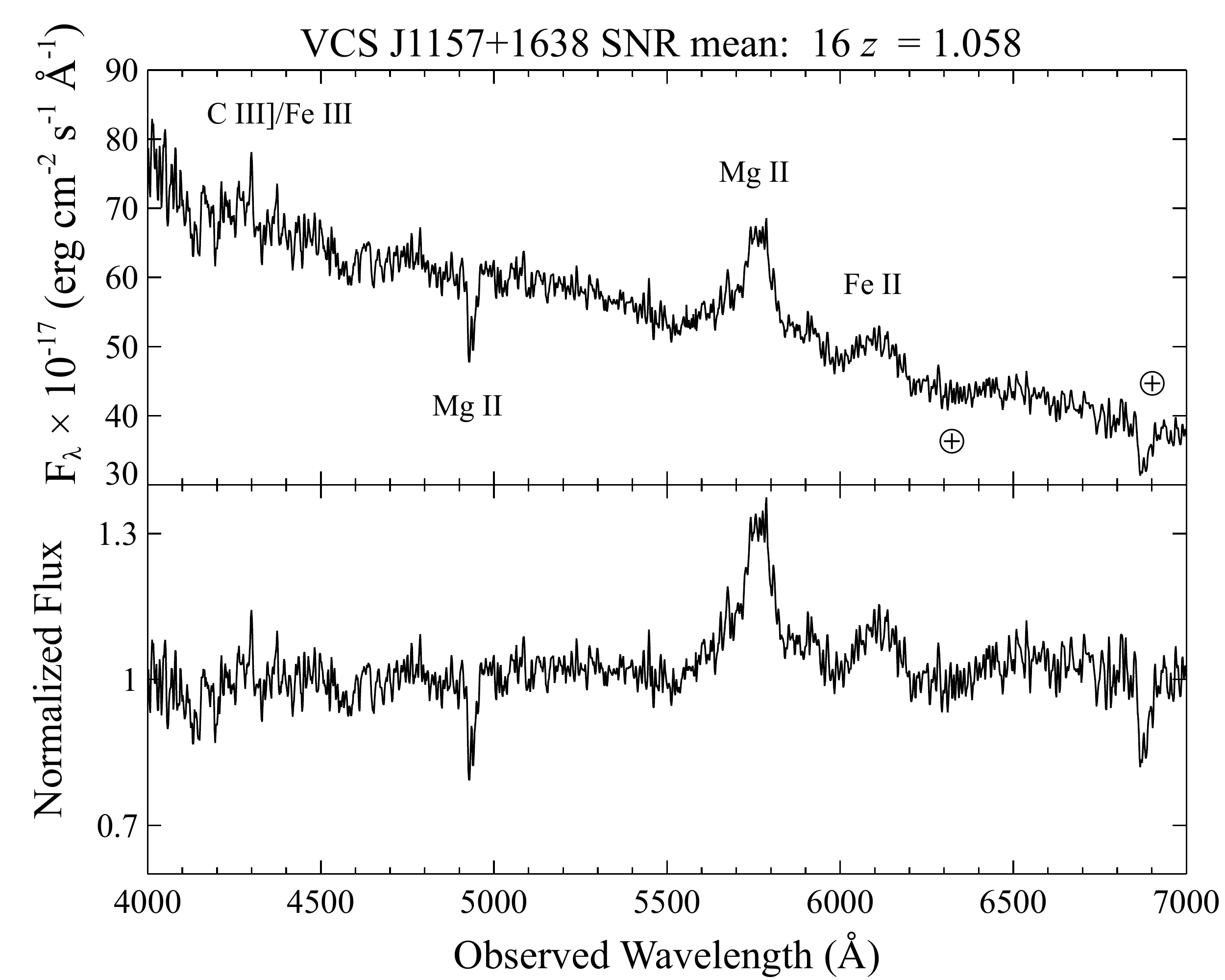} &
\includegraphics[clip=true, width=7cm]{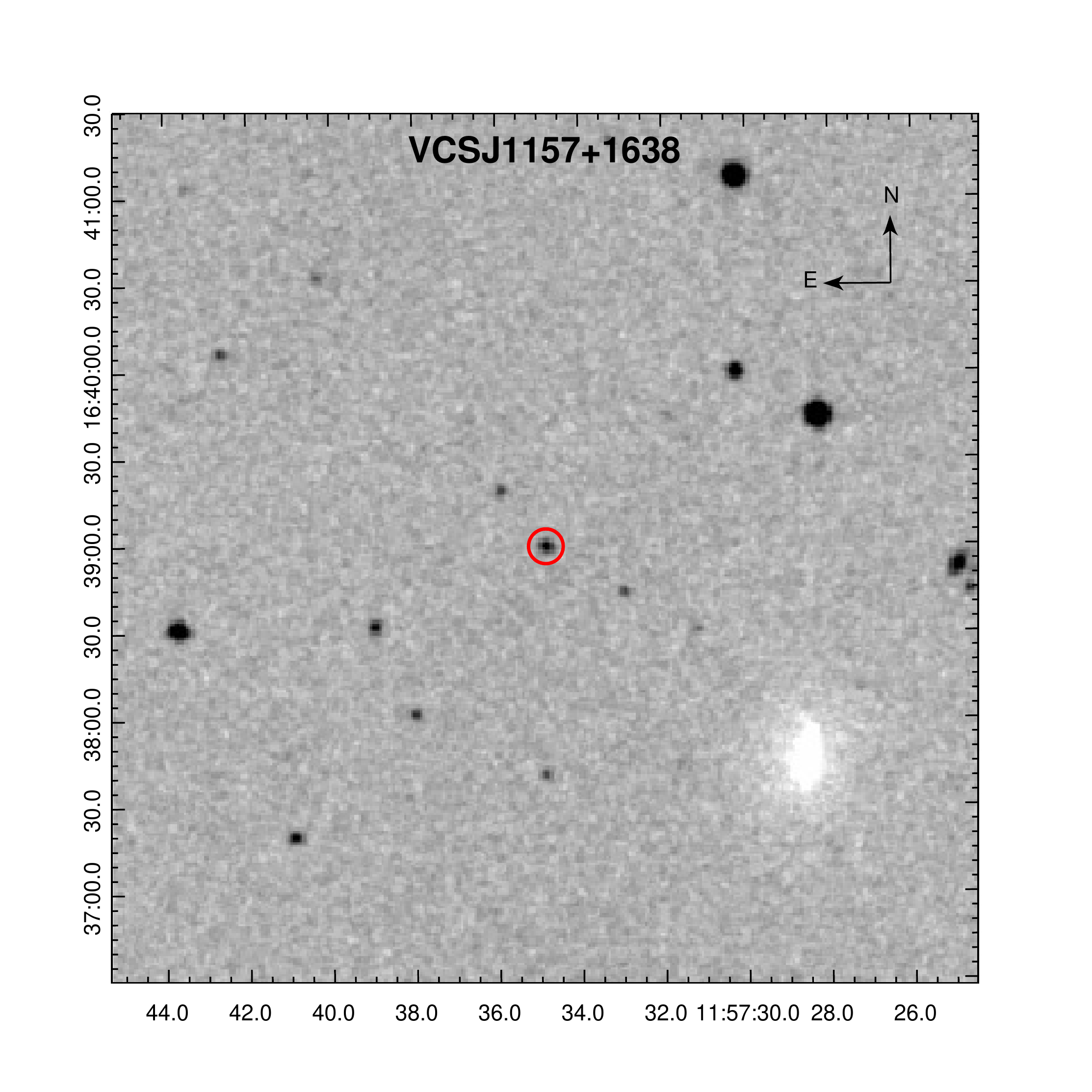} \\
\end{array}$
\end{center}
\caption{(Left panel) Optical spectrum of  VCS J1157+1638 associated with FL8Y J1157.5+1639, in the upper part it is shown the Signal-to-Noise Ratio of the spectrum. (Right panel) The finding chart ( $5'\times 5'$ ) retrieved from the Digitized Sky Survey (DSS) highlighting the location of the counterpart:   VCS J1157+1638 (red circle).}
\label{fig:J1157}
\end{figure*}

\begin{figure*}{}
\begin{center}$
\begin{array}{cc}
\includegraphics[width=\mywidth]{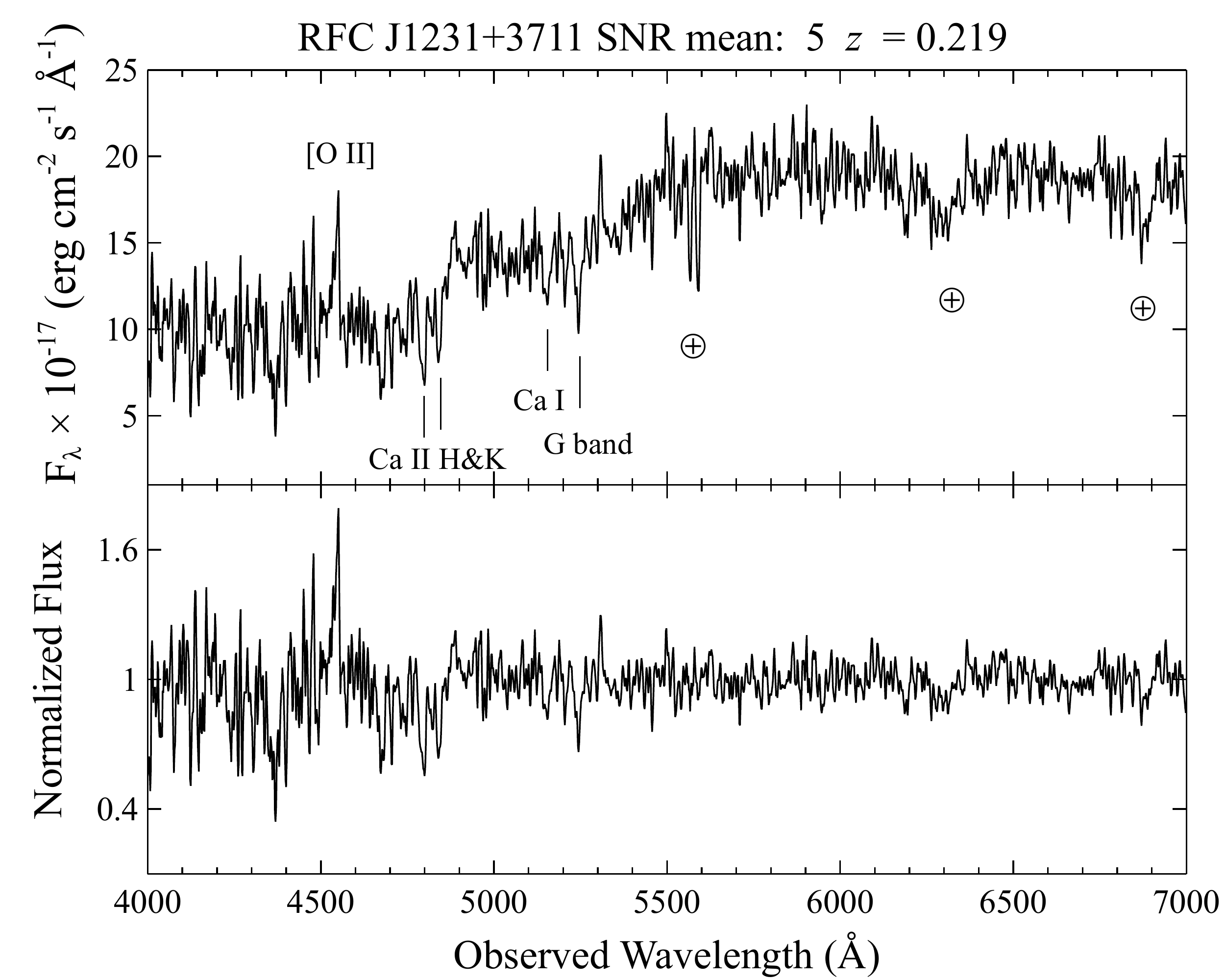} &
\includegraphics[clip=true, width=7cm]{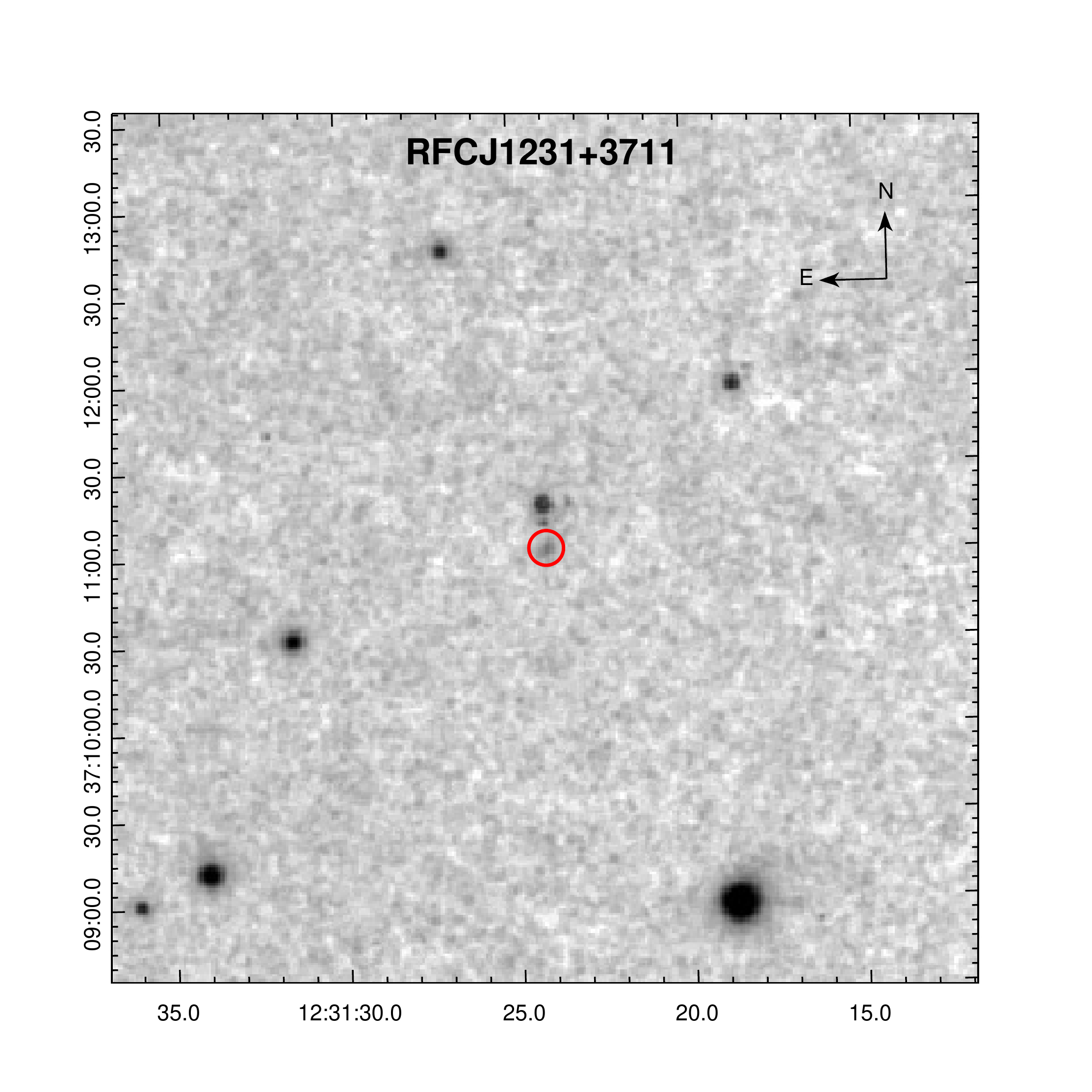} \\
\end{array}$
\end{center}
\caption{(Left panel) Optical spectrum of RFC J1231+3711 associated with FL8Y J1231.1+3711, in the upper part it is shown the Signal-to-Noise Ratio of the spectrum. (Right panel) The finding chart ( $5'\times 5'$ ) retrieved from the Digitized Sky Survey (DSS) highlighting the location of the counterpart:   RFC J1231+3711 (red circle).}
\label{fig:J1231}
\end{figure*}

\begin{figure*}{}
\begin{center}$
\begin{array}{cc}
\includegraphics[width=\mywidth]{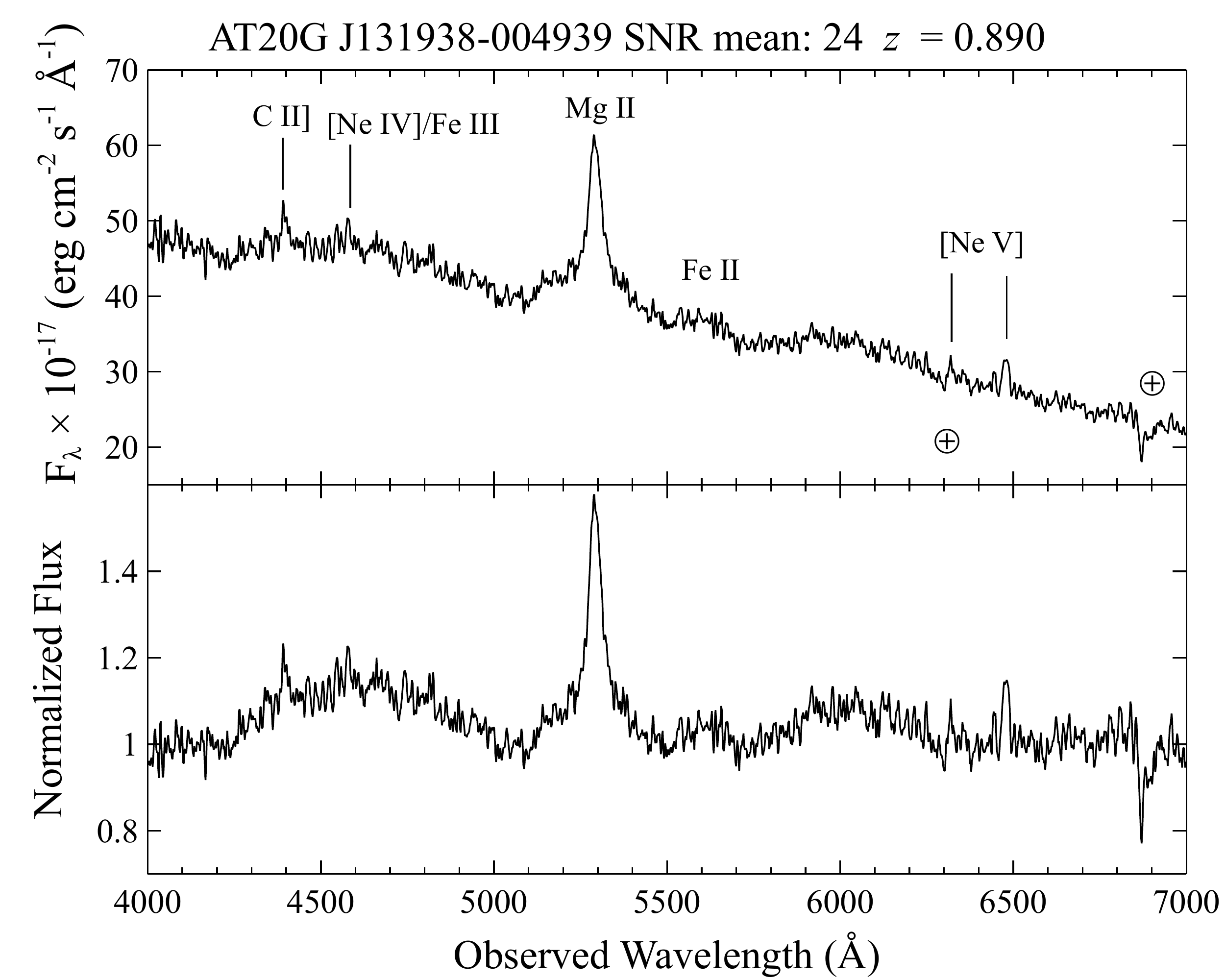} &
\includegraphics[clip=true, width=7cm]{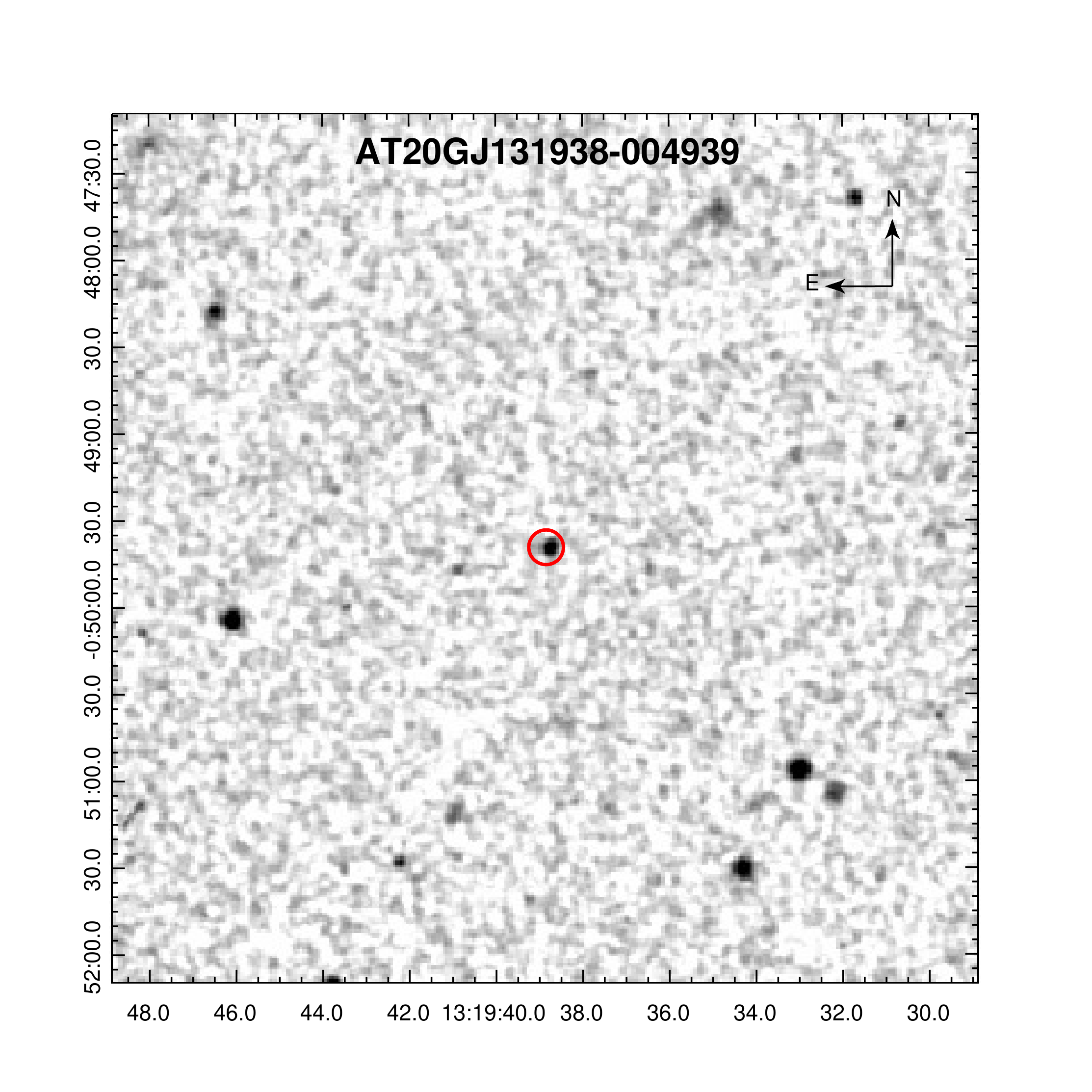} \\
\end{array}$
\end{center}
\caption{(Left panel) Optical spectrum of AT20G J131938-004939 potential candidate of the UGS with FL8Y J1319.5-0046, in the upper part it is shown the Signal-to-Noise Ratio of the spectrum. (Right panel) The finding chart ( $5'\times 5'$ ) retrieved from the Digitized Sky Survey (DSS) highlighting the location of the potential counterpart:  AT20G J131938-004939 (red circle).}
\label{fig:J1319}
\end{figure*}

\begin{figure*}{}
\begin{center}$
\begin{array}{cc}
\includegraphics[width=\mywidth]{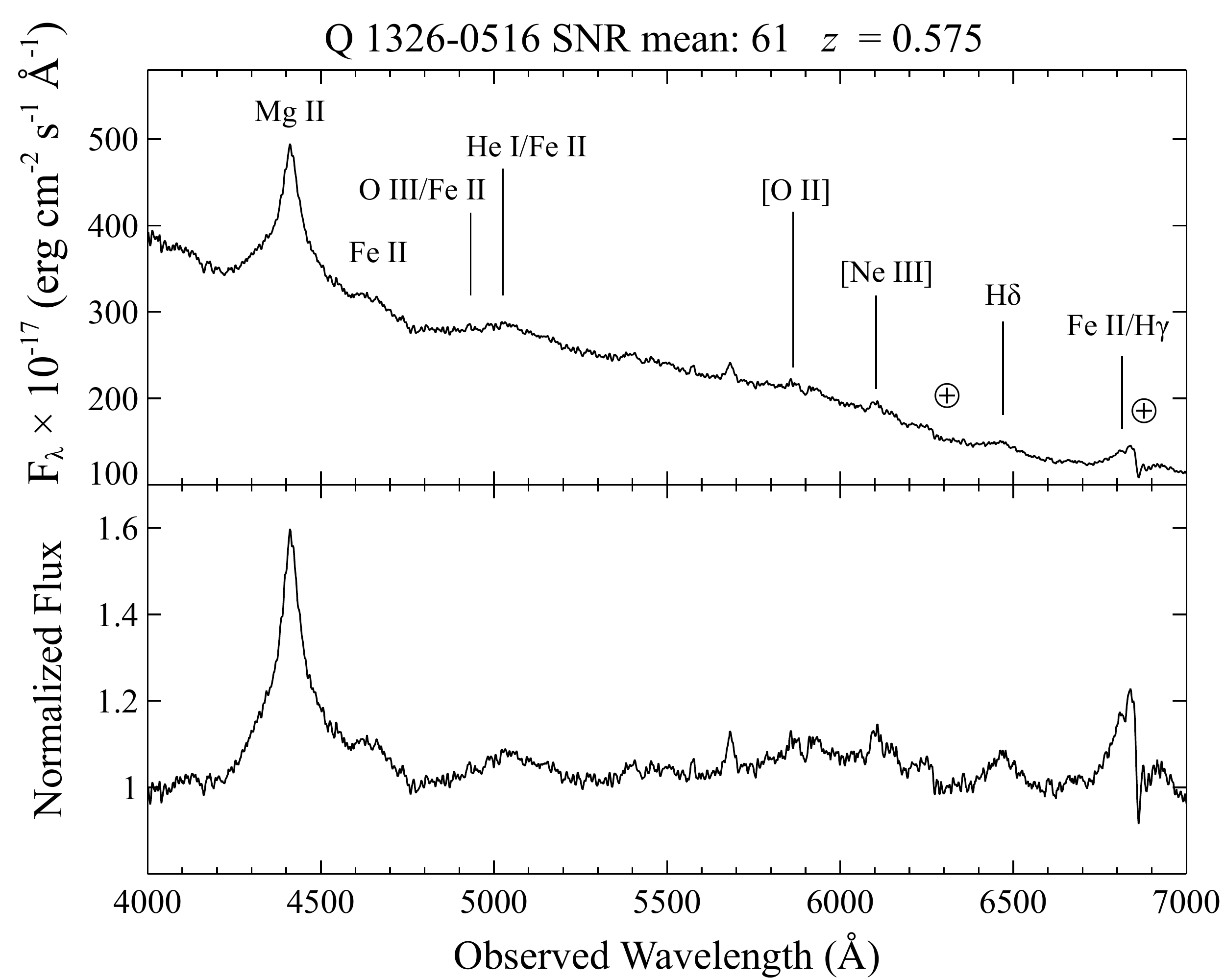} &
\includegraphics[clip=true, width=7cm]{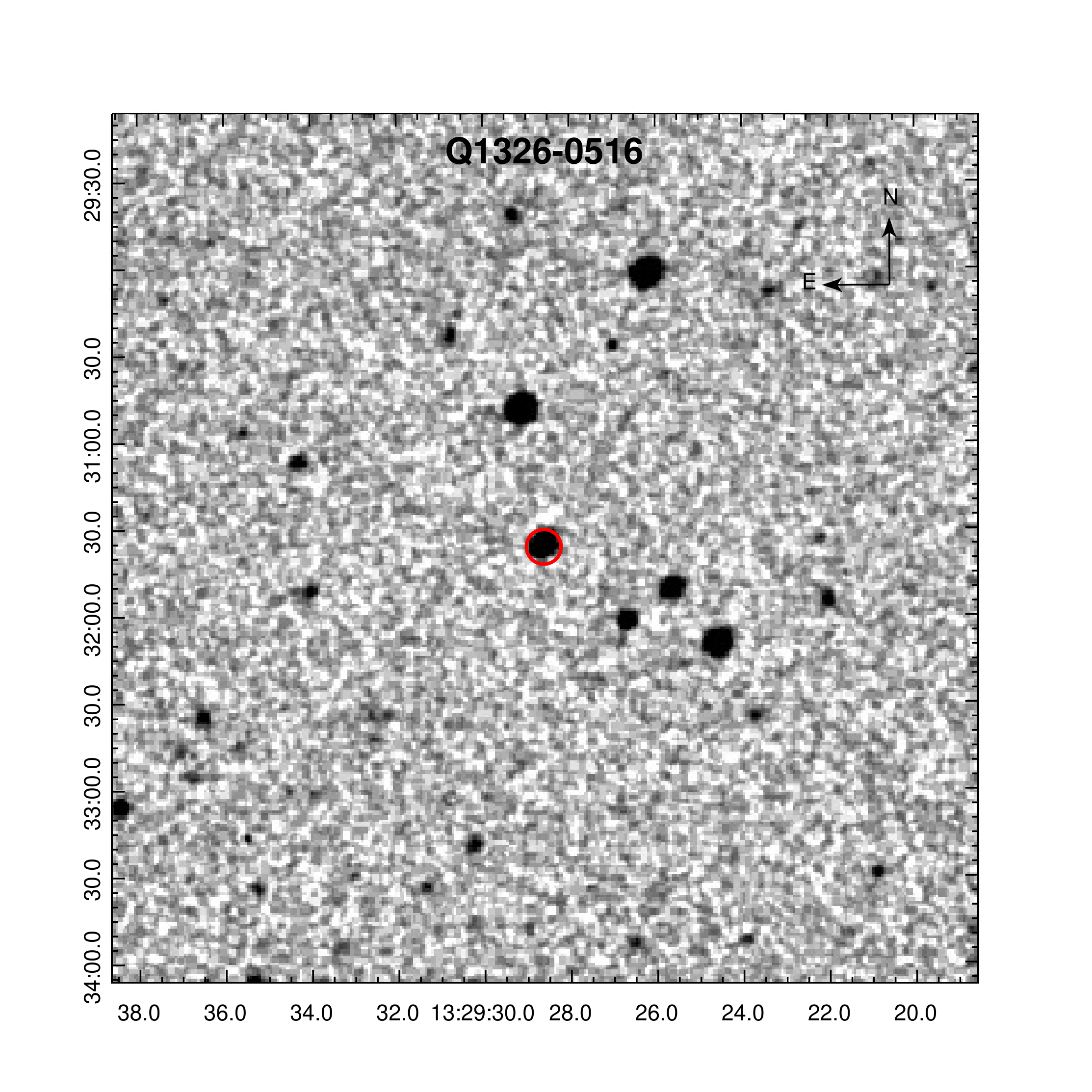} \\
\end{array}$
\end{center}
\caption{(Left panel) Optical spectrum of Q 1326-0516 associated with FL8Y J1329.4-0530, in the upper part it is shown the Signal-to-Noise Ratio of the spectrum. (Right panel) The finding chart ( $5'\times 5'$ ) retrieved from the Digitized Sky Survey (DSS) highlighting the location of the counterpart:  Q 1326-0516 (red circle).}
\label{fig:Q1326}
\end{figure*}

\begin{figure*}{}
\begin{center}$
\begin{array}{cc}
\includegraphics[width=\mywidth]{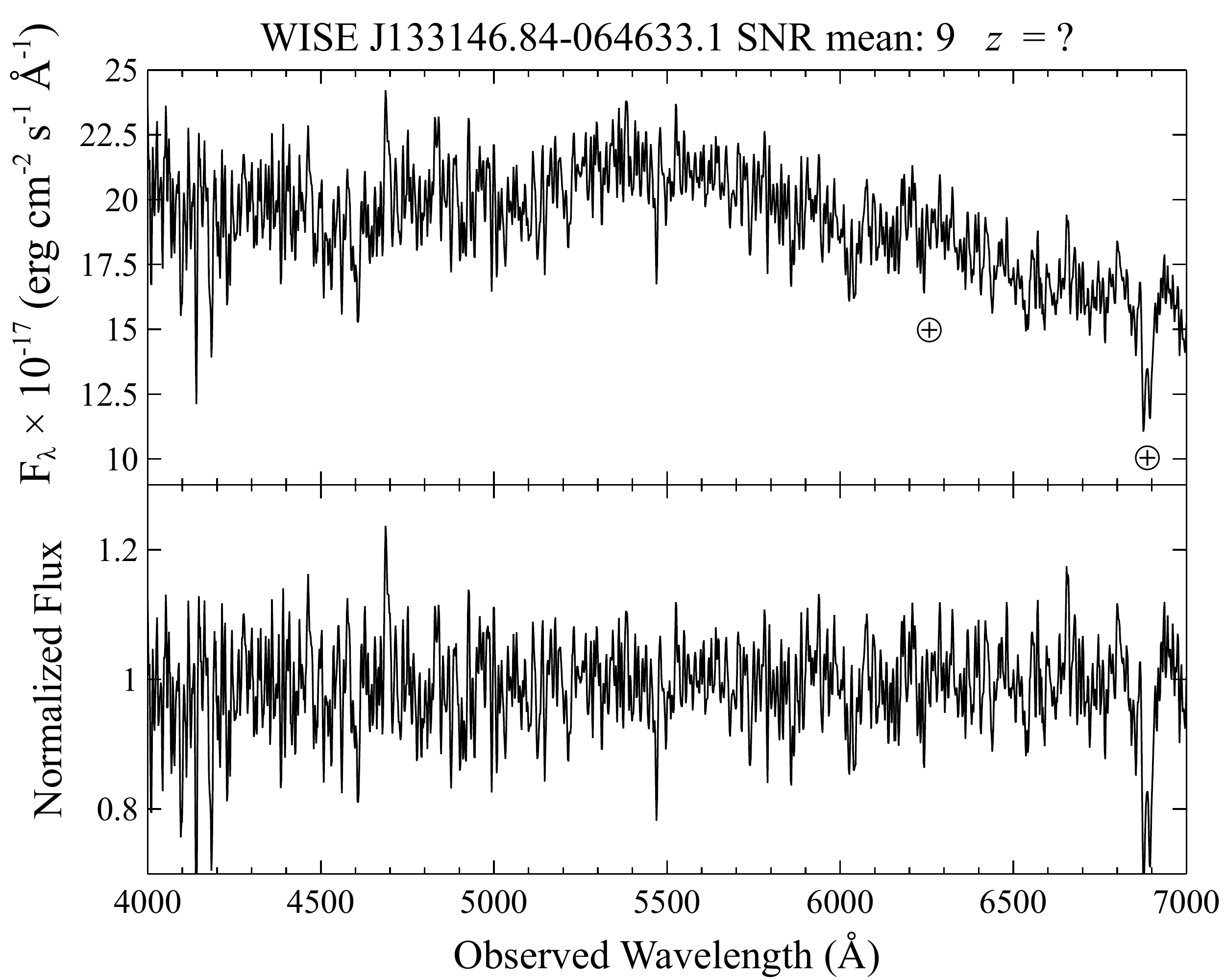} &
\includegraphics[clip=true, width=7cm]{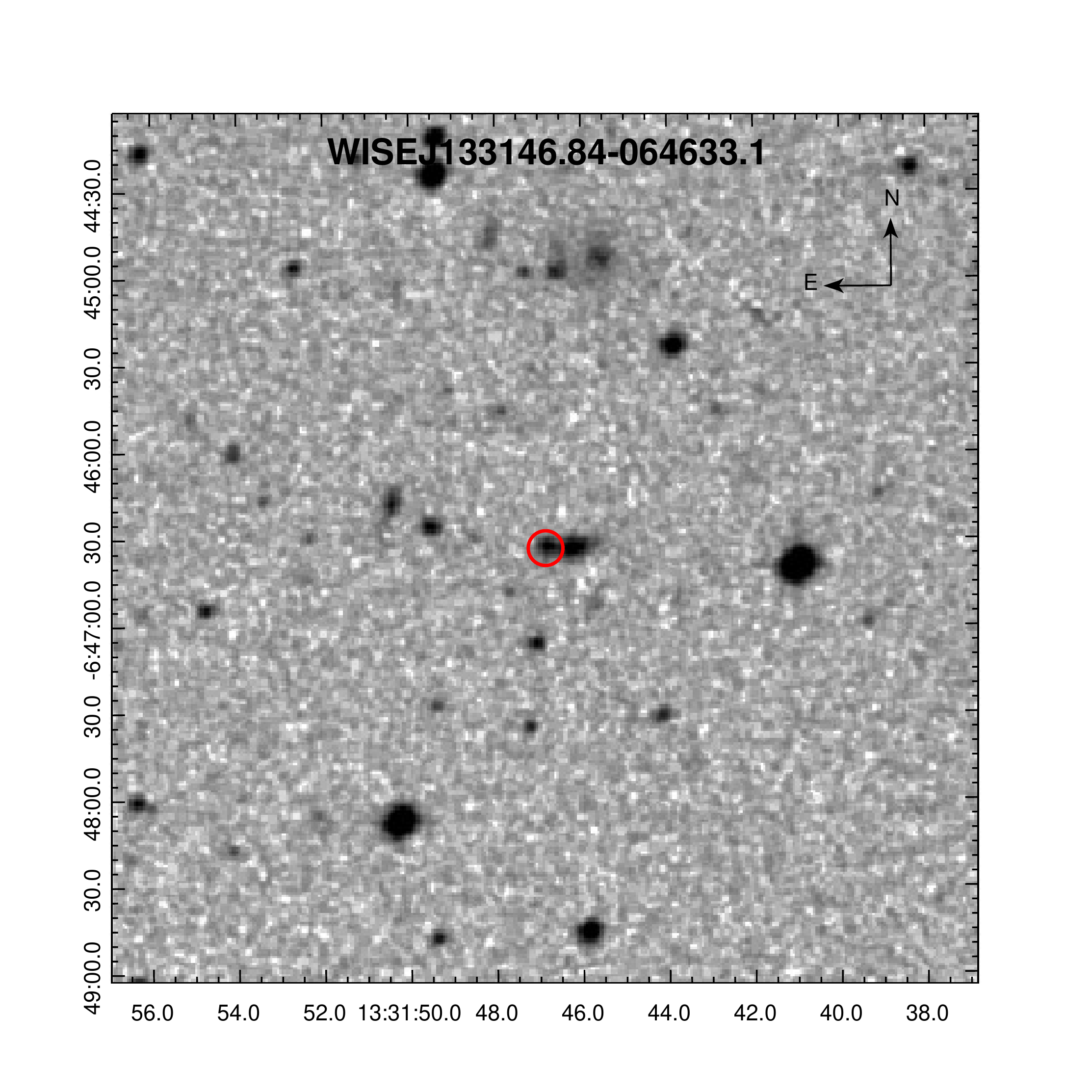} \\
\end{array}$
\end{center}
\caption{(Left panel) Optical spectrum of WISE J133146.84-064633.1 associated with FL8Y J1331.7-0647, in the upper part it is shown the Signal-to-Noise Ratio of the spectrum. (Right panel) The finding chart ( $5'\times 5'$ ) retrieved from the Digitized Sky Survey (DSS) highlighting the location of the counterpart:  WISE J133146.84-064633.1 (red circle).}
\label{fig:J13314}
\end{figure*}

\begin{figure*}{}
\begin{center}$
\begin{array}{cc}
\includegraphics[width=\mywidth]{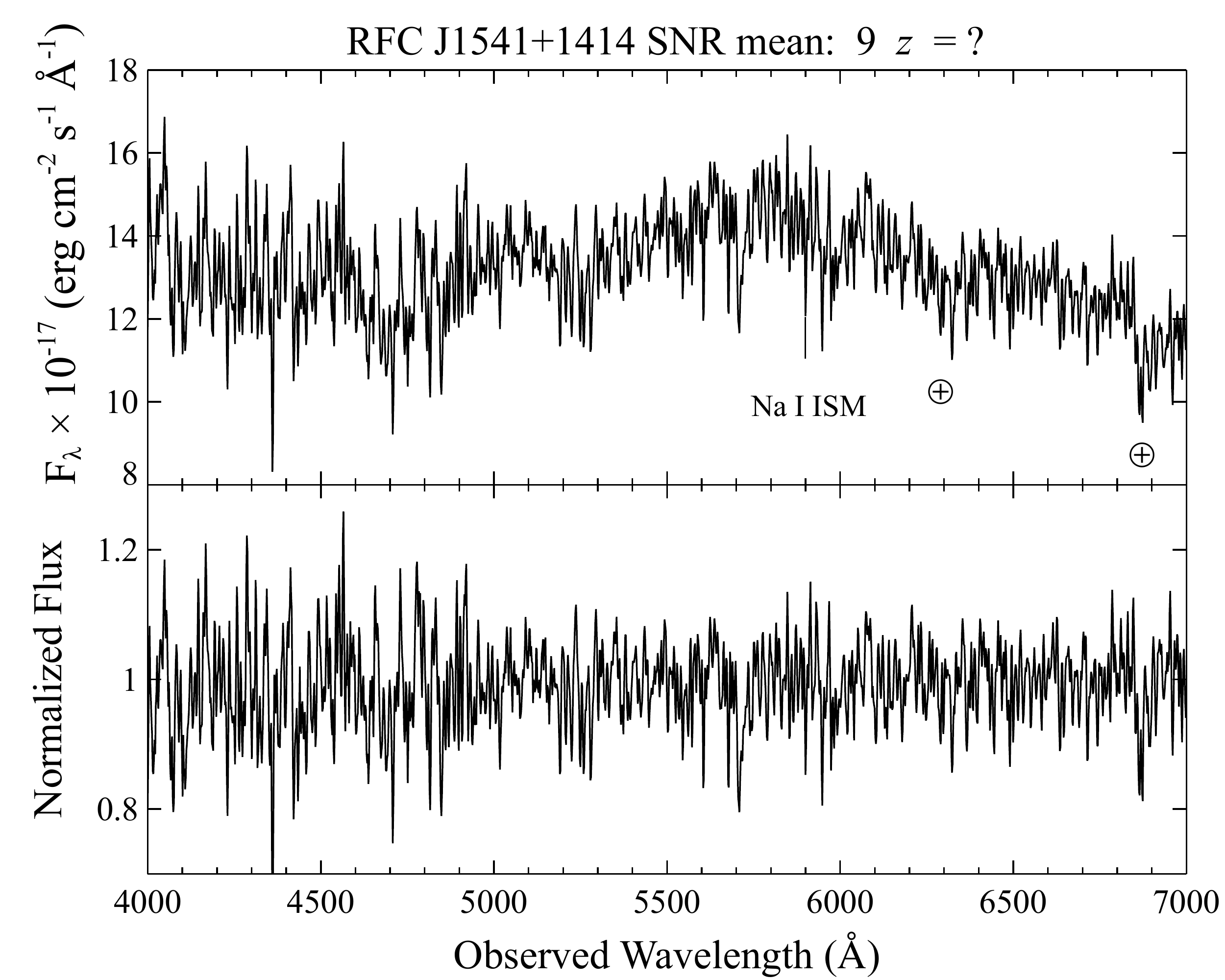} &
\includegraphics[clip=true, width=7cm]{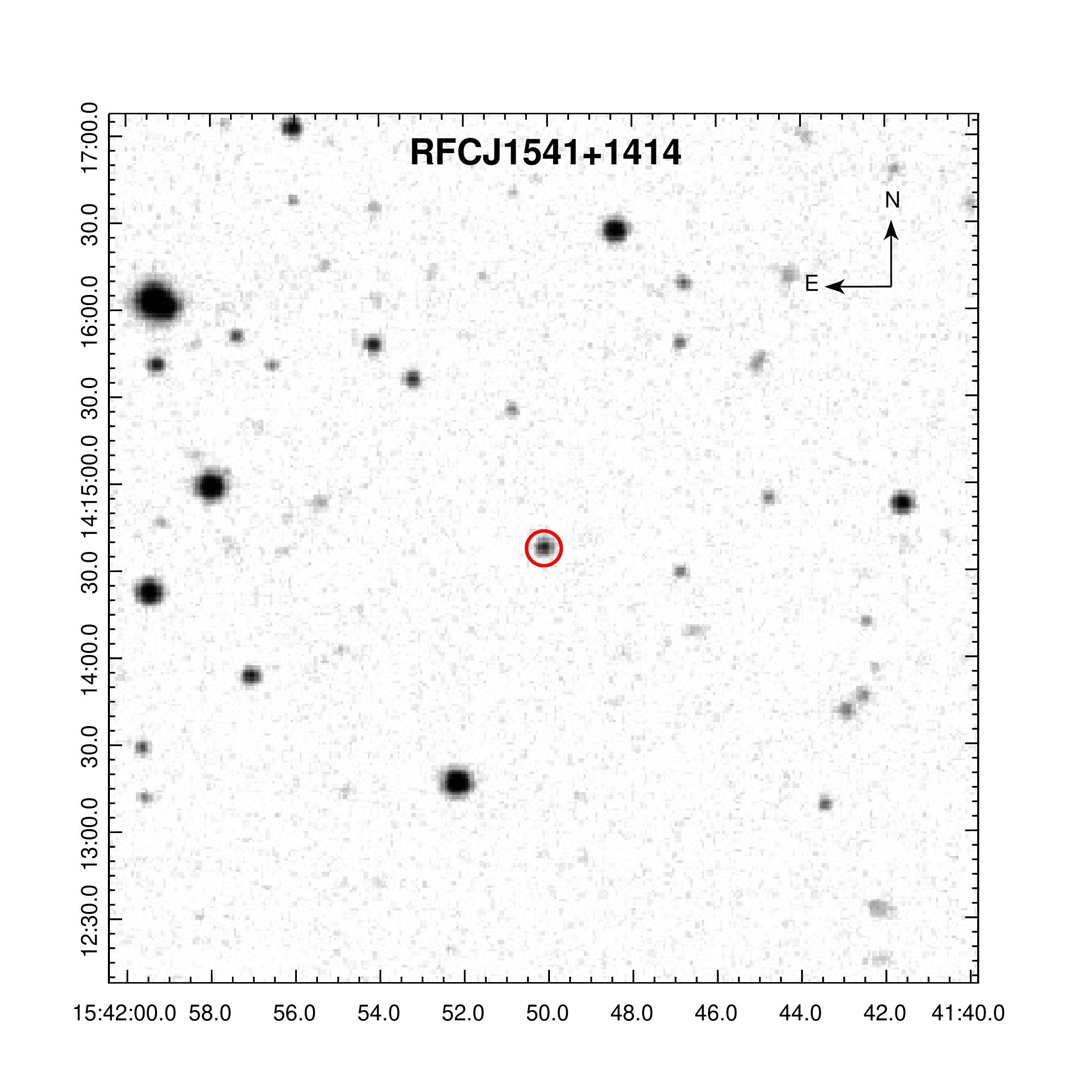} \\
\end{array}$
\end{center}
\caption{(Left panel) Optical spectrum of RFC J1541+1414 associated with FL8Y J1541.7+1413, in the upper part it is shown the Signal-to-Noise Ratio of the spectrum. (Right panel) The finding chart ( $5'\times 5'$ ) retrieved from the Digitized Sky Survey (DSS) highlighting the location of the counterpart: RFC J1541+1414 (red circle).}
\label{fig:J1541}
\end{figure*}

\begin{figure*}{}
\begin{center}$
\begin{array}{cc}
\includegraphics[width=\mywidth]{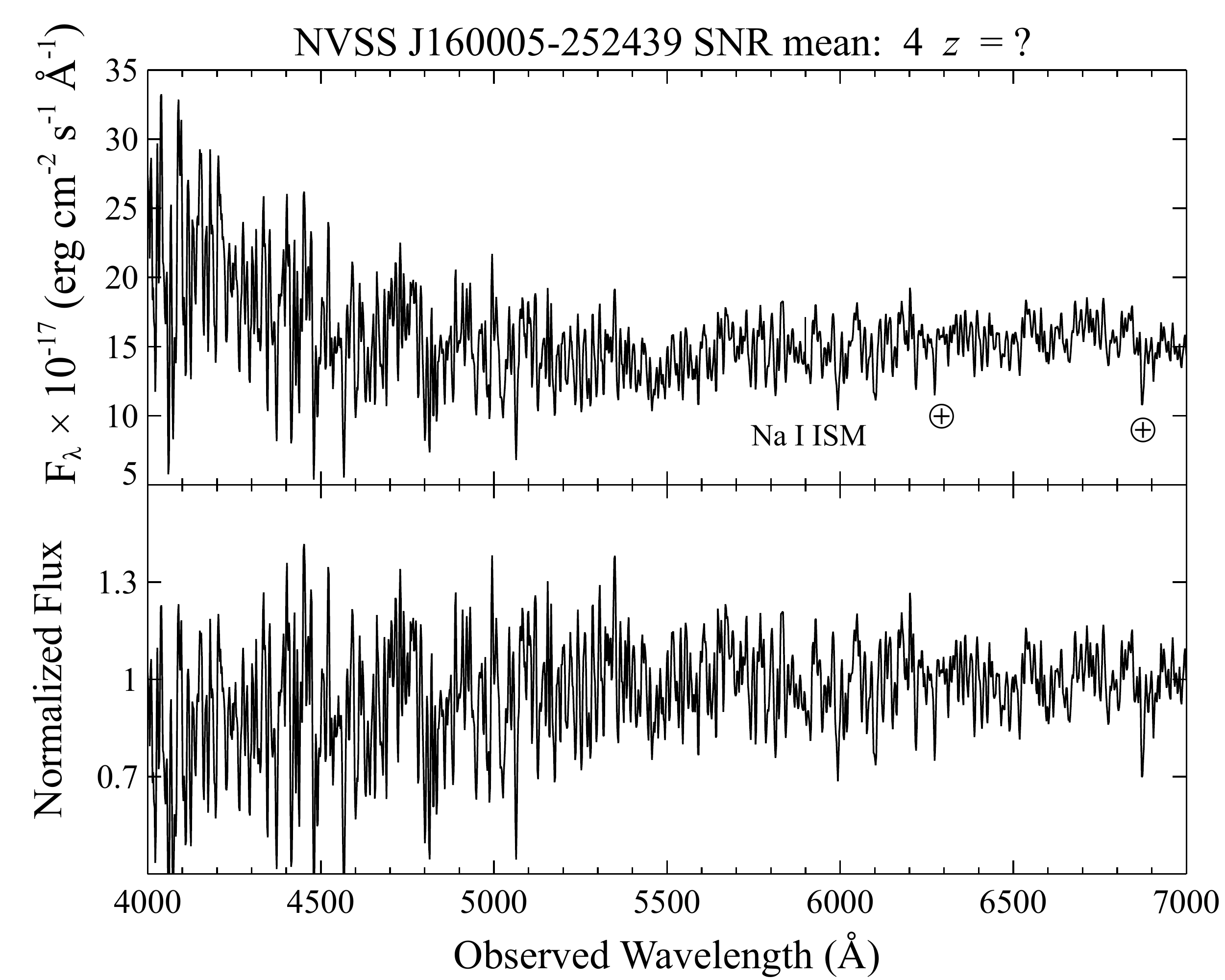} &
\includegraphics[clip=true, width=7cm]{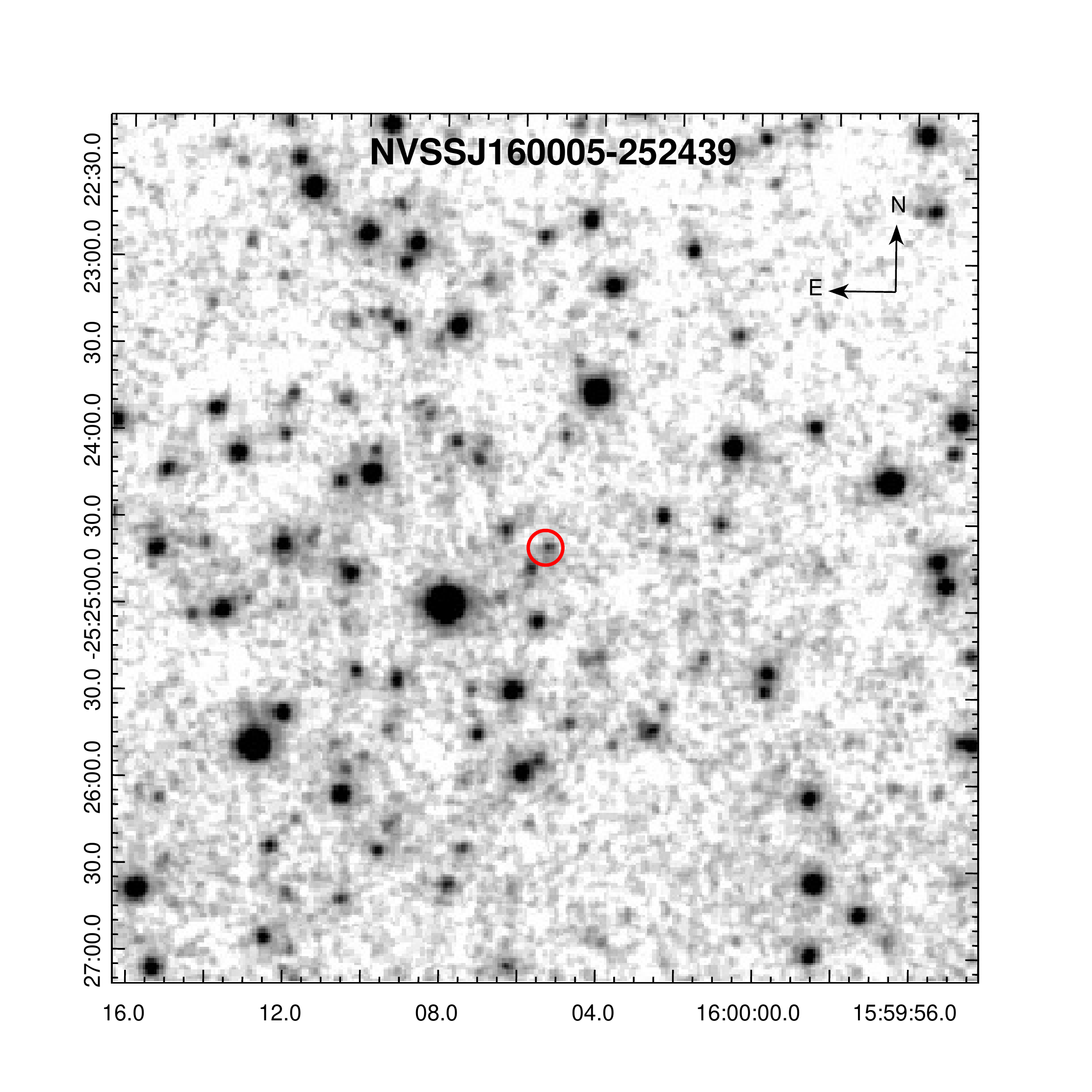} \\
\end{array}$
\end{center}
\caption{(Left panel) Optical spectrum of NVSS J160005-252439 associated with FL8Y J1559.8-2525, in the upper part it is shown the Signal-to-Noise Ratio of the spectrum. (Right panel) The finding chart ( $5'\times 5'$ ) retrieved from the Digitized Sky Survey (DSS) highlighting the location of the counterpart: NVSS J160005-252439 (red circle).}
\label{fig:J1600-2524}
\end{figure*}

\begin{figure*}{}
\begin{center}$
\begin{array}{cc}
\includegraphics[width=\mywidth]{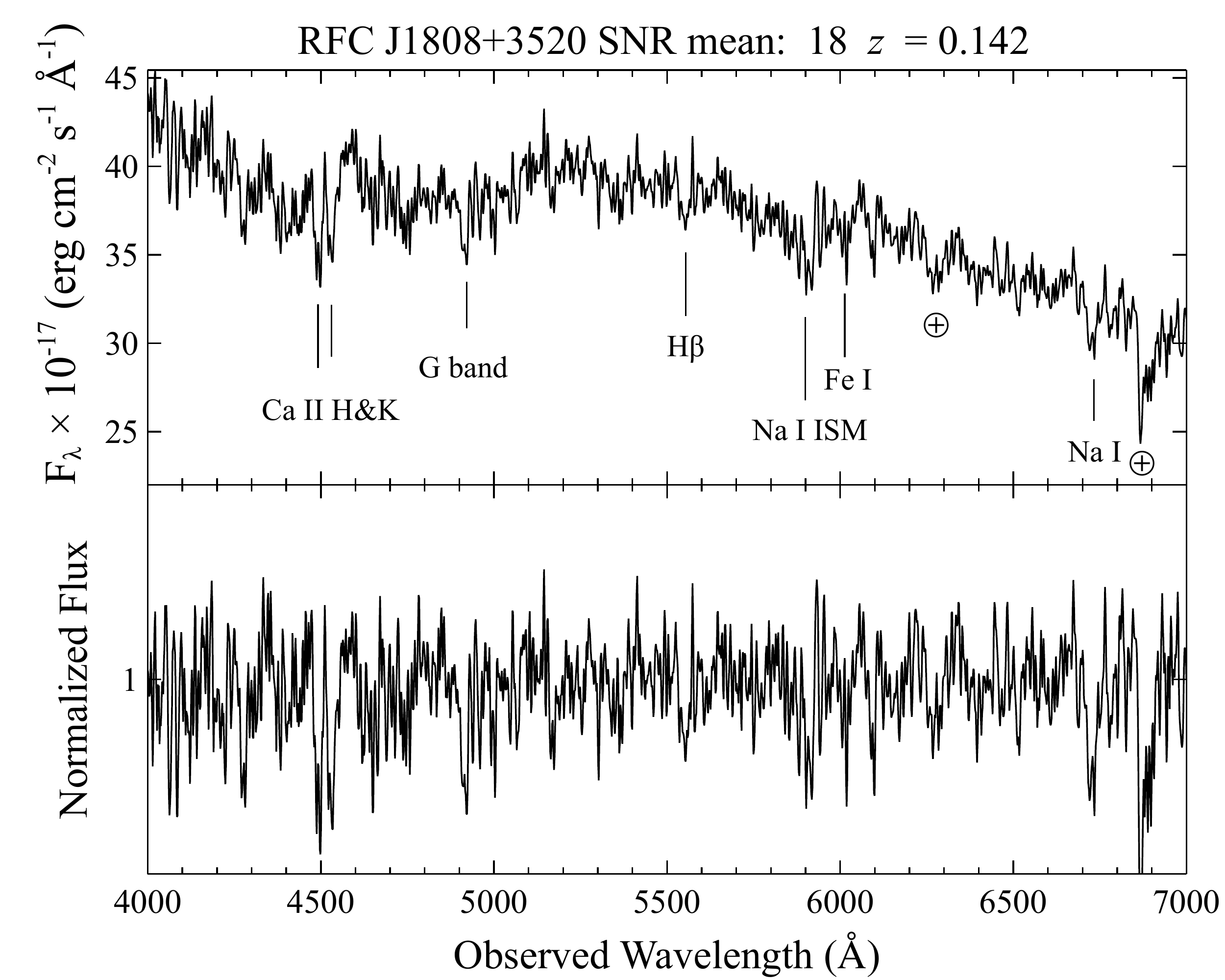} &
\includegraphics[clip=true, width=7cm]{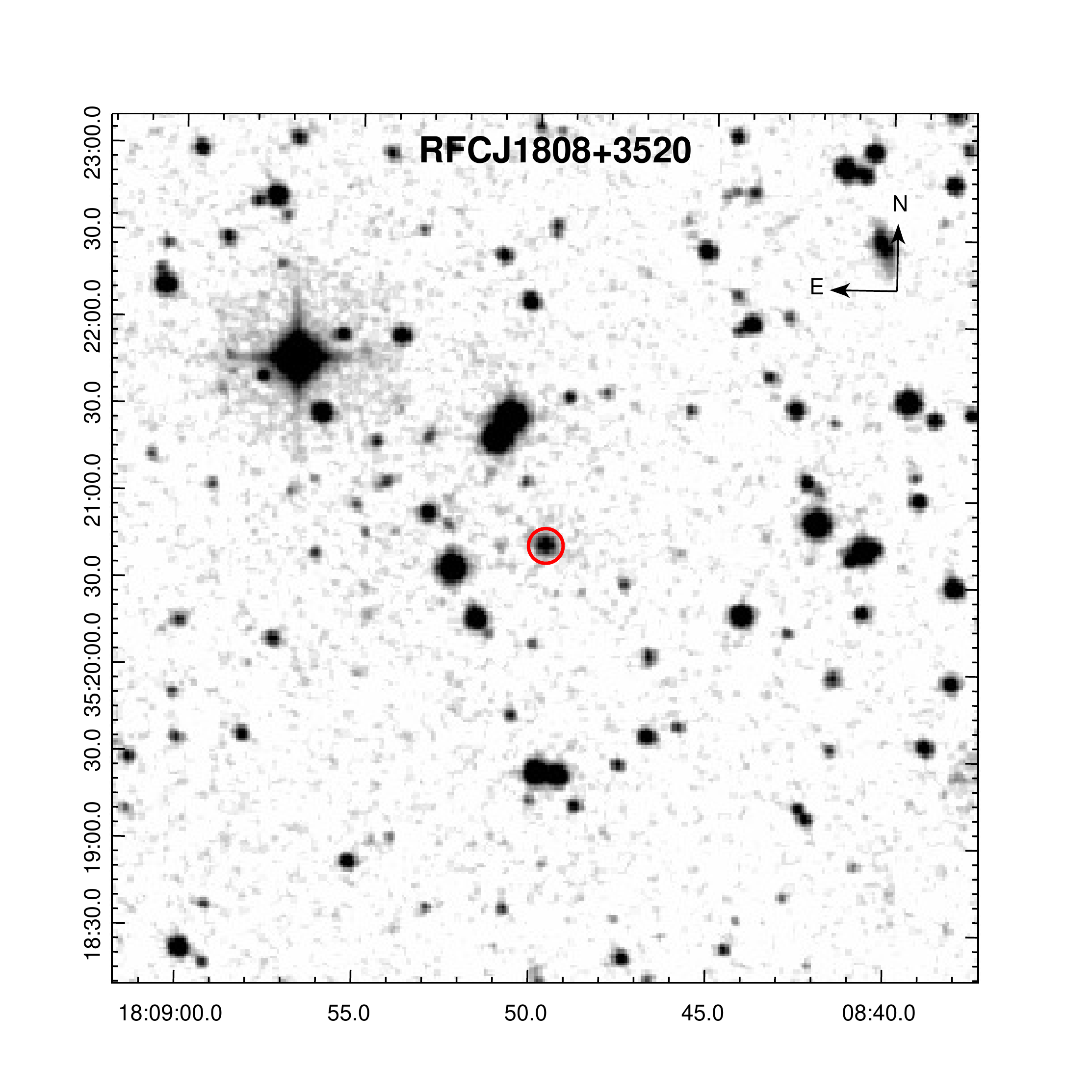} \\
\end{array}$
\end{center}
\caption{(Left panel) Optical spectrum of RFC J1808+3520 associated with FL8Y J1808.9+3522, in the upper part it is shown the Signal-to-Noise Ratio of the spectrum. (Right panel) The finding chart ( $5'\times 5'$ ) retrieved from the Digitized Sky Survey (DSS) highlighting the location of the counterpart: RFC J1808+3520 (red circle).}
\label{fig:J1808}
\end{figure*}

\begin{figure*}{}
\begin{center}$
\begin{array}{cc}
\includegraphics[width=\mywidth]{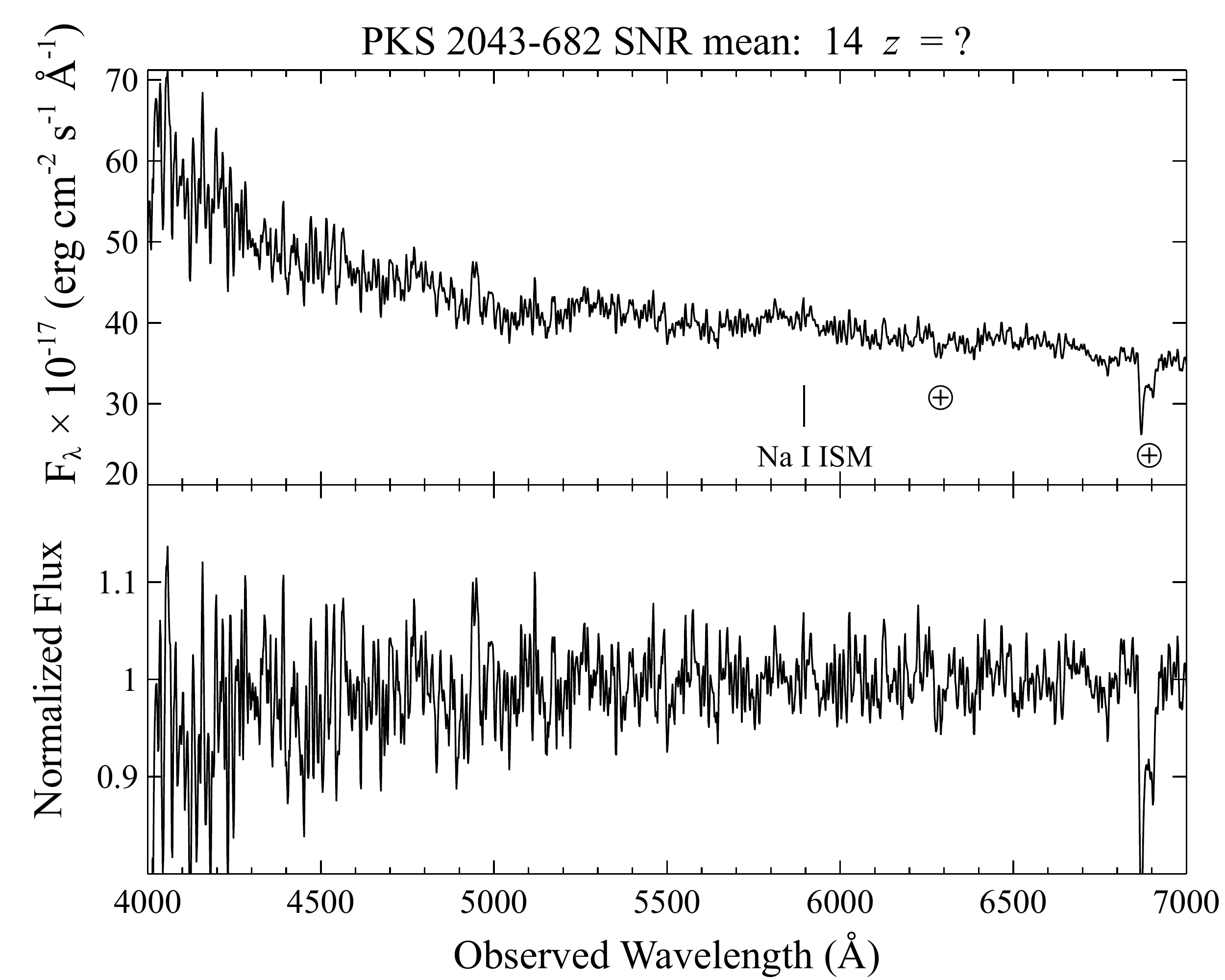} &
\includegraphics[clip=true, width=7cm]{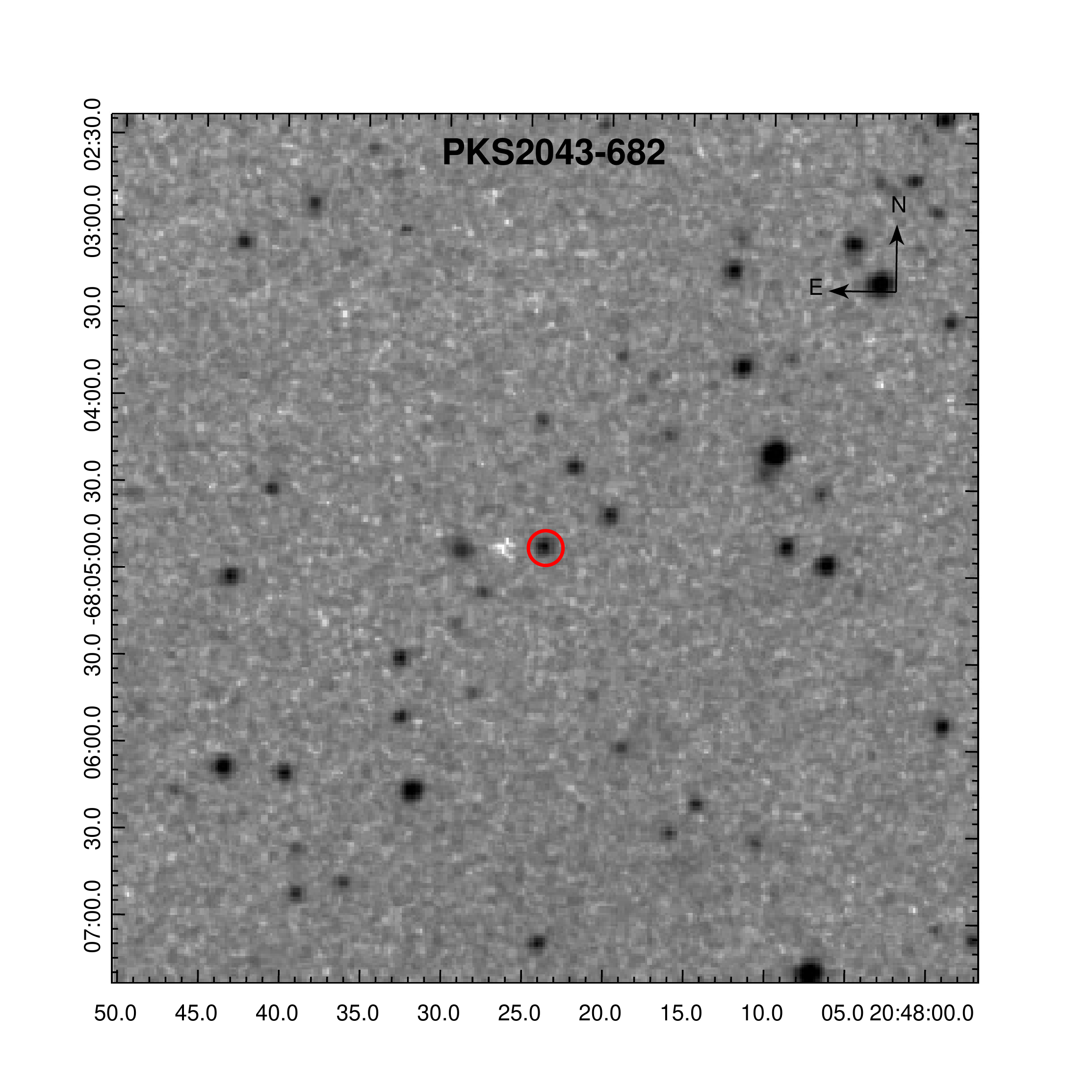} \\
\end{array}$
\end{center}
\caption{(Left panel) Optical spectrum of PKS 2043-682 associated with FL8Y J2048.6-6804, in the upper part it is shown the Signal-to-Noise Ratio of the spectrum. (Right panel) The finding chart ( $5'\times 5'$ ) retrieved from the Digitized Sky Survey (DSS) highlighting the location of the counterpart: PKS 2043-682 (red circle).}
\label{fig:PKS2043}
\end{figure*}

\begin{figure*}{}
\begin{center}$
\begin{array}{cc}
\includegraphics[width=\mywidth]{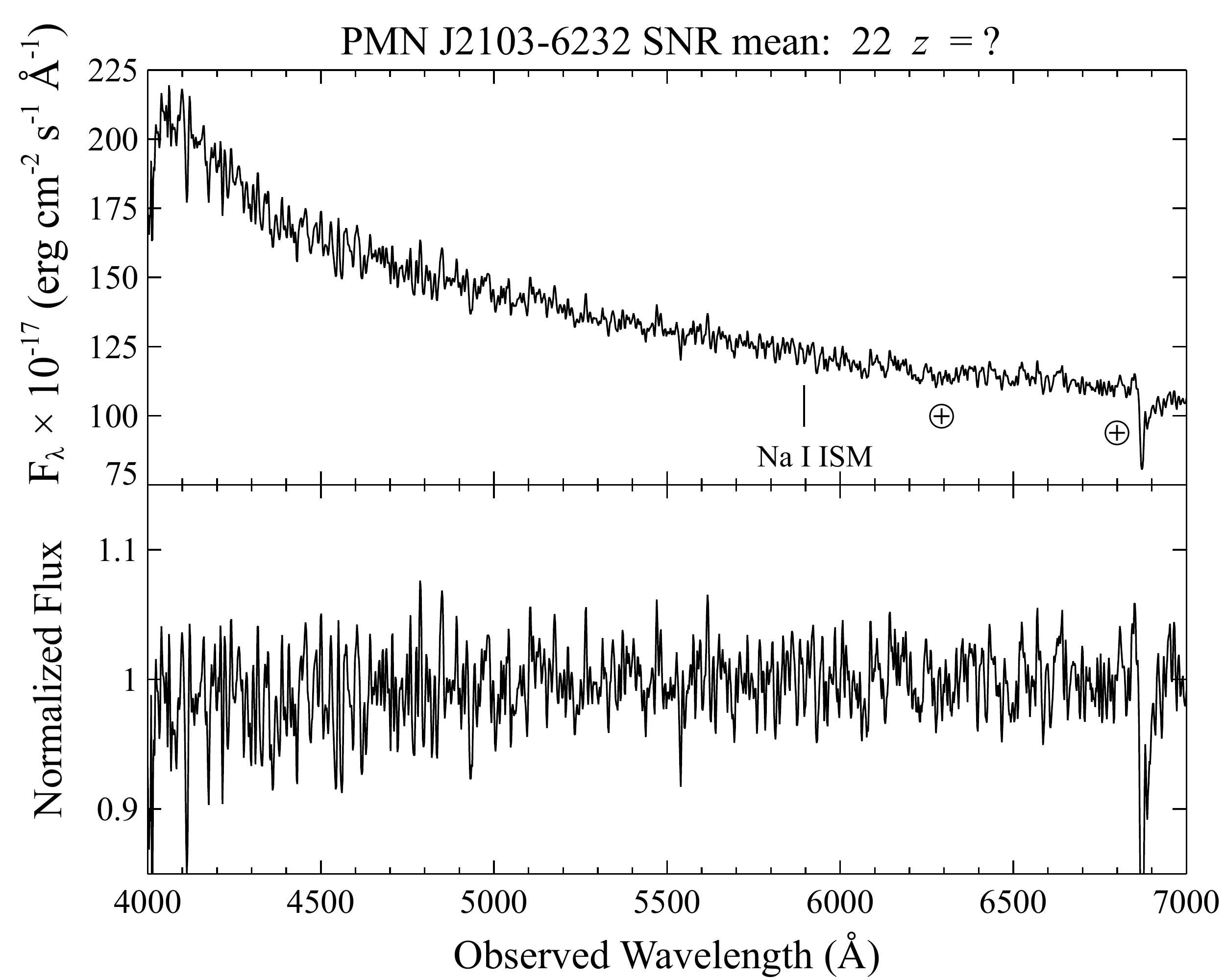} &
\includegraphics[clip=true, width=7cm]{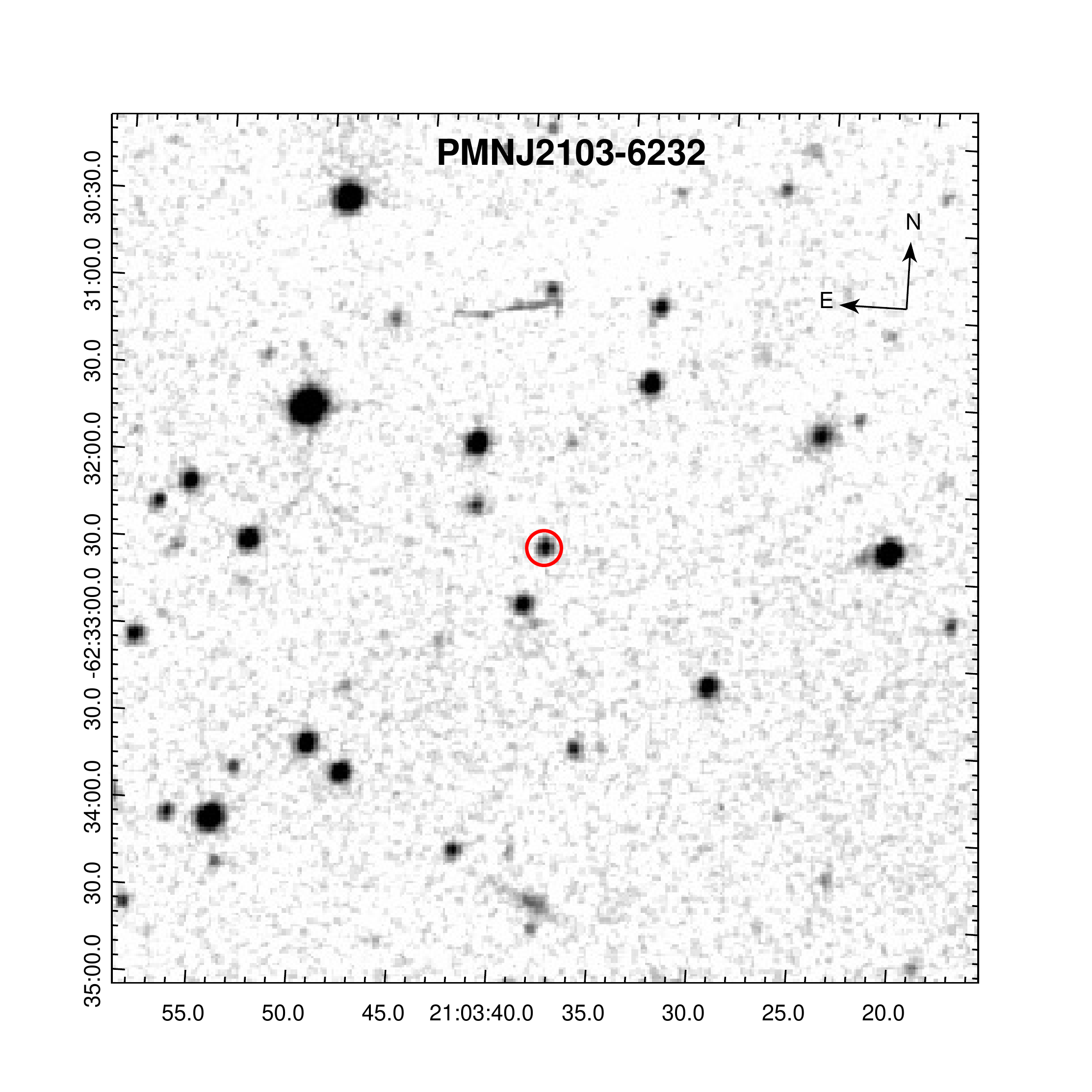} \\
\end{array}$
\end{center}
\caption{(Left panel) Optical spectrum of PMN J2103-6232 associated with FL8Y J2103.8-6233, in the upper part it is shown the Signal-to-Noise Ratio of the spectrum. (Right panel) The finding chart ( $5'\times 5'$ ) retrieved from the Digitized Sky Survey (DSS) highlighting the location of the counterpart: PMN J2103-6232 (red circle).}
\label{fig:PMNJ2103}
\end{figure*}

\begin{figure*}{}
\begin{center}$
\begin{array}{cc}
\includegraphics[width=\mywidth]{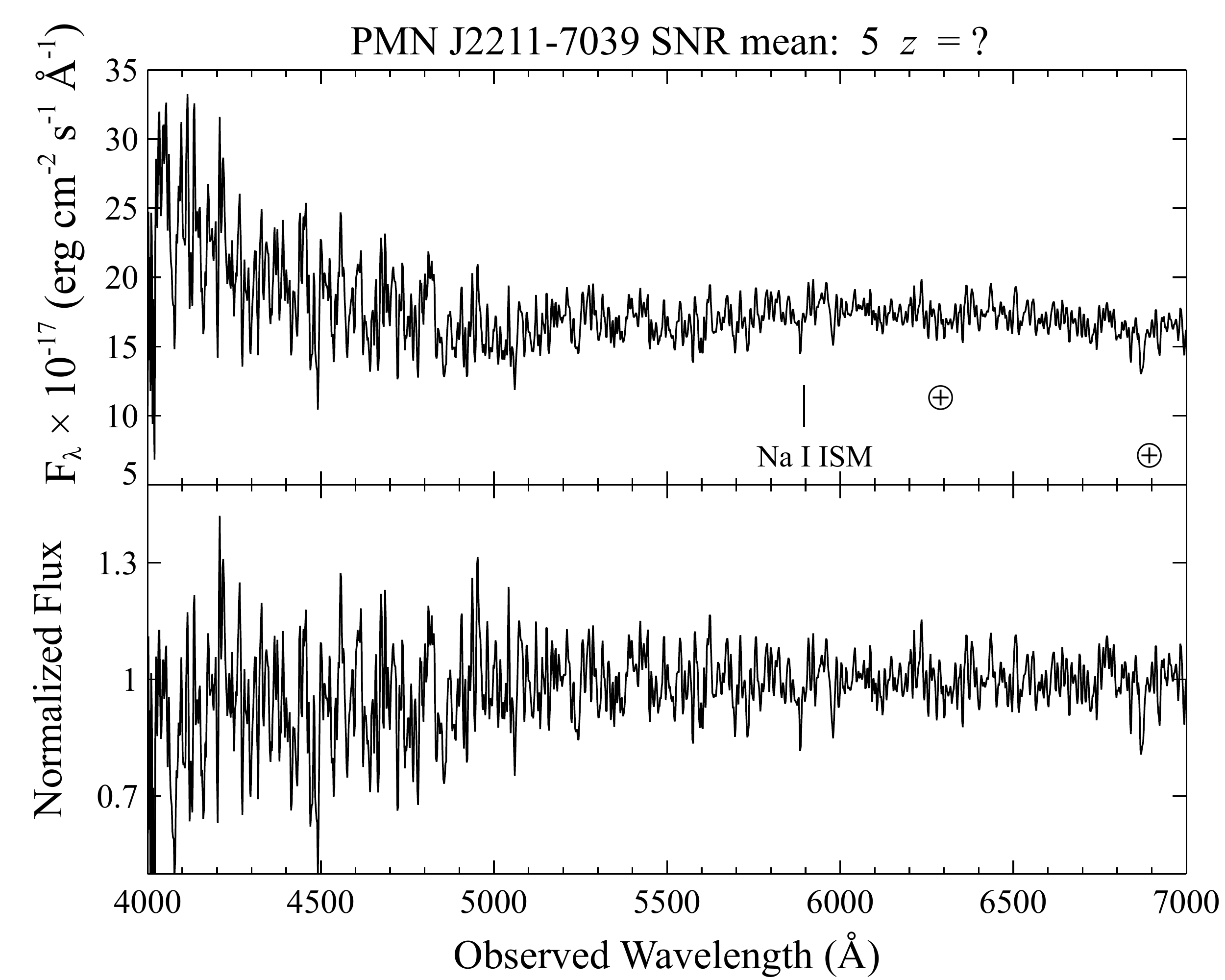} &
\includegraphics[clip=true, width=7cm]{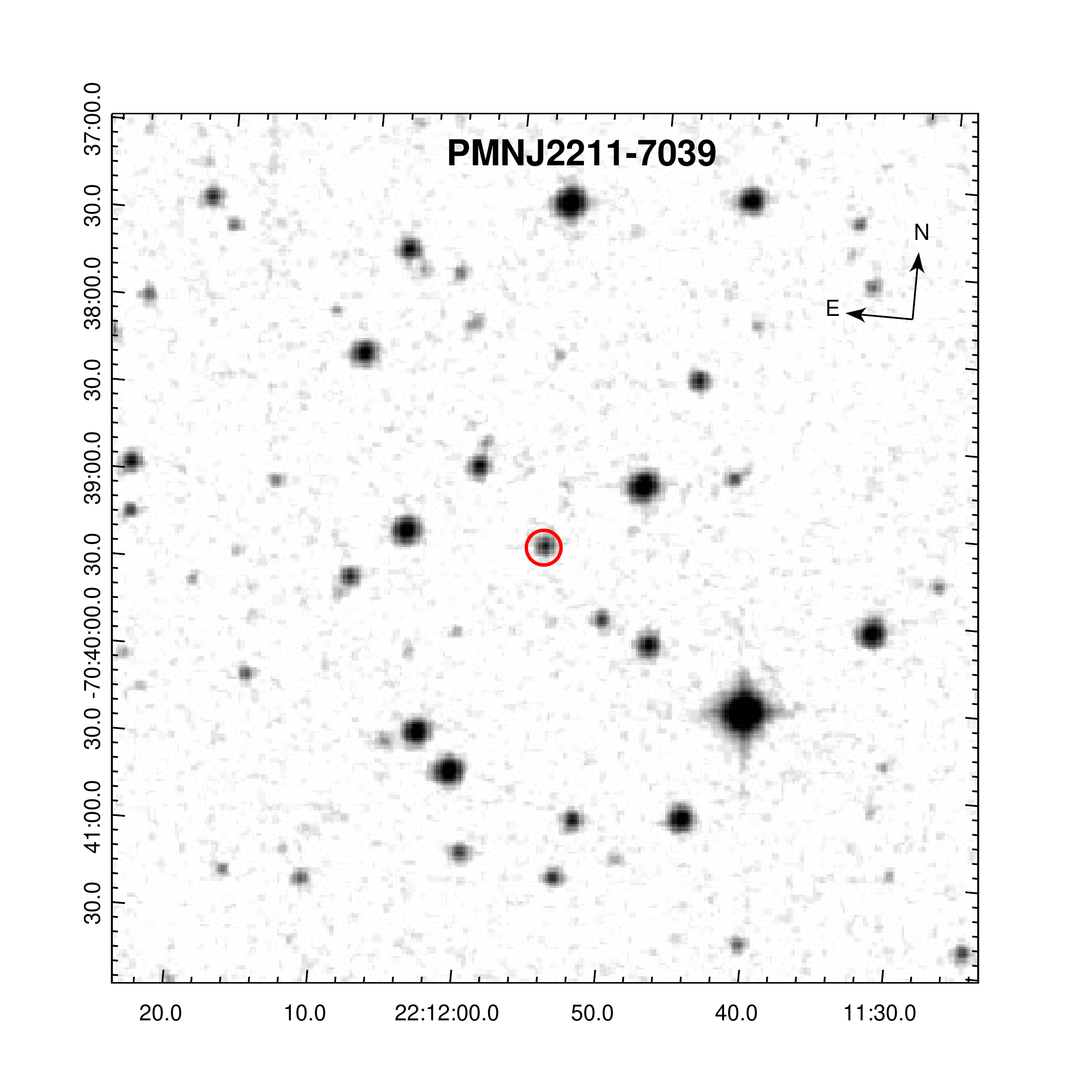} \\
\end{array}$
\end{center}
\caption{(Left panel) Optical spectrum of PMN J2211-7039 associated with 3FGL J2212.3-7039, in the upper part it is shown the Signal-to-Noise Ratio of the spectrum. (Right panel) The finding chart ( $5'\times 5'$ ) retrieved from the Digitized Sky Survey (DSS) highlighting the location of the counterpart: PMN J2211-7039 (red circle).}
\label{fig:PMNJ2211}
\end{figure*}

\end{document}